# Improving intermolecular interactions in DFTB3 using extended polarization from chemical-potential equalization

Anders S. Christensen,*,† Marcus Elstner,‡ and Qiang Cui*,†

*University of Wisonsin-Madison, Department of Chemsitry, 1101 University Ave, Madison, WI 53706, USA, and Universität Karlsruhe, Theoretische Chemische Biologie, Kaiserstr. 12, 76131 Karlsruhe, Germany*

E-mail: andersx@chem.wisc.edu; cui@chem.wisc.edu

**Abstract**

Semi-empirical quantum mechanical methods traditionally expand the electron density in a minimal, valence-only electron basis set. The minimal-basis approximation causes molecular polarization to be underestimated, and hence intermolecular interaction energies are also underestimated, especially for intermolecular interactions involving charged species. In this work, the third-order self-consistent charge density functional tight-binding method (DFTB3) is augmented with an auxiliary response density using the chemical-potential equalization (CPE) method and an empirical dispersion correction (D3). The parameters in the CPE and D3 models are fitted to high-level CCSD(T) reference interaction energies for a broad range of chemical species, as well as dipole moments calculated at the DFT level; the impact of including polarizabilities of molecules in the parameterization is also considered. Parameters for the elements H, C, N, O and S are presented. The RMSD interaction energy is improved from 6.07 kcal/mol to 1.49 kcal/mol for interactions with one charged specie, whereas the RMSD is improved from 5.60 kcal/mol to 1.73 for a set of 9 salt bridges, compared to uncorrected DFTB3. For large water clusters and complexes that are dominated by dispersion interactions, the already satisfactory performance of the DFTB3-D3 model is retained; polarizabilities of neutral molecules are also notably improved. Overall, the CPE extension of DFTB3-D3 provides a more balanced description of different types of non-covalent interactions than NDDO type of semi-empirical methods (e.g., PM6-D3H4) and PBE-D3 with modest basis sets.

*To whom correspondence should be addressed
†University of Wisonsin-Madison
‡Universität Karlsruhe

# Introduction

Semi-empirical (SE) quantum mechanical (QM) methods have enabled QM to be used where *ab initio* methods are too computationally expensive. Conceptually, the SE methods are approximations to *ab initio* QM methods, but introduce parameters that must be fitted empirically based on either *ab initio* or experimental data. SE methods has been discussed and benchmarked thoroughly, as most recently reviewed in Refs. 1–5.

In the NDDO/MNDO-based methods, the formalism is derived from Hartree-Fock theory, but with several approximations in both the matrix algebra and integral calculation.[6,7] In the density functional tight-binding (DFTB) methods,[8,9] the formalism is derived from a Taylor expansion of the DFT energy in terms of density fluctuation with respect to a reference, and the matrix elements are calculated from first-principles DFT.[10,11] The basic DFTB method has recently been expanded to second- and third-order monopole charge expansions of the density fluctuation, leading to DFTB2 (referred to also as SCC-DFTB), and DFTB3, respectively.[12,13]

Both the NDDO/MNDO and DFTB methods discard three- and four-center electron-repulsion integrals, and hence the computational bottleneck of these methods lies in solving a set of secular equations. Traditionally the methods employ minimal, valence electron-only basis sets to keep the computational cost at a minimum. One downside to the use of a minimal basis set is that intermolecular polarization is underestimated by about 25%, which leads to poor accuracy for the description of intermolecular forces, especially for polar interactions, such as hydrogen bonds and interactions to ionic groups.[14] Furthermore, the minimal basis also limits the accuracy of computed Pauli-repulsion within the current DFTB framework.[13]

This well-known problem has recently led to a plethora of hydrogen-bond corrections to be added as post-SCF terms to the electronic energy calculated by the SE methods. Within the last decade the H, H2, H2X, H+ H4, and H4X hydrogen bond corrections have been published.[1,15–20] While such mechanical post-SCF corrections greatly increase the accuracy of SE methods, they do not directly address the fundamental problem at the QM level, as they do not alter the electron density at all. Therefore, the transferability of these corrections is likely limited (see discussions below).

Several improvements for the DFTB3 method are currently under development. In the context of non-covalent interactions, these involve extending the monopole expansion to include dipoles and quadrupoles in the density fluctuation,[21] as well as improvements of the description of Pauli repulsion. This work addresses the low accuracy of DFTB3 for intermolecular interactions that implicate highly polarizable moieties, while keeping the increase in computational cost at a minimum. Such effort is important for the enhancement of not only the accuracy of DFTB3 but also its transferability to the analysis of chemical events in different environments.[22,23]

It has previously been proposed to lift the rudimentary restriction of minimal, valence-only basis sets in several NDDO/MNDO-based SE methods. For instance, the minimal-basis SINDO1 method[24] was augmented with *p*-function on hydrogen atoms, which greatly improved the accuracy of dimerization energies and hydrogen bonding geometries.[25] Likewise, the MNDO/d and PM6 methods[26,27] add *d*-functions to several main-group and transition metals, for more accurate descriptions of these elements. More recently, the PMO and PMO2 methods[28–30] were developed with a focus on molecular polarizabilities. As observed for the SINDO1 method, the PMO polarization functions on hydrogen atoms increase the accuracy of predicted hydrogen bonding geometries,[29] and additionally reduces the PMO2 error in predicted molecular polarizabilities by about 80%.[30] Similarly to the NDDO/MNDO based SE methods, the tight-binding based DFTB2 method has also been previously augmented with *p*-orbitals on hydrogen with similar improvements in energetics and geometries.[31,32]

For biological molecules, the addition of polarization function to hydrogen atoms would increase the size of the DFTB Hamiltonian and overlap matrices by about a factor of two. Since the diagonalization step formally scales as $\mathcal{O}(N^3)$, the valence polarized method would be eight times slower, which is not desirable for simulation studies.

An alternative approach to increasing the size of the basis set was explored in the SCP-NDDO, where an NDDO density matrix is augmented by an additional, self-consistent polarizing (SCP) multipole density matrix.[33,34] A related approach is the chemical-potential equalization (CPE) method, in which the density is augmented by an additional, polarizable response density.[35,36] The CPE approach has previously been combined with the MNDO/d method[36] and more recently the DFTB3 method.[37] In both cases the addition of the CPE response density increased the accuracy of molecular dipoles and polarizabilities, but the effect on intermolecular binding has not been addressed.

In the present work we discuss the implementation of the CPE method in the framework of DFTB3, leading to a combined model we refer to as DFTB3/CPE. The new, additional parameters are obtained by fitting DFTB3/CPE calculated data to reference values obtained from high-level QM data. We note that DFTB is in itself unable to describe dispersion due to the use of PBE as the functional. Thus, we also derive the relevant parameters for DFTB3/CPE using the empirical two- and three-body dispersion interaction corrections due to Grimme.[38]

# Theory

## Third-order SCC-DFTB (DFTB3)

The third-order SCC-DFTB (DFTB3) method is thoroughly discussed elsewhere, here we introduce the necessary equations to help present the combined implementation of DFTB3 and CPE (DFTB3/CPE) and its variants. In DFTB3, the total energy is given as,[13]

$$E_{\mathrm{dftb3}} = \sum_{i}^{\mathrm{occ}} n_i \sum_{\mu\nu} C_{\mu i} C_{\nu i} H^0_{\mu\nu} + \frac{1}{2}\sum_{ab} \Delta q_a \Delta q_b \gamma_{ab} + \frac{1}{3}\sum_{ab} \Delta q_a^2 q_b \Gamma_{ab} + \frac{1}{2}\sum_{ab} V^{\mathrm{rep}}_{ab}. \quad (1)$$

The matrix elements $H^0_{\mu\nu}$ are calculated numerically using the PBE functional and tabulated in the Slater-Koster files. Expressions for the second-order kernel, $\gamma_{ab}$, and its charge derivative, $\Gamma_{ab}$, are derived in previous work,[12,13] and the pair-wise repulsive potentials $V^{\mathrm{rep}}_{ab}$ are fitted empirically and tabulated in the form of splines.[39,40]

The LCAO-MO orbital coefficients, $C_{\mu i}$ and $C_{\nu i}$, and the partial Mulliken charges, $\Delta q_a$ and $\Delta q_b$, are obtained by solving the secular Kohn-Sham equations:

$$\sum_{\nu} C_{\mu i} \left(H_{\mu\nu} - \varepsilon_i S_{\mu\nu}\right) = 0, \quad (2)$$

where $S_{\mu\nu}$ is the overlap matrix.

The Mulliken charges in turn enter the Hamiltonian matrix elements,

$$H_{\mu\nu} = H^0_{\mu\nu} + S_{\mu\nu} \sum_{c} \Delta q_c \left( \frac{1}{2}\left(\gamma_{ac} + \gamma_{bc}\right) + \frac{1}{3}\left(\Delta q_a \Gamma_{ac} + \Delta q_b \Gamma_{bc}\right) + \frac{\Delta q_c}{6}\left(\Gamma_{ca} + \Gamma_{cb}\right) \right), \quad (3)$$

and Eqn. 2 must be iteratively solved until self-consistency is reached.

## Chemical-potential Equalization

In the chemical-potential equalization method,[35] the electron density is augmented with an (additional) polarization response density, $\delta\rho_{\text{cpe}}$, described by a set of atom-centered basis functions:

$$\delta\rho_{\text{cpe}} = \sum_i c_i \varphi_i^{\text{cpe}}(\mathbf{r}). \tag{4}$$

The response basis functions are taken to be $p$-type Gaussian functions of the following form:

$$\varphi_i^{\text{cpe}}(\mathbf{r}) = -2\zeta_i^2 \left(\frac{\zeta_i^2}{\pi}\right)^{3/2} (k - K_i) e^{-\zeta_i^2 |\mathbf{r}-\mathbf{R}_i|^2}, \tag{5}$$

where the atom is centered at $\mathbf{R}_i$, and $k$ and $K_i$ are the $x$-, $y$- or $z$-components of $\mathbf{r}$ and $\mathbf{R}_i$, respectively; $\zeta_i$ is a basis-set exponent. As suggested by Giese and York,[36] the $\zeta$-exponent takes into account fluctuations in the partial charge, via an exponential factor, i.e.,:

$$\zeta_i = Z_i \ \exp\left(B_i \ \Delta q_i\right). \tag{6}$$

The parameters $Z_i$ and $B_i$ are element-specific parameters, the value of which must be calculated or fitted from empirical or *ab initio* data.

In this work, the CPE basis functions interact by means of a simple kernel[35] in the form of a Coulomb integral:

$$N_{ij} = \iint \frac{\varphi_i^{\text{cpe}}(\mathbf{r})\varphi_j^{\text{cpe}}(\mathbf{r}')}{|\mathbf{r}-\mathbf{r}'|} \mathrm{d}^3\mathbf{r} \ \mathrm{d}^3\mathbf{r}'. \tag{7}$$

The interaction between the DFTB3 basis and the CPE basis is likewise described by an approximate Coulomb integral:

$$M_{ij} = f(R_{ij}) \iint \frac{\varphi_i^{\text{cpe}}(\mathbf{r})\varphi_j^{\text{dftb3}}(\mathbf{r}')}{|\mathbf{r}-\mathbf{r}'|} \mathrm{d}^3\mathbf{r} \ \mathrm{d}^3\mathbf{r}'. \tag{8}$$

A screening function $f(R_{ij})$ is applied to the DFTB3-CPE interaction term as an empirical term to account for the missing kinetic energy component and also dampens the short-range interaction with the DFTB3 density which seems well-described already.[36,37]

$$f(R_{ij}) = \begin{cases} 1, & \text{if } R_{ij} > R_u \\ 0, & \text{if } R_{ij} < R_l \\ 1 - 10x^3 + 15x^4 - 6x^5, & \text{otherwise,} \end{cases} \tag{9}$$

where

$$x = \frac{R_u - R_{ij}}{R_u - R_l}, \tag{10}$$

and $R_u = R_{i,u} + R_{j,u}$ and $R_l = R_{i,l} + R_{j,l}$. The parameters $R_{i,u}$ and $R_{i,l}$ are empirical parameters that relate to atom $i$. These parameters are fitted element-wise. This approach has previously been used successfully to improve the description of molecular dipole moments and polarizabilities in a DFTB3+CPE framework.[37]

The correction to the total energy in the combined DFTB3 and CPE model (DFTB3/CPE) is then:

$$E_{\text{dftb3/cpe}} = E_{\text{dftb3}}\left[\rho\right] + E_{\text{cpe}}\left[\mathbf{q}, \mathbf{c}\right], \tag{11}$$

where

$$E_{\text{cpe}}\left[\mathbf{q}, \mathbf{c}\right] = \mathbf{c}^T \cdot \mathbf{M} \cdot \mathbf{q} + \frac{1}{2}\mathbf{c}^T \cdot \mathbf{N} \cdot \mathbf{c}. \tag{12}$$

The variational implicit dependence of the response density coefficients, $\mathbf{c}$, which must be added to the DFTB3 Hamiltonian matrix (Eqn. 3) is (derived in the **Supporting Information**):

$$\Delta H_{\mu\nu}^{(\text{cpe})} = \frac{1}{2} S_{\mu\nu} \left( \frac{\partial E_{\text{cpe}}\left[\mathbf{q}, \mathbf{c}\right]}{\partial q_a} + \frac{\partial E_{\text{cpe}}\left[\mathbf{q}, \mathbf{c}\right]}{\partial q_b} \right) \qquad \mu \in a, \nu \in b \tag{13}$$

where

$$\frac{\partial E_{\text{cpe}}\left[\mathbf{q}, \mathbf{c}\right]}{\partial q_i} = \mathbf{c}^{\mathbf{T}} \cdot \left( \frac{\partial \mathbf{M}}{\partial q_i} \right) \cdot \mathbf{q} + \left[\mathbf{c}^{\mathbf{T}} \cdot \mathbf{M}\right]_i + \frac{1}{2}\mathbf{c}^T \cdot \left( \frac{\partial \mathbf{N}}{\partial q_i} \right) \cdot \mathbf{c}. \tag{14}$$

The set of coefficients of the CPE response density basis that variationally minimizes the total DFTB3/CPE energy in Eqn. 11 is given by:

$$\mathbf{c} = -\mathbf{N}^{-1} \cdot \mathbf{M} \cdot \mathbf{q}. \tag{15}$$

These are re-calculated in each SCF cycle. In practice 2-3 cycles are sufficient to converge the response density coefficients, whereas the DFTB3 method typically requires around 10-20 SCF cycles to converge the Mulliken charges. If the CPE basis set exponents depend on the Mulliken charges, the inversion of **N** must be performed each cycle. If the CPE model is assumed to be independent of the Mulliken charges, i.e., by setting $B = 0$ in Eqn. 6, the inversion is only performed once. This inversion is fortunately much cheaper than solving the Kohn-Sham equations, and the resulting overhead is about 5% for a charge-independent CPE model and up to 30% for the charge-dependent model.[37]

The DFTB3/CPE gradient contribution is derived in the **Supporting Information**, and the relevant equations necessary to calculate molecular dipoles moments and polarizabilities in DFTB3 and DFTB3/CPE is also derived in the **Supporting Information**.

## D3 dispersion correction

To avoid fitting the parameters of the CPE model such that the response density is implicitly compensating for the missing dispersion in the DFTB3 Hamiltonian, we augment DFTB3 energy with the two-body D3 dispersion due to Grimme[38,41] and three-body Axilrod-Teller-Muto dispersion correction also due to Grimme.[38] The two-body D3 term is given by:

$$E_{\mathrm{d3(bj)}} = -\sum_{a<b} s_6 \frac{C_{6,ab}}{r_{ab}^6 + [f(r_{ab})]^6} + s_8 \frac{C_{8,ab}}{r_{ab}^8 + [f(r_{ab})]^8}, \tag{16}$$

where $r_{ab}$ is the interatomic distance and $f(r_{ab})$ is the Becke-Johnson damping function:[41]

$$f(r_{ab}) = a_1 r_{0,ab} + a_2. \tag{17}$$

In this scheme, all parameters besides $s_8$, $a_1$ and $a_2$ are determined from first principles. A value of $s_6 = 1.0$ is used as suggested by Grimme.[38]

Furthermore, we include the Axilrod-Teller-Muto three-body dispersion term given by:

$$E_{\mathrm{abc}} = -\sum_{a<b<c} f_{d,3}(r_{abc}) \frac{C_{9,abc}\left(3\cos\theta_a \cos\theta_b \cos\theta_c + 1\right)}{\left(r_{ab} r_{bc} r_{ca}\right)^3}, \tag{18}$$

where $\theta_a$, $\theta_b$ and $\theta_c$ are the angles formed by the triangle formed by the atoms $a$, $b$ and $c$, and $r_{ab}$, $r_{bc}$ and $r_{ca}$ are the corresponding interatomic distances, and finally $C_{9,abc}$ is a constant calculated from first principles. $f_{d,3}(r_{abc})$ is a damping function described in Ref. 38.

The total energy of the dispersion corrected DFTB3/CPE-D3 model is then:

$$E_{\mathrm{dftb3/cpe-d3}} = E_{\mathrm{dftb3/cpe}} + E_{\mathrm{d3(bj)}} + E_{\mathrm{abc}}. \tag{19}$$

## Optimization of parameters

We employ an optimization approach based on Bayesian inference to find the most likely set of parameters given the data available in our training set. Since the chemical nature of the data sets is diverse and the reference interaction energies span two orders of magnitude, we also apply different weights to the restraints in our optimization. Rather than hand-picking these weights we optimize them on-the-fly during the parameterization by including these in our probability function.

## Cost function

The probability of a set of unknown parameters $\{P_j\}$ and the unknown relative weights $\{\sigma_i\}$ of the reference data sets, given the set of input reference data sets $\{D_i\}$ is given from Bayes' theorem by the following relation:

$$p\left(\{P_j\}, \{\sigma_i\} | \{D_i\}\right) \propto \mathcal{L}\left(\{D_i\} | \{P_j\}, \{\sigma_i\}\right) p\left(\{P_j\}\right) p\left(\{\sigma_i\}\right) \tag{20}$$

Here $j$ is the index of each parameter in the model and $i$ is the index of each data set available. The probability of observing a particular set of reference data given a set of parameters and a set of weights is give as

$$\begin{aligned} \mathcal{L}\left(\{D_i\}|\{P_j\},\{\sigma_i\}\right) &= \prod_i \mathcal{L}\left(D_i|\{P_j\},\sigma_i\right) \\ &\propto \prod_i \sigma_i^{-N_i} \exp\left(\frac{-\chi^2}{2\sigma_i^2}\right), \end{aligned} \tag{21}$$

Where $\chi^2$ is the chi-squared agreement between the reference data and the corresponding data, calculated using a particular set of parameters. This assumes that the error in the reference and model data follows a Gaussian distribution. According to the principle of maximum entropy this is the least biasing choice.[42] Since all that is known about the parameters and weights is that they are positive numbers, these are described using Jeffery's prior as the least biasing uninformative prior, in this case[43]

$$p(x) \propto \frac{1}{x}. \tag{22}$$

Maximizing the probability in Eqn. 20 is in practice carried out by minimizing the following equivalent cost function:

$$\mathcal{E} = -\beta^{-1} \ln\left(p\left(\{P_j\},\{\sigma_i\}|\{D_i\}\right)\right) \tag{23}$$

$$= \beta^{-1}\left[\sum_j \ln(P_j) + \sum_i \left((N_i+1)\ln(\sigma_i) + \frac{\chi_i^2}{2\sigma_i^2}\right)\right], \tag{24}$$

where $\beta$ is an artificial simulation temperature factor.

Likely parameter sets are generated by running a Monte Carlo Metropolis-Hastings[44] simulation at $\beta = 0.25$ for 10,000 to 25,000 steps. From this simulation a number of parameter sets with high likelihood are picked and a greedy optimization is performed on these, until the cost function has converged into a minimum.

## Akaike Information Criterion

The CPE model introduces a considerable number of free parameters (see Table 1) that are fitted to a relatively limited set of data. The validity of adding each parameter is evaluated using the Akaike Information Criterion[45] which is a measure of the relative Kullback-Liebler distance between a collection of models and a possible "true" model.

For the data presented in this paper The AIC can be calculated as:

$$\mathrm{AIC} = 2k + 2\sum_i N_i \ln\left(\mathrm{RMSD}_i\right), \tag{25}$$

where $k$ is the number of parameters in the model, $i$ is the index of each data set, $N_i$ is the number of data points for the $i$'th data set and $\mathrm{RMSD}_i$ is the root mean squared deviation between the model calculated and reference data for the $i$'th data set. See the **Supporting Information** for a detailed derivation of this expression for the AIC.

In cases where $k^2 \ll N$ ($N = \sum_i N_i$) is not true – when the data is somewhat sparse compared to the number of fitting parameters – the AIC is slightly biased towards more parameters.[46] In these cases, the corrected AIC (AICc) can be used to correct for finite-size effects, by adding a slightly heavier penalty on more parameters.[47] The AIC and AICc are asymptotically equivalent for $N \to \infty$, and also for $k \to 0$. The AICc can be derived as an added correction to the AIC:

$$\mathrm{AICc} = \mathrm{AIC} + \frac{2k\left(k+1\right)}{N-k-1}. \tag{26}$$

The Akaike weight ($w_i$) is a measure of the relative likelihood for each model amongst a collection of candidate models.[46] In the following, we use the AICc values to calculate the Akaike weights as

$$w_i = \frac{\exp\left(-\frac{1}{2}\Delta\mathrm{AICc}_i\right)}{\sum_r \exp\left(-\frac{1}{2}\Delta\mathrm{AICc}_r\right)} \tag{27}$$

where $\Delta\mathrm{AICc}_i = \mathrm{AICc}_i - \mathrm{AICc}_{\mathrm{min}}$, i.e., the difference between the AICc for a particular model minus the lowest AICc observed across all models.

# Computational methodology

The dispersion correction models and the CPE model are added to the SCCDFTB module of CHARMM version 40a1.[48] All DFTB3 and DFTB3/CPE calculations are carried out in CHARMM using the 3OB parameter set and the X-H correction.[39,40] DFT calculations are carried out in Gaussian 09.[49] CCSD(T) and MP2 calculations are carried out in MOLPRO 2012.1.17.[50,51]

## Reference Data sets

The data sets used to parameterize and test the DFTB3/CPE method are described here; among them, **S22** is not used in the parameterization and serves as a test set for the final models, while all other datasets are included in the training set. We note that there is some overlap between S22 and S66 (5 complexes are found in both sets). All interaction energies are calculated at the CCSD(T) level of theory. If the counter poise approximation is used to compensate for basis set-superposition error, this is noted as (cp) in the following.

### S22

The S22 data set consists of 22 small organic molecules with a mix of polarization and dispersion dominated interactions.[52] We use the updated energies for the S22 data set given by Takatani *et al.* in Ref. 53. Energies are calculated at the CCSD(T)/CBS(cp)//MP2/cc-pVTZ level of theory.

### S66

The S66 data set consists of 66 small organic molecules with a mix of polarization and dispersion dominated interactions.[54] Energies are calculated at the CCSD(T)/CBS(cp)//MP2/cc-pVTZ level of theory.

### Charged interactions ("C15")

This dataset[1] consists of 15 dimer complexes where one specie is charged, and is thus dominated by strong polarization interactions. This dataset is referred to as the C15 dataset in this paper. Energies are calculated at the CCSD(T)/CBS(cp)//MP2/cc-pVTZ level of theory.

During parameter optimization, it was discovered that one complex in the C15 dataset, namely the imidazolium-methylamine complex had a discrepancy of 12 kcal/mol between the CCSD(T)/CBS and DFTB3 interaction energy. This complex is excluded from the fitting data and also from the statistics presented in the Results section, as the large error reflects an intrinsic limitation of the DFTB3 model for treating (nitrogen) lone-pairs rather than issues related to the limited polarizability (see additional discussions below).

### Sulfur ("S14")

This data set consists of 14 dimer complexes where one specie contains a sulfur atom, and is very similar in nature to the S22, S66 and C15 datasets.[55] This dataset is referred to as the S14 dataset in this work. Energies are calculated at the CCSD(T)/CBS(cp)//MP2/cc-pVTZ level of theory.

### Ionic bonds ("I9")

This dataset is created similarly to the S22, S66 and C15 datasets, and consists of 9 salt-bridge dimer complexes, using combinations of guanidinium, imidazolium and methyl ammonium as cations and methyl acetate, thiomethoxide and methoxide as anions. Details about this dataset, including coordinates and interaction energies, are described in the **Supporting Information**. This dataset is referred to as the I9 dataset in this paper. Energies are calculated at the CCSD(T)/CBS(cp)//MP2/cc-pVTZ level of theory.

### Charged water clusters

This dataset consists of 9 water clusters, where each complex contains one hydronium and one to three water molecules. Details about this dataset, including coordinates and interaction energies, are described in the **Supporting Information**. Energies are calculated at the CCSD(T)/CBS(cp)//MP2/cc-pVTZ level of theory.

### Charged water dimers ("W2")

This dataset consists of two dimer complexes: The hydronium-water complex and the hydroxide-water complex. Details about this dataset, including coordinates and interaction energies, are described in the **Supporting Information**. This dataset is referred to as the W2 dataset in this paper. Energies are calculated at the CCSD(T)/CBS(cp)//MP2/cc-pVTZ level of theory.

### Large water

This dataset consists of 15 water clusters, ranging from 6 to 17 water molecules.[56] Interaction energies are calculated at the CCSD(T)/aug-cc-pVTZ//MP2/aug-cc-pVTZ level of theory.

### L7

This dataset consists of 7 large organic complexes that are dominated by large dispersion forces.[57] Interaction energies are calculated at the DLPNO-CCSD(T)/CBS(cp)//TPSS/TZVP level of theory, and are given in the **Supporting Information**.

### Polarizabilities

The geometries for 133 molecules are taken from the QCRNA database,[58] and dipole moments and molecular polarizabilities are re-calculated at the B3LYP/aug-cc-pVTZ level of theory.[59] This data set is divided into 87 neutral molecules, 27 anions and 19 cations.

# Results

The optimization process outlined above results in four different models discussed below. They differ in the number of parameters.

- DFTB3/CPE($U$)-D3*: In this model the values of the parameters in the D3 dispersion model are fixed to those found in Ref. 3. Additionally, the charge dependence of the CPE basis functions is set to $B = 0$ in Eqn. 6. Furthermore, the value of $Z$ is set to the Hubbard $U$,[12] but scaled for all elements by a single adjustable parameters.

- DFTB3/CPE($U$)-D3: This model additionally relaxes the parameters of the D3 dispersion model.

- DFTB3/CPE($\zeta$)-D3: This model relaxes all parameters, except that the charge dependence of the CPE basis functions is ignored by setting $B = 0$ in Eqn. 6. Two versions of this model are parametrized: one based on only interaction energies, and one additionally using polarizabilities for neutral molecules.

- DFTB3/CPE($q$)-D3: In this model all parameters are optimized. Two versions of this model are parametrized: one based on only interaction energies, and one additionally using polarizabilities for neutral molecules.

The final parameters of the four models are summarized in Table 1, and the final RMSD values of the fitting datasets and the S22 test set are presented in Table 2, which also includes comparison to several other semi-empirical methods as well as PBE calculations. A graphic overview for the performance of several methods is presented on Fig. 1. We note that these parameters are rather different from those optimized in Ref. 37 based on polarizabilities. Indeed, parameters from Ref. 37 would lead to very poor intermolecular interaction energies for some systems (see Table 2 and discussion below).

We start by examining the results of DFTB3, DFTB3-D3 and DFTB3-D3H4 models[1,3,39,40] as a reference to gauge the performance of the DFTB3/CPE models. As shown in

Table 1: Parameters with the highest likelihood for the DFTB3/CPE-D3 models parameterized from interaction energies. Models marked with (pol) are parametrized using additional polarization data for a set of 87 neutral molecules

| **Method** | DFTB3 | DFTB3i-D3[a] | DFTB3/CPE($U$)-D3[b,c] | DFTB3/CPE($U$)-D3[c] | DFTB3/CPE($\zeta$)-D3 | DFTB3/CPE($q$)-D3 | DFTB3/CPE($\zeta$)-D3 (pol) | DFTB3/CPE($q$)-D3 (pol) |
|---|---|---|---|---|---|---|---|---|
| $a_1$ [au] | | 0.5719 | 0.5719 | 0.1227 | 0.3772 | 0.3942 | 0.3863 | 0.3045 |
| $a_2$ [au] | | 3.6017 | 3.6017 | 5.2156 | 4.3174 | 3.7047 | 3.5912 | 0.0000 |
| $s_8$ [au] | | 0.5883 | 0.5883 | 0.0166 | 0.0179 | 0.0139 | 0.0128 | 4.1738 |
| H $Z$ [au] | | | 1.8557 | 2.1040 | 1.3356 | 2.2551 | 2.3933 | 2.8005 |
| H $B$ [au] | | | 0.0000 | 0.0000 | 0.0000 | 0.8566 | 0.0000 | 0.4084 |
| H $r_l$ [au] | | | 0.0624 | 0.1398 | 0.1315 | 0.3796 | 0.1449 | 0.4029 |
| H $r_u$ [au] | | | 5.1978 | 4.5281 | 5.3714 | 0.3796 | 2.2003 | 0.4030 |
| C $Z$ [au] | | | 1.6133 | 1.8292 | 1.2331 | 1.4783 | 2.4025 | 1.9271 |
| C $B$ [au] | | | 0.0000 | 0.0000 | 0.0000 | 0.0048 | 0.0000 | 0.0111 |
| C $r_l$ [au] | | | 2.2399 | 3.0349 | 2.1469 | 1.0862 | 0.4482 | 1.5431 |
| C $r_u$ [au] | | | 6.9382 | 5.8196 | 6.5002 | 2.3530 | 1.6382 | 1.9163 |
| N $Z$ [au] | | | 2.1914 | 2.4847 | 5.3497 | 2.0292 | 28.867 | 2.1352 |
| N $B$ [au] | | | 0.0000 | 0.0000 | 0.0000 | 0.3238 | 0.0000 | 0.2542 |
| N $r_l$ [au] | | | 6.2019 | 6.3024 | 5.8490 | 1.6511 | 6.0026 | 2.0131 |
| N $r_u$ [au] | | | 6.2023 | 6.3027 | 5.8496 | 2.2921 | 6.0028 | 2.0321 |
| O $Z$ [au] | | | 1.9061 | 2.1612 | 53.419 | 4.3227 | 58.602 | 9.7552 |
| O $B$ [au] | | | 0.0000 | 0.0000 | 0.0000 | 0.0451 | 0.0000 | 0.0965 |
| O $r_l$ [au] | | | 3.0359 | 3.0606 | 3.5507 | 3.4832 | 3.4609 | 3.4807 |
| O $r_u$ [au] | | | 3.7043 | 3.6479 | 3.6175 | 3.6050 | 4.3822 | 3.5745 |
| S $Z$ [au] | | | 1.4545 | 1.6491 | 1.4068 | 3.2853 | 1.4895 | 2.9192 |
| S $B$ [au] | | | 0.0000 | 0.0000 | 0.0000 | 1.8661 | 0.0000 | 1.7258 |
| S $r_l$ [au] | | | 3.0731 | 3.2127 | 3.1834 | 17.555 | 2.4655 | 16.577 |
| S $r_u$ [au] | | | 3.0731 | 3.2127 | 3.1836 | 1884.98 | 2.4659 | 2752.47 |
| # Parameters | 0 | 3 | 11 | 14 | 18 | 23 | 18 | 23 |

[a] D3 parameters from Gerit et al 2014.
[b] The D3 parameters are not fitted for this model.
[c] The values of $Z$ for each element is set to the value of the Hubbard $U$ times the globally fitted constant.

Table 2, DFTB3-D3 and DFTB3-D3H4 are major improvements over the original DFTB3/3OB for dispersion dominated datasets; for datasets where no charged molecules are present, the RMSD is typically around 1 kcal/mol for smaller complexes and 2.31 kcal/mol and 2.61 kcal/mol, respectively, for the L7 dataset, as opposed to the value of 15.92 kcal/mol for DFTB3. However for the charged C15 and I9 data sets, the degree of improvement is notably smaller. The RMSD values are 4.99 kcal/mol and 3.91 kcal/mol for the DFTB3-D3 model, and 4.10 kcal/mol and 4.66 kcal/mol for the DFTB3-D3H4 model; the corresponding values are 6.07 and 5.60 kcal/mol for DFTB3. We also note that for the large water dataset of Truhlar et al.,[56] the average binding energy is greatly underestimated by DFTB3, which gives a large RMSD of 14.16 kcal/mol. With the inclusion of D3 dispersion, the RMSD drops significantly to a remarkable value of 2.04 kcal/mol, supporting discussions in the literature regarding the importance of dispersion to bulk water properties.[34,56,60–62] Interestingly, the binding energies of large water clusters are severely overestimated by DFTB3-D3H4, with a RMSD of 23.88 kcal/mol. This is likely caused by the lack of cooperative hydrogen-bonding effects in the D3H4 model, which is molecular mechanical in nature.

Table 2: RMSD and mean error for 10 data sets using various methods. Reference energies are calculated at the CCSD(T)/CBS(cp)//MP2/cc-pVTZ(cp) level of theory, and polarizabilities using B3LYP/aug-cc-pVTZ. Values are given in kcal/mol for energies and $bohr^3$ for polarizabilities. Models marked with (pol) are parametrized using additional polarizability data for neutral molecules. The model marked (orig) uses the CPE parameter set of Kaminski *et al.*[37] and the D3 parameters of Ref. 3.

| | **S22** | | **S66** | | **C15** | | **I9** | | **S14** | |
|---|---|---|---|---|---|---|---|---|---|---|
| **Method** | **RMSD** | **Mean** | **RMSD** | **Mean** | **RMSD** | **Mean** | **RMSD** | **Mean** | **RMSD** | **Mean** |
| DFTB3 | 4.12 | 3.38 | 2.99 | 2.74 | 6.07 | 4.82 | 5.60 | 4.74 | 2.04 | 1.60 |
| DFTB3-D3 | 1.45 | 0.68 | 1.07 | 0.36 | 4.99 | 3.59 | 3.91 | 2.56 | 1.08 | 0.46 |
| DFTB3-D3H4 | 1.24 | 0.48 | 0.89 | 0.29 | 4.10 | 2.22 | 4.66 | 2.69 | 1.48 | 0.81 |
| DFTB3/CPE($U$)-D3* | 1.18 | 0.30 | 0.85 | 0.04 | 2.27 | 1.27 | 3.02 | 1.44 | 1.00 | 0.17 |
| DFTB3/CPE($U$)-D3 | 1.19 | 0.44 | 0.84 | 0.19 | 2.37 | 1.40 | 3.09 | 1.71 | 0.98 | 0.29 |
| DFTB3/CPE($\zeta$)-D3 | 1.21 | 0.48 | 0.80 | 0.16 | 1.78 | 0.41 | 2.58 | 0.20 | 1.01 | 0.14 |
| DFTB3/CPE($q$)-D3 | 1.13 | 0.51 | 0.63 | -0.02 | 1.49 | 0.62 | 1.73 | 0.51 | 0.85 | 0.07 |
| DFTB3/CPE($\zeta$)-D3 (pol) | 1.17 | 0.24 | 0.92 | 0.04 | 2.22 | 1.40 | 2.36 | 0.24 | 0.68 | -0.06 |
| DFTB3/CPE($q$)-D3 (pol) | 1.15 | 0.52 | 0.60 | -0.09 | 2.17 | 1.08 | 2.41 | 1.80 | 0.93 | 0.22 |
| DFTB3/CPE($q$)-D3 (orig) | 1.23 | 0.23 | 0.00 | 0.16 | 3.12 | 1.96 | 3.49 | 1.51 | 0.98 | 0.32 |
| PM6 | 4.18 | 3.38 | 2.99 | 2.65 | 4.57 | 4.27 | 9.13 | 8.50 | 1.74 | 1.35 |
| PM6-D3H4 | 0.83 | 0.38 | 0.64 | 0.17 | 1.48 | 0.80 | 6.05 | 5.61 | 1.19 | 0.56 |
| PBE/6-31G(d) | 3.07 | 0.51 | 2.14 | 0.30 | 4.57 | -4.30 | 12.96 | -12.13 | 1.38 | -0.55 |
| PBE-D3/6-31G(d) | 2.82 | -2.18 | 2.34 | -2.04 | 5.71 | -5.46 | 14.90 | -14.22 | 1.90 | -1.72 |
| PBE/def2-QZVP | 3.71 | 2.55 | 2.65 | 2.05 | 0.67 | 0.05 | 2.39 | -1.88 | 1.08 | 0.63 |
| PBE-D3/def2-QZVP | 0.79 | -0.14 | 0.52 | -0.29 | 1.25 | -1.11 | 4.31 | -3.97 | 0.61 | -0.54 |

| | **Large Water** | | **Charged Water** | | **L7** | | **W2** | | **Polarizability** | |
|---|---|---|---|---|---|---|---|---|---|---|
| **Method** | **RMSD** | **Mean** | **RMSD** | **Mean** | **RMSD** | **Mean** | **RMSD** | **Mean** | **RMSD** | **Mean** |
| DFTB3 | 14.16 | 11.05 | 5.75 | 5.61 | 15.92 | 14.10 | 6.04 | -0.81 | 19.08 | -18.30 |
| DFTB3-D3 | 2.04 | -1.43 | 3.77 | 3.72 | 2.31 | -1.36 | 6.04 | -1.39 | | |
| DFTB3-D3H4 | 23.88 | -20.31 | 2.51 | 2.46 | 2.61 | 0.75 | 6.03 | -0.69 | | |
| DFTB3/CPE($U$)-D3* | 3.59 | -3.14 | 2.04 | 2.01 | 4.23 | -2.90 | 5.78 | -1.88 | | |
| DFTB3/CPE($U$)-D3 | 3.44 | -2.95 | 2.01 | 1.97 | 3.72 | -1.89 | 5.75 | -1.87 | | |
| DFTB3/CPE($\zeta$)-D3 | 2.51 | -1.57 | 1.21 | 1.14 | 4.03 | -1.79 | 5.49 | -2.23 | 92.37 | 82.29 |
| DFTB3/CPE($q$)-D3 | 3.04 | -1.89 | 2.97 | 2.94 | 2.11 | -0.55 | 5.63 | -1.79 | 23.27 | 21.29 |
| DFTB3/CPE($\zeta$)-D3 (pol) | 2.46 | -0.78 | 2.50 | 2.41 | 2.40 | -1.51 | 5.48 | -1.78 | 3.64 | 1.53 |
| DFTB3/CPE($q$)-D3 (pol) | 2.75 | 0.92 | 4.04 | 3.97 | 2.08 | -0.81 | 5.75 | -1.40 | 4.75 | 3.41 |
| DFTB3/CPE($q$)-D3 (orig) | 223.69 | -156.88 | 3.32 | -2.06 | 2.18 | -0.93 | 6.37 | -2.35 | 3.67 | -3.00 |
| PM6 | 34.81 | 27.89 | 14.01 | 12.24 | 12.83 | 10.92 | 11.50 | 6.63 | | |
| PM6-D3H4 | 11.04 | 8.85 | 9.34 | 8.22 | 3.42 | -1.06 | 11.73 | 6.92 | | |
| PBE/6-31G(d) | 69.57 | -58.30 | 15.86 | -13.76 | 14.53 | 11.49 | 14.58 | -13.02 | 9.88 | -9.46 |
| PBE-D3/6-31G(d) | 82.69 | -68.79 | 17.42 | -15.09 | 5.02 | -4.35 | 14.98 | -13.51 | | |
| PBE/def2-QZVP | 4.11 | -3.69 | 3.90 | -3.45 | 17.88 | 15.56 | 4.41 | -3.85 | | |
| PBE-D3/def2-QZVP | 16.05 | -14.17 | 5.43 | -4.78 | 1.79 | -0.29 | 4.81 | -4.35 | | |

* denotes that the D3 parameters are not fitted for this model

Regarding the DFTB3/CPE models, it is seen from Table 2 that both adjusting the D3 dispersion and improving the response properties of DFTB3 are required to achieve a satisfactory description for intermolecular interactions of different nature. Without adjusting the D3 model, for example, large errors are seen for dispersion dominated cases such as L7. On the other hand, including the CPE component is essential to bringing down errors for polar cases such as C15, I9 and charged water clusters. For C15 and I9, for example, the RMSD values for DFTB3/D3 and DFTB3/D3H4 are in the range of 4-5 kcal/mol, while DFTB3/CPE-D3 models have RMSD errors on the order of 2 kcal/mol. Among the four DFTB3/CPE-D3 models introduced here, as the number of parameters in the fitting procedure is increased, the RMSD for the fitting datasets is lowered. Therefore, which model to choose requires a statistical analysis as we present in the next section. Finally, we note that the impact of geometry optimization on the performance of the DFTB3/CPE models is rather modest (see **Supporting Information**).

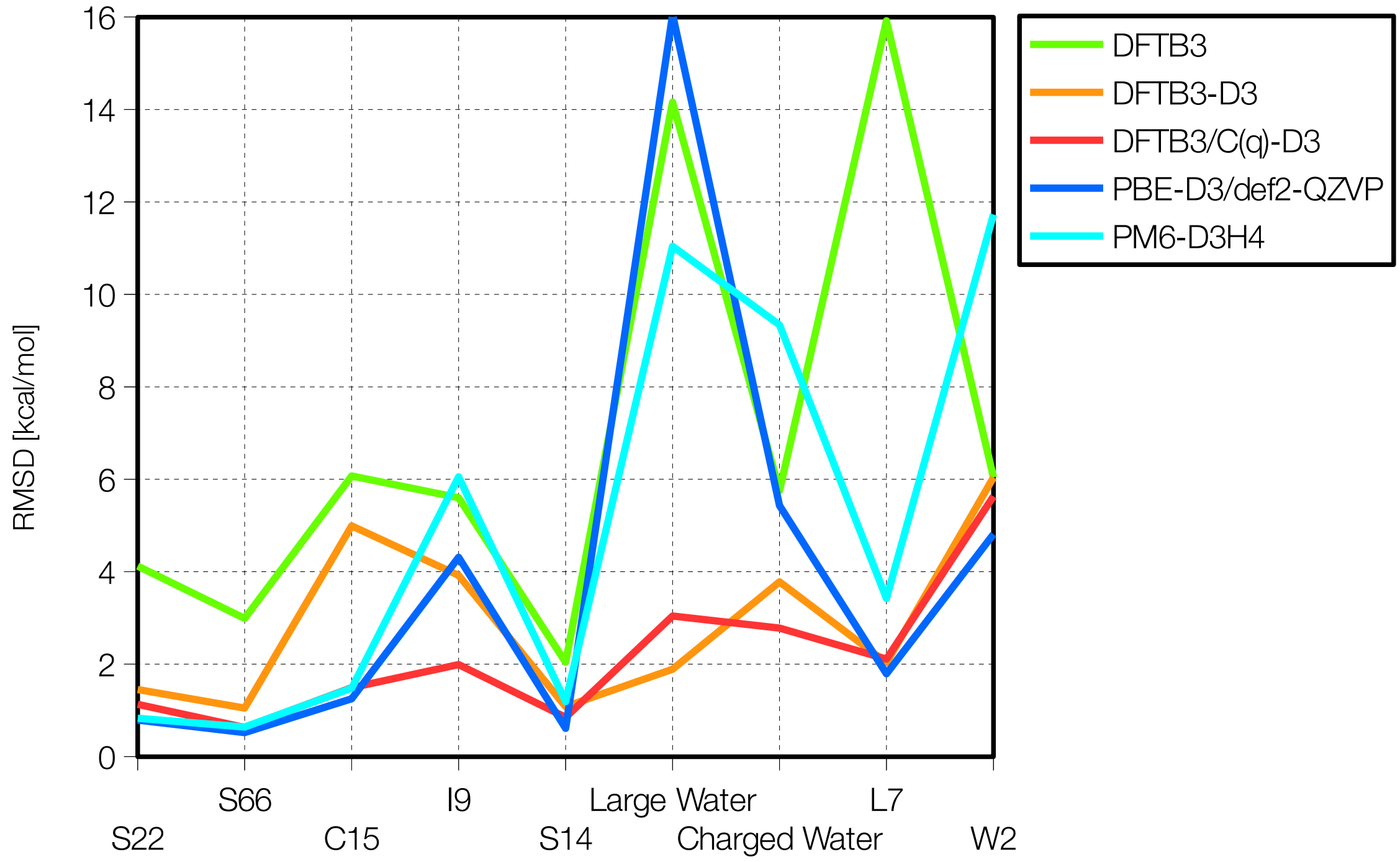

Figure 1: The RMSD interaction energy relative to CCSD(T) for 9 data sets (see text) using five different methods.

As a comparison to representative NDDO models, we focus on PM6 and its recent extension, PM6-D3H4. As shown in Table 2, PM6 has lower RMSD interaction energies than DFTB3 for the smaller complexes, but somewhat higher RMSD interaction energies for the C15 and I9 datasets. The addition of the -D3H4 correction makes the PM6-D3H4 model rather accurate for the S22, S66 and C14 datasets, but only improves I9 from 9.13 kcal/mol to 6.05 kcal/mol. Additionally, the RMSD values for the water clusters are rather large for the PM6 models: for the “Large water” dataset of Truhlar et al.,[56] for example, the RMSD is 34.81 kcal/mol for PM6 and 11.04 kcal/mol for PM6-D3H4. The latter can be compared to DFTB-D3 (i.e., without any hydrogen bond correction), for which the same RMSD value is only 2.04 kcal/mol. Again, we attribute this to the lack of cooperativity in the D3H4 model, which results in a loss of accuracy when the molecular cluster size is scaled towards the condensed phase.

As emphasized in our recent discussions,[2,39] it is worthwhile comparing DFTB3 models to “first principles” DFT methods, especially when modest basis sets are used, because typical DFT based molecular dynamics simulations require the use of modest basis sets. Accordingly, we show in Table 2 also RMSD values for PBE and PBE-D3 DFT models with a small basis set, 6-31G(d), and a large basis set, def2-QZVP. The PBE functional with the 6-31G(d) basis set gives very large RMSD error across all the test sets, presumably due to basis set superposition errors, while using the large def2-QZVP yields very accurate interaction energies for all the smaller complexes. For various water clusters, even the PBE-D3/def2-QZVP method gives large errors in some cases (see Fig. 1); for the “Large water” dataset, for example, the RMSD is 16.05 kcal/mol.

We note from Table 2 (also see Fig. 2) that DFTB3 with no CPE correction gives an underpolarization for neutral molecules in the gas phase of ∼18 bohr$^3$ on average. If the DFTB3/CPE models are parametrized using only interaction energies, the combined model actually overpolarizes greatly by 82 bohr$^3$ on average for the charge-independent CPE model and by ∼21 bohr$^3$ for the charge dependent model. These numbers can be improved by including polarization data in the parametrization to 1.5 and 3.4 bohr$^3$, respectively. This, however, increases the RMSD interaction energy error by between 0.1 to 3.4 kcal/mol, except for the charged water data set, which improves by about 1 kcal/mol. For comparison, PBE with the modest sized 6-31G(d) basis set underpolarizes by 10 bohr$^3$ on average.

Using DFTB3/CPE($q$) with the original parameters from Kaminski *et al.*[37] and the standard D3 parameters, predicted interaction energies are generally improved compared to DFTB3-D3. However, for the large water clusters there is a catastrophic overpolarization, and the binding energy is greatly overestimated by up to around 200 kcal/mol. As the CPE from Kaminski *et al.* parameters were fitted using the MIO parameter set, we re-did the same calculation with the MIO parameters set with the same observations. The fact that a single set of CPE parameters is not optimal for simultaneously describing intermolecular interactions and polarizabilities to high accuracy reflects the semi-empirical nature of the current DFTB3 model; i.e., some of the errors are compensated empirically during the fitting process, which emphasizes on energetics. On the other hand, the fact that a reparameterization of essentially the same DFTB3/CPE model has alleviated the over polarization problem in large water clusters highlights the importance of considering larger clusters than dimer models for the calibration and parameterization of intermolecular interactions for condensed phase applications.

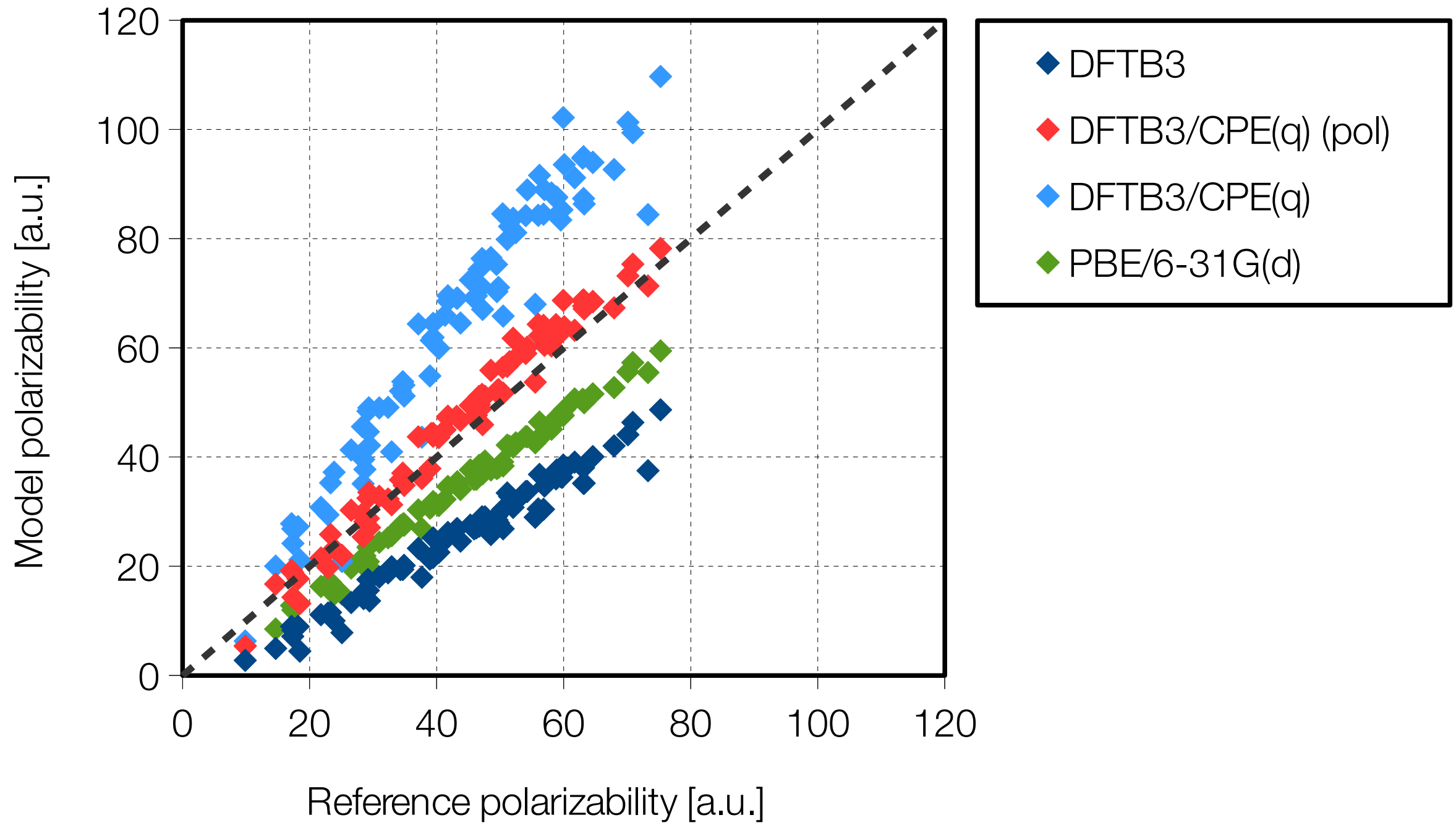

Figure 2: Gas-phase polarizability for 87 neutral molecules calculated with four different methods, compared to a B3LYP/aug-cc-pVTZ reference. DFTB3/CPE($q$) is parameterized without polarizabilities in the training data, while for DFTB3/CPE($q$) (pol) polarizabilities are included in the training data.

In the **Supporting Information** we present the performance of the models on anions and cations. Generally, all the new DFTB3/CPE models as well as PBE/6-31G(d) fail to reproducing the gas-phase polarizability of anions accurately; this is hardly surprising because for anions much more diffuse functions are needed in the gas phase. However, we note that anions in solution are generally much more electronically restricted and, hence, much less polarizable. For example, in the **Supporting Information** we show for one of the worst outliers (thiomethoxide), that when embedded in a droplet of water, the predicted DFTB3/CPE polarizability is close to that of a higher level DFT calculation. These results further underline the great importance and difficulty of extrapolating from properties of gas-phase model systems to an accurate description of solution-phase behavior.

Before concluding this section, we note that in the C15 dataset, the imidazolium-methylamine complex has an discrepancy of 12 kcal/mol between the CCSD(T)/CBS and DFTB3 interaction energies. Hence we chose to leave this complex out of the optimization procedure, as otherwise the cost function was dominated by this single outlier. Also the methylammonium-methylamine complex differs by almost 10 kcal/mol between CCSD(T)/CBS and DFTB3 results. In fact, we note a general trend in S66: complexes that include a nitrogen atom with a lone pair have a larger deviation between CCSD(T)/CBS and DFTB3 compared other cases (see **Supporting Information**). These observations suggest a limitation of the current DFTB3/3OB model,[39] which likely reflects the inadequacy of a monopole charge model for treating strong interactions involve a lone-pair. We do not expect the CPE model to solve this problem, apart from fact that the CPE parameters will be fitted such that they may implicitly compensate for this issue. A more physical improvement requires including multipoles in the charge fluctuations[21] and is being pursued independently. Similarly, although the performance of DFTB3 for water clusters is rather encouraging once dispersion is included, the coupled DFTB3/CPE approach remains to have sizable errors for charged water clusters, especially for water-hydroxide interactions (W2 set); along this line, we note that a recent study that included an “on-site” integral correction[63] seemed to reduce the error for charged water cluster considerably.

### Model selection

Since the CPE model introduces a number of new parameters, we estimate the effect of the increasing number of parameters in the models against the relative likelihood of each DFTB3/CPE model using the corrected Akaike Information Criterion (AICc).

The baseline model is the introduction of 0 parameters, i.e., the DFTB3/3OB model with no further corrections. The AICc for this model is 359.0. As expected from the higher accuracy of the DFTB3-D3 model, the three parameters introduced in this model vastly decrease the AICc to 131.5. A graphical overview of the ΔAICc values vs. number of parameters is displayed on Fig. 3. AIC values for the DFTB3/CPE models are presented in Table 3. The lowest scoring (i.e., favorable) models are the DFTB3/CPE($\zeta$)-D3 and DFTB3/CPE($q$)-D3 models. The AIC weight clearly favors a DFTB3/CPE model over DFTB3 ($w_i = 0.0\%$) and DFTB3-D3 ($w_i = 0.0\%$). Additionally, it also seems very beneficial to optimize the parameters in the dispersion model along with the parameters in the CPE model. The lowest AICc is found for the model with the most parameters, namely the DFTB3/CPE($q$)-D3 model ($w_i = 99.999\%$).

The RMSD for the S22 test set, which is not included in the training data, is improved both by addition of the D3 model and the CPE model. Among the DFTB3/CPE models, the RMSD varies by a few percent, although such small variations may also reflect the fact that the S22 test set does not include charged molecules.

## Concluding Remarks

We have augmented the DFTB3 method with a CPE response density and the D3 dispersion correction in a combined methodology termed DFTB3/CPE-D3. Depending on the number of free parameters, four different DFTB3/CPE-D3 models are parametrized using a broad range of molecular complexes of biological interest. Compared to DFTB3-D3, the accuracy is largely unchanged for small, neutral complexes, where as the accuracy is clearly improved for charged complexes. Compared to the D3H4 corrected DFTB3 and PM6 models, the accuracy of DFTB3/CPE-D3 models is comparable for small, neutral complexes, but the scaling to larger clusters and larger complexes is notably improved. Compared to PBE and PBE-D3 models with modest (double-zeta quality) or large (def2-QZVP) basis sets, the DFTB3/CPE-D3 models are also competitive, especially for large water clusters. The statistical analysis using the AICc favors the DFTB3/CPE-D3 models over the DFTB3-D3 and DFTB3 models.

Table 3: Akaike Information Criterion

| Method | $k$ | AIC | AICc | ΔAICc | $w_i$ | S22 RMSD |
|---|---|---|---|---|---|---|
| DFTB3 | 0 | 359.0 | 359.0 | 295.29 | 0.000% | 4.12 kcal/mol |
| DFTB3-D3 | 3 | 131.3 | 131.5 | 67.72 | 0.000% | 1.45 kcal/mol |
| DFTB3/CPE($U$)-D3* | 11 | 100.7 | 102.9 | 39.16 | 0.000% | 1.18 kcal/mol |
| DFTB3/CPE($U$)-D3 | 14 | 102.8 | 106.5 | 42.73 | 0.000% | 1.18 kcal/mol |
| DFTB3/CPE($\zeta$)-D3 | 18 | 79.9 | 86.1 | 22.34 | 0.001% | 1.21 kcal/mol |
| DFTB3/CPE($q$)-D3 | 23 | 53.2 | 63.8 | 0.00 | 99.999% | 1.13 kcal/mol |

$^*$ denotes that the D3 parameters are not fitted for this model

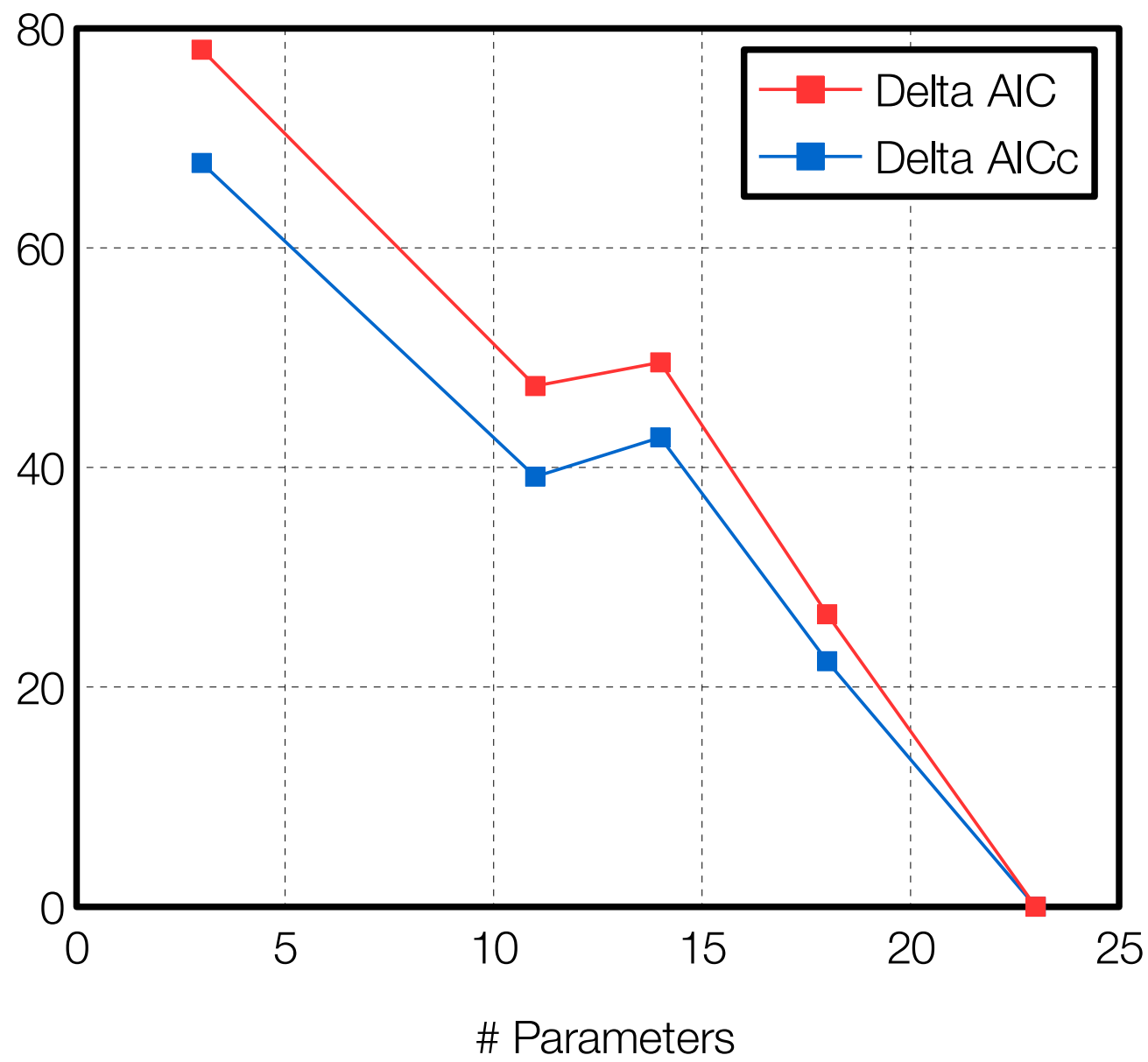

Figure 3: AIC values for the DFTB-D3 model and four DFTB3/CPE models with an increasing number of parameters (see also Table 3).

Despite these encouraging observations, we emphasize that the DFTB3 methodology requires further development for a generally robust and accurate treatment of non-covalent interactions in different environments.[23] As discussed in our and related studies, including multipoles and improved description of Pauli repulsion are of the highest priority; this is also underlined by the large errors for several cases involving nitrogen lone-pairs noted in this study. What we hope to illustrate in this work is that improving the response properties of DFTB3 is a viable approach to improve intermolecular interactions involving charged and highly polarizable molecules. Ultimately, these developments need to be integrated together to form an efficient and robust computational framework for condensed phase properties, especially reactive events in polar liquids and biomolecules. Along this line, developing test cases beyond the relatively small clusters used here and most benchmark studies is also crucial, as illustrated by the different performances of the previous[37] and current DFTB3/CPE models for large water clusters. Nevertheless, to echo the point of the recent work of Grimme et al.,[3] the future of describing non-covalent interactions using the DFTB3 methodology seems bright.

## Acknowledgement

This work is supported by NIH grant R01-GM106443 to QC. The authors thank Prof. Dr. Stefan Grimme for the reference energies for the DPLNO-CCSD(T)/CBS interaction energies for the L7 data set, and for providing the FORTRAN routines to calculate the dispersion correction. Discussions with Drs. T. Giese and K. Welke are also acknowledged. Computational resources from the Extreme Science and Engineering Discovery Environment (XSEDE), which is supported by NSF grant number OCI-1053575, are greatly appreciated; computations are also supported in part by NSF through a major instrumentation grant (CHE-0840494) to the Chemistry department.

## Supporting Information

A more detailed discussion of the Akaike Information Criterion (AIC), calculation procedure for the new data sets (I9, CHW9 and W2), the corresponding interaction energies and Cartesian coordinates are included. Also included are the detailed RMSD values for all the methods discussed here for the collection of benchmark and test sets. The minor impact of geometry optimization on the DFTB3/CPE-D3 results is also briefly discussed. Detailed derivations for DFTB3/CPE, especially for gradient and polarizability calculations, are included.

# Supporting information for: Improving intermolecular interactions in DFTB3 using extended polarization from chemical-potential equalization

Anders S. Christensen,[*,†] Marcus Elstner,[‡] and Qiang Cui[*,†]

*University of Wisonsin-Madison, Department of Chemsitry, 1101 University Ave, Madison, WI 53706, USA, and Universität Karlsruhe, Theoretische Chemische Biologie, Kaiserstr. 12, 76131 Karlsruhe, Germany*

E-mail: andersx@chem.wisc.edu; cui@chem.wisc.edu

## S1: Calculation of AIC and AICc

The calculations of the AIC and AICc require the maximum likelihood estimates (MLE) of the parameters that maximize the following likelihood:

$$\begin{aligned}\mathcal{L}\left(\{D_i\}|\{P_j\},\{\sigma_i\}\right) &= \prod_i \mathcal{L}\left(D_i|\{P_j\},\sigma_i\right) \\ &\propto \prod_i \sigma_i^{-N_i} \exp\left(\frac{-\chi^2}{2\sigma_i^2}\right) \end{aligned} \quad (1)$$

which is discussed in the text.

[*]To whom correspondence should be addressed
[†]University of Wisonsin-Madison
[‡]Universität Karlsruhe

For the MLE estimates of the parameters, we naturally use the best parameters found during the optimization procedure. For the unknown weight parameters, we can calculate the MLE estimate by minimizing the log-likelihood function with respect to $\sigma$, i.e. by solving:

$$\begin{aligned}
\frac{\mathrm{d}}{\mathrm{d}\sigma_i}\ln\left(\sigma_i^{-N_i}\exp\left(\frac{-\chi^2}{2\sigma_i^2}\right)\right) &= 0 && (2)\\
\frac{\mathrm{d}}{\mathrm{d}\sigma_i}\ln\left(\sigma_i^{-N_i}\right) &= -\frac{\mathrm{d}}{\mathrm{d}\sigma_i}\ln\left(\exp\left(\frac{-\chi^2}{2\sigma_i^2}\right)\right) && (3)\\
-N\frac{1}{\sigma_i} &= -\frac{\chi^2}{\sigma_i^3} && (4)\\
\sigma_i^2 &= \frac{\chi^2}{N} && (5)
\end{aligned}$$

The value of $\sigma_i^{\mathrm{MLE}}$ for which the above is true can be simplified as:

$$\sigma_i^{\mathrm{MLE}} = \sqrt{\frac{\chi^2}{N_i}} = \sqrt{\frac{\sum\left(\Delta x\right)^2}{N_i}} = \mathrm{RMSD}_i \tag{6}$$

Furthermore, an expression for the minimized total log-likelihood is required. Inserting the $\sigma_i^{\mathrm{MLE}}$ into the log-likelihood, we obtain the following expression for the logarithm of the maximized likelihood, $\hat{\mathcal{L}}$:

$$\begin{aligned}
\ln\left(\hat{\mathcal{L}}\right) &= \ln\left(\prod_i \sigma_{\mathrm{MLE},i}^{-N_i}\exp\left(\frac{-\chi^2}{2\sigma_{\mathrm{MLE},i}^2}\right)\right) && (7)\\
&= \ln\left(\prod_i \mathrm{RMSD}_i^{-N_i}\exp\left(\frac{-\chi^2}{2\ \mathrm{RMSD}_i^2}\right)\right) && (8)\\
&= -\sum_i N_i\ln\left(\mathrm{RMSD}_i\right) + \frac{N_i}{2} && (9)
\end{aligned}$$

Neglecting all constant term (i.e. $\sum_i \frac{N_i}{2}$), this simplifies to:

$$\ln\left(\hat{\mathcal{L}}\right) = -\sum_i N_i\ln\left(\mathrm{RMSD}_i\right) \tag{10}$$

## S1.1: Akaike Information Criterion (AIC)

We derive the Akaike Information Criterion[S1] from its definition, using the ML estimate obtained in the previous section:

$$\begin{aligned} \mathrm{AIC} &= 2k - 2\ln\left(\hat{\mathcal{L}}\right) & (11) \\ &= 2k + 2\sum_i N_i \ln\left(\mathrm{RMSD}_i\right), & (12) \end{aligned}$$

where $k$ is the number of parameters in the model.

## S1.2: Corrected Akaike Information Criterion (AICc)

In cases where $k^2 \ll N$ is not true - that is, when the data is somewhat sparse compared to the number of fitting parameters - the AIC is slightly biased towards more parameters. In these cases, the corrected AIC (AICc) can be used to correct for finite-size effects,[S2] by adding a slightly heavier penalty on more parameters.[S3] The AIC and AICc are asymptotically equivalent for $N \to \infty$, and also for $k \to 0$. The AICc can be derived as an added correction to the AIC:

$$\mathrm{AICc} = \mathrm{AIC} + \frac{2k\left(k+1\right)}{N-k-1}, \tag{13}$$

where $k$ is the number of parameters in the model and $N$ is the total number of data points in the fitting data.

## S2: Data sets

The creation of these data sets closely follows the methodology used to create the S66×8 data set.[S4] All interaction energies are calculated using the counterpoise approximation, e.g. for a two-body system:

$$\Delta E_{\text{interaction}} = E_{12(12)} - E_{1(12)} - E_{2(12)}, \tag{14}$$

where the subscript denotes the presence of the nuclei and electrons of each monomer (1 and 2), and the presence of their respective basis functions in the calculation in parentheses.

The CCSD(T)/CBS interaction energies are given by:

$$E(\text{CCSD(T)/CBS}) = E(\text{HF/CBS}) + E(\text{CCSD(T)/aug} - \text{cc} - \text{pVDZ}) \tag{15}$$

$$-E(\text{MP2/aug} - \text{cc} - \text{pVDZ}) + E_{\text{corr}}(\text{MP2/CBS}) \tag{16}$$

where $E_{\text{corr}}$(MP2/CBS) is extrapolated using the cc-pVTZ and cc-pVQZ basis sets, and $E$(HF/CBS) is extrapolated using the cc-pVQZ and cc-pV5Z basis sets.

The complexes are optimized at the MP2/cc-pVTZ level of theory (not using the counterpoise approximation). All HF, MP2 and CCSD(T) calculations use the density fitting approximation. In the I9×8 dataset, several complexes would exhibit a proton transfer upon optimization. In these complexes, the hydrogen atoms were constrained to the position found in the monomer equilibrium geometry.

For the I9 and W2 data sets, dissociation curves are created by displacing the monomers along the vector between the center of mass of both monomers. The vector is multiplied by a factor of $x$ where $x \in \{0.90, 0.95, 1.00, 1.05, 1.10, 1.25, 1.50, 2.00\}$.

## S2.1: Ionic bonds (I9×8)

Table S1: Interaction energies in kcal/mol for the I9×8 dataset as a function of displacement.

| ID | Description | E(0.90) | E(0.95) | E(1.00) | E(1.05) | E(1.10) | E(1.25) | E(1.50) | E(2.00) |
|---|---|---|---|---|---|---|---|---|---|
| 01 | Guanidinium ... Methyl acetate | -91.63 | -126.08 | -134.31 | -131.67 | -124.81 | -101.36 | -74.87 | -50.11 |
| 02 | Guanidinium ... Thiometoxide | -112.02 | -122.09 | -125.30 | -124.30 | -120.85 | -105.29 | -81.56 | -54.84 |
| 03 | Guanidinium ... Methoxide | -157.41 | -176.35 | -181.95 | -179.37 | -172.23 | -142.87 | -103.34 | -69.76 |
| 04 | Imidazolium ... Methyl acetate | -91.49 | -115.61 | -120.39 | -117.15 | -110.91 | -90.82 | -68.14 | -46.25 |
| 05 | Imidazolium ... Thiometoxide | -88.88 | -100.44 | -103.97 | -103.09 | -99.94 | -86.61 | -67.59 | -46.43 |
| 06 | Imidazolium ... Methoxide | -100.78 | -116.28 | -120.35 | -118.53 | -113.93 | -96.60 | -74.29 | -51.26 |
| 07 | Methyl ammonium ... Methyl acetate | -122.75 | -140.00 | -145.10 | -143.67 | -138.91 | -119.19 | -91.63 | -61.83 |
| 08 | Methyl ammonium ... Thiometoxide | -112.13 | -116.98 | -117.60 | -115.76 | -112.54 | -100.21 | -81.53 | -58.75 |
| 09 | Methyl ammonium ... Methoxide | -120.99 | -129.70 | -132.43 | -131.57 | -128.62 | -115.18 | -93.80 | -69.13 |

**I9/01: Guanidinium ... Methyl Acetate, equilibrium geometry**

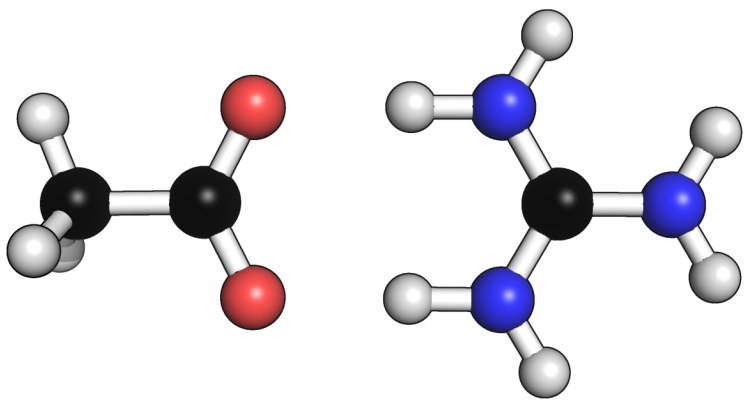

```
17

 C        1.923803467700      -0.017759142900       0.000000000000
 N        1.252409495800      -1.153082639800       0.000000000000
 N        3.280756017700      -0.026187563500       0.000000000000
 N        1.270506131500       1.129882851300       0.000000000000
 H        3.797978734300       0.830097081700       0.000000000000
 H        3.786548440800      -0.889261268500       0.000000000000
 H        1.766006307100      -2.014875162400       0.000000000000
 H        0.155974368200      -1.141295455600       0.000000000000
 H        1.794553240500       1.984974403800       0.000000000000
 H        0.183172416200       1.132821954000       0.000000000000
 C       -3.393358909900      -0.014224469600       0.000000000000
 H       -3.744724464200      -0.555490169800       0.875958756700
 H       -3.744724464200      -0.555490169800      -0.875958756700
 C       -1.878308544300       0.002067479300       0.000000000000
 H       -3.794755240200       0.993456679100       0.000000000000
 O       -1.304896269600      -1.126614917800       0.000000000000
 O       -1.302710427400       1.126346320700       0.000000000000
```

**I9/02: Guanidinium ... Thiomethoxide, equilibrium geometry**

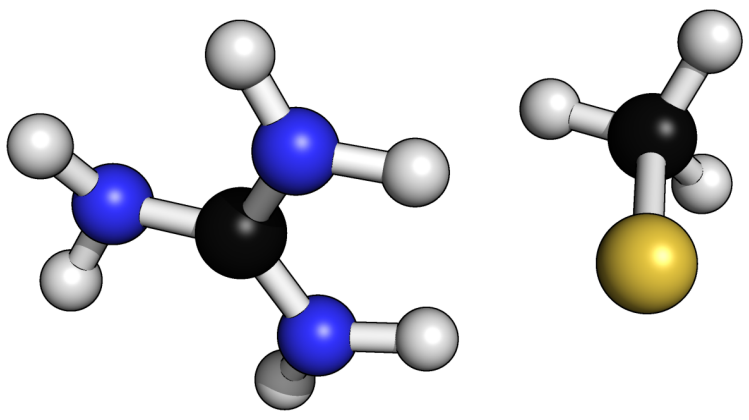

```
15

 C        0.024471742900     -1.489383581400      0.000000000000
 N        0.260302785000     -0.856872652000     -1.135945958000
 N       -0.474248393600     -2.759643634900      0.000000000000
 N        0.260302785000     -0.856872652000      1.135945958000
 H       -0.342089655800     -3.290836515200      0.843147167100
 H       -0.342089655800     -3.290836515200     -0.843147167100
 H       -0.021049595400     -1.274046235800     -2.004978985800
 H        0.484178351000      0.194544493200     -1.019322588700
 H       -0.021049595400     -1.274046235800      2.004978985800
 H        0.484178351000      0.194544493200      1.019322588700
 C       -1.148613093168      2.448953283218      0.000000000000
 H       -1.671583124668      1.488961144018      0.000000000000
 H       -1.468503129868      3.003020420918     -0.880544849600
 S        0.664779818432      2.227454735318      0.000000000000
 H       -1.468503129868      3.003020420918      0.880544849600
```

### I9/03: Guanidinium ... Methoxide, equilibrium geometry

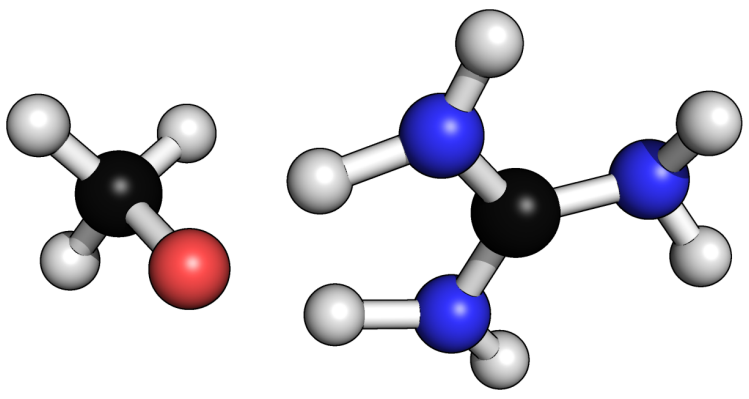

```
15

 C        0.010370994400       1.110889967000       0.000000000000
 N        0.197195295400       0.446277876400       1.128590383500
 N       -0.408134561800       2.424517989400       0.000000000000
 N        0.197195295400       0.446277876400      -1.128590383500
 H       -0.187145369700       2.937703853100      -0.836695341700
 H       -0.187145369700       2.937703853100       0.836695341700
 H        0.002835480000       0.913989047000       1.996010153900
 H        0.438967409300      -0.684775759600       0.850010593000
 H        0.002835480000       0.913989047000      -1.996010153900
 H        0.438967409300      -0.684775759600      -0.850010593000
 C       -0.519149198453      -2.801032400755      -0.000000000000
 H       -1.444092076753      -2.208833177255      -0.000000000000
 H       -0.539391708953      -3.446423499955       0.880962273200
 H       -0.539391708953      -3.446423499955      -0.880962273200
 O        0.630080538747      -1.986279416655      -0.000000000000
```

**I9/04: Imidazolium ... Methyl Acetate, equilibrium geometry**

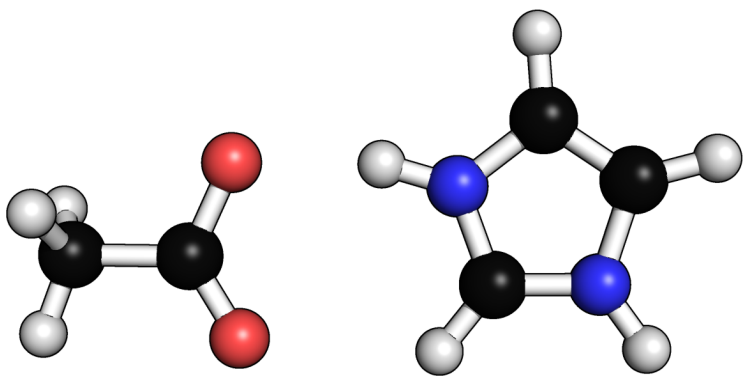

```
17

 N        -2.562614830000        0.964587890000        0.000000000000
 N        -0.901460800000       -0.455702670000        0.000000000000
 C        -1.223181340000        0.833717200000        0.000000000000
 C        -2.071682980000       -1.175489200000        0.000000000000
 C        -3.122200400000       -0.303048400000        0.000000000000
 H        -0.458572760000        1.608402630000        0.000000000000
 H        -2.100015530000       -2.252594890000        0.000000000000
 H        -4.187260780000       -0.465710020000        0.000000000000
 H        -3.068632440000        1.839862030000        0.000000000000
 H         0.024433810000       -0.833475110000        0.000000000000
 C         4.104563039280       -0.096512208597        0.000000000000
 H         4.348578739280       -0.683867768597        0.925237700000
 H         4.591470749280        0.876870461403        0.000000000000
 C         2.597226309280        0.078185771403        0.000000000000
 H         4.348578739280       -0.683867768597       -0.925237700000
 O         2.069847469280        1.187666711403        0.000000000000
 O         1.928906489280       -1.054175078597        0.000000000000
```

### I9/05 Imidazolium ... Thiomethoxide, equilibrium geometry

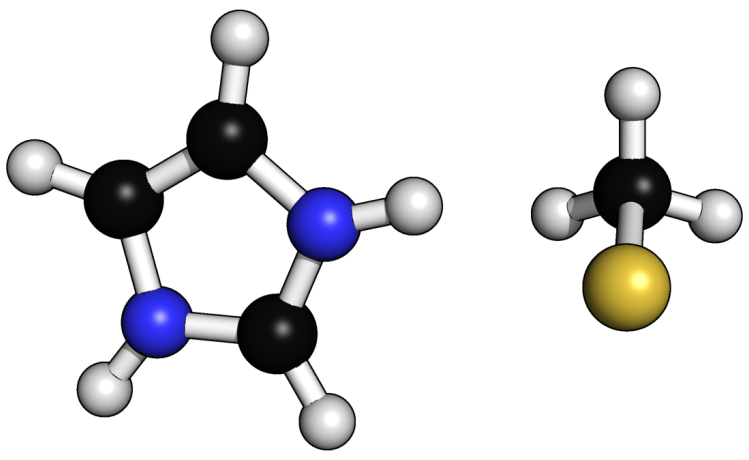

```
15

 N       -0.208558970000       0.043929690000       0.318609930000
 N       -2.010102970000      -0.985925310000      -0.463509070000
 C       -0.668732970000      -1.072534310000      -0.218506070000
 C       -2.426005970000       0.271534690000      -0.056321070000
 H       -2.594702970000      -1.707650310000      -0.883455070000
 C       -1.295944970000       0.887865690000       0.424522930000
 H       -0.082264970000      -1.956937310000      -0.450059070000
 H       -3.453559970000       0.594046690000      -0.152161070000
 H       -1.193909970000       1.880025690000       0.830125930000
 H        0.733628590000       0.237568340000       0.592081430000
 C        3.105082811872       0.831282379653      -0.859343069423
 H        3.030795811872       1.835695379653      -0.428316069423
 H        2.319971811872       0.702579379653      -1.607781069423
 S        2.988466811872      -0.466017620347       0.439892930577
 H        4.081719811872       0.722544379653      -1.341704069423
```

**I9/06: Imidazolium ... Methoxide, equilibrium geometry**

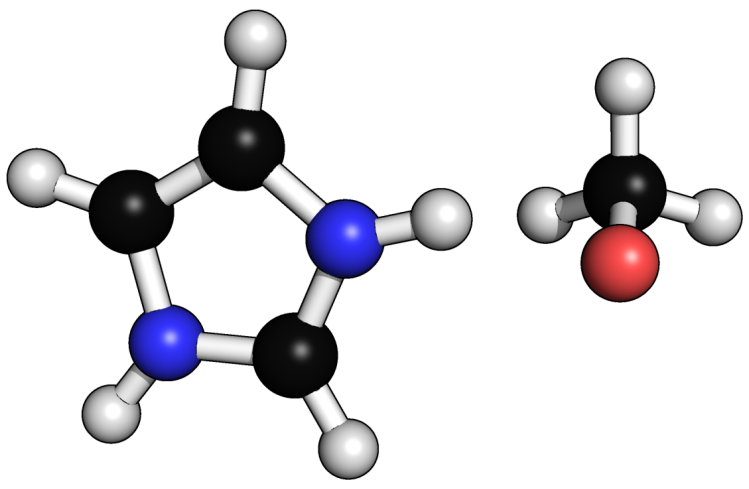

```
15


 N       -0.208558970000       0.043929690000       0.318609930000
 N       -2.010102970000      -0.985925310000      -0.463509070000
 C       -0.668732970000      -1.072534310000      -0.218506070000
 C       -2.426005970000       0.271534690000      -0.056321070000
 H       -2.594702970000      -1.707650310000      -0.883455070000
 C       -1.295944970000       0.887865690000       0.424522930000
 H       -0.082264970000      -1.956937310000      -0.450059070000
 H       -3.453559970000       0.594046690000      -0.152161070000
 H       -1.193909970000       1.880025690000       0.830125930000
 H        0.733628590000       0.237568340000       0.592081430000
 C        2.594473659335       0.902281711510      -0.696273875095
 H        2.520186659335       1.906694711510      -0.265246875095
 H        1.809362659335       0.773578711510      -1.444711875095
 H        3.571110659335       0.793543711510      -1.178634875095
 O        2.510802009335      -0.028527398490       0.235924304905
```

**I9/07: Methyl Ammonium ... Methyl Acetate, equilibrium geometry**

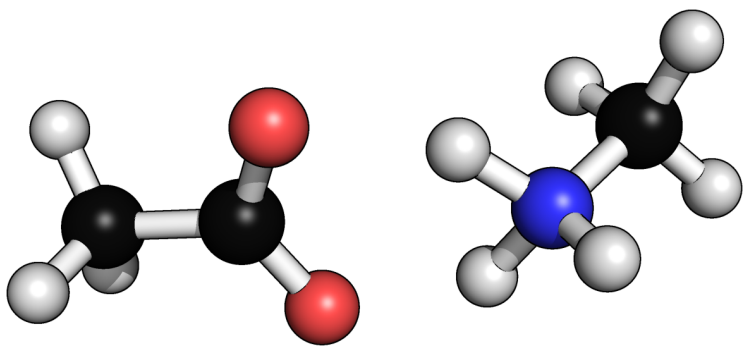

```
15

 N        -0.243667639400      -1.754375754500       0.000000000000
 C         1.051835353300      -2.445074005300       0.000000000000
 H         1.156342921500      -3.061880728700       0.887281335000
 H         1.828497275900      -1.686651263200       0.000000000000
 H         1.156342921500      -3.061880728700      -0.887281335000
 H        -1.037884210400      -2.381004691100       0.000000000000
 H        -0.299919193300      -0.984762835500      -0.771002209100
 H        -0.299919193300      -0.984762835500       0.771002209100
 C        -0.275767195681       2.916616523013       0.000000000000
 H         0.207333131919       3.295587820113       0.895115256100
 C        -0.277830777781       1.404843345913       0.000000000000
 H         0.207333131919       3.295587820113      -0.895115256100
 O        -0.291795758881       0.820978014013      -1.128777198300
 O        -0.291795758881       0.820978014013       1.128777198300
 H        -1.309749091681       3.259817997813       0.000000000000
```

**I9/08: Methyl Ammonium ... Thiomethoxide, equilibrium geometry**

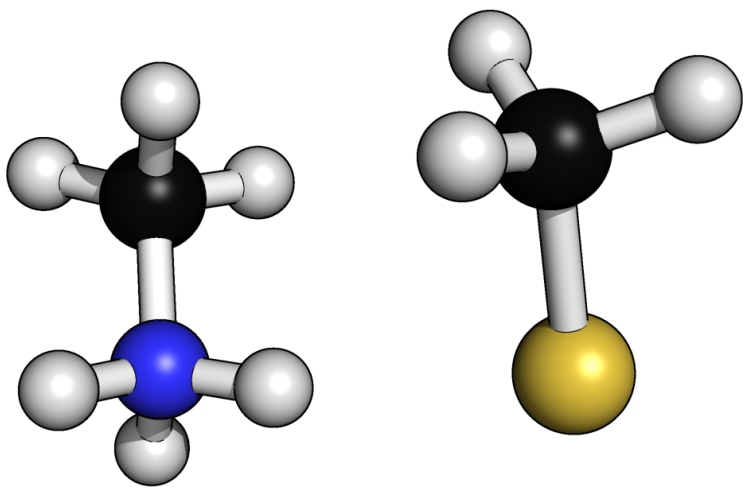

```
13

 N        -1.871728000000        0.590372000000       -0.409427000000
 C        -2.211908000000       -0.607269000000        0.366391000000
 H        -3.290849000000       -0.800379000000        0.474364000000
 H        -1.777031000000       -0.515398000000        1.366626000000
 H        -1.756203000000       -1.480607000000       -0.110636000000
 H        -2.292946460000        1.392500770000        0.013851220000
 H        -2.211824870000        0.489597950000       -1.344402240000
 H        -0.878517500000        0.705429620000       -0.426592450000
 C         1.508119864122       -1.098242666717       -0.355707818837
 H         0.854838864122       -1.714612666717        0.267630181163
 H         2.543640864122       -1.395055666717       -0.173706818837
 S         1.369005864122        0.679416333283        0.079135181163
 H         1.277662864122       -1.268349666717       -1.410659818837
```

**I9/09: Methyl Ammonium ... Methoxide, equilibrium geometry**

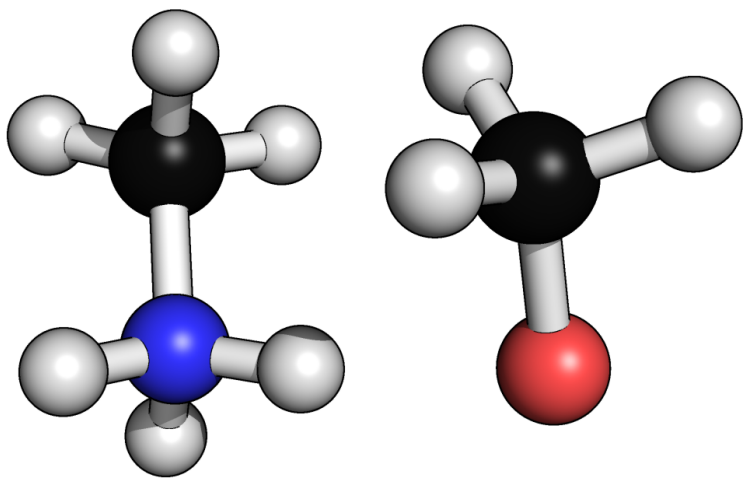

```
13

 N        -1.871728000000      0.590372000000     -0.409427000000
 C        -2.211908000000     -0.607269000000      0.366391000000
 H        -3.290849000000     -0.800379000000      0.474364000000
 H        -1.777031000000     -0.515398000000      1.366626000000
 H        -1.756203000000     -1.480607000000     -0.110636000000
 H        -2.292946460000      1.392500770000      0.013851220000
 H        -2.211824870000      0.489597950000     -1.344402240000
 H        -0.878517500000      0.705429620000     -0.426592450000
 C         0.939686306881     -0.620510570333     -0.380590381131
 H         0.286405306881     -1.236880570333      0.242747618869
 H         1.975207306881     -0.917323570333     -0.198589381131
 H         0.709229306881     -0.790617570333     -1.435542381131
 O         0.831296666881      0.764539269667     -0.041785681131
```

## S2.2: Charged Water (CHW9)

Table S2: Interaction energies in kcal/mol for the CHW9 data set.

| ID | Complex | $E$ |
|---|---|---|
| 1 | 1 Water ... 1 Hydronium | -33.93 |
| 2 | 2 Water ... 1 Hydronium | -57.42 |
| 3 | 2 Water ... 1 Hydronium | -57.37 |
| 4 | 3 Water ... 1 Hydronium | -77.14 |
| 5 | 3 Water ... 1 Hydronium | -77.08 |
| 6 | 3 Water ... 1 Hydronium | -73.06 |
| 7 | 3 Water ... 1 Hydronium | -72.69 |
| 8 | 3 Water ... 1 Hydronium | -73.13 |

**CHW9/01: 1 Water, 1 Hydronium**

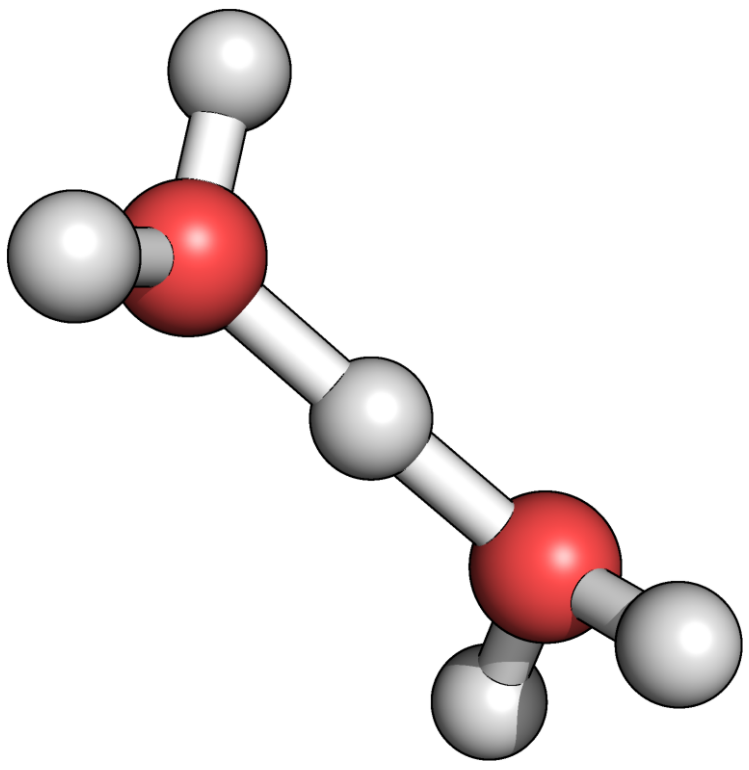

```
   7


O          1.6130818479      0.9577977382      3.8060481114
H          1.8692950753      1.4847112432      4.5750813314
H          1.6328918384      1.5223814517      3.0207483336
O          2.4646947897     -1.2462689662      3.4720602043
H          3.3581971853     -1.4875410872      3.7515926317
H          1.8574942836     -1.9568808991      3.7214197016
H          2.0873672799     -0.1322535806      3.6833121859
```

### CHW9/02: 2 Water, 1 Hydronium

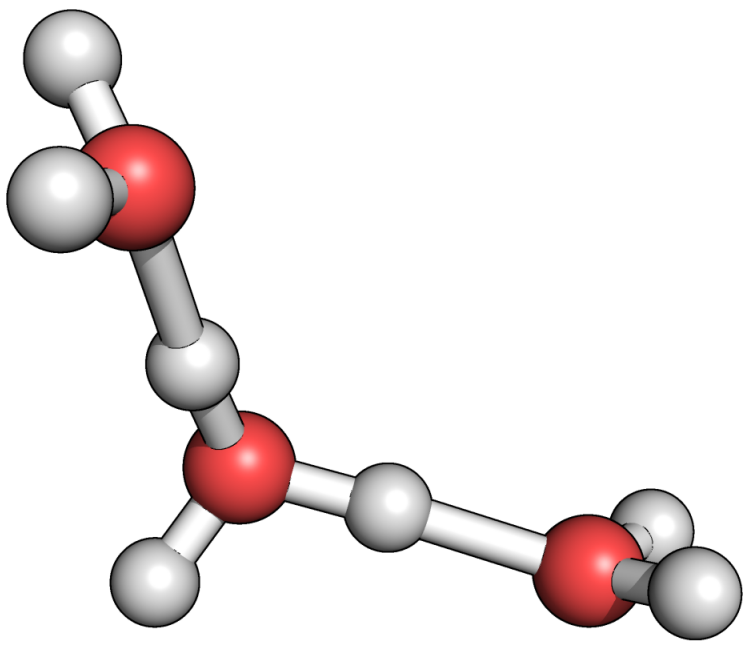

```
  10

O            1.5291752404     -0.1470902635      1.5890245371
H            1.0588273488      0.5519950471      1.1220747256
H            1.9016793768      0.2549460675      2.3806251525
O            2.3463345459     -4.1370806378      0.4289564820
H            3.2950906265     -4.1728516787      0.5873288414
H            2.1964485260     -4.5106547429     -0.4459576100
O            0.7396795832     -2.4889330600      1.3457929899
H            1.4557977884     -3.1420618863      0.9718586951
H            0.3048362999     -2.8565130072      2.1258365085
H            1.0656783641     -1.5181567383      1.5041697778
```

**CHW9/03: 2 Water, 1 Hydronium**

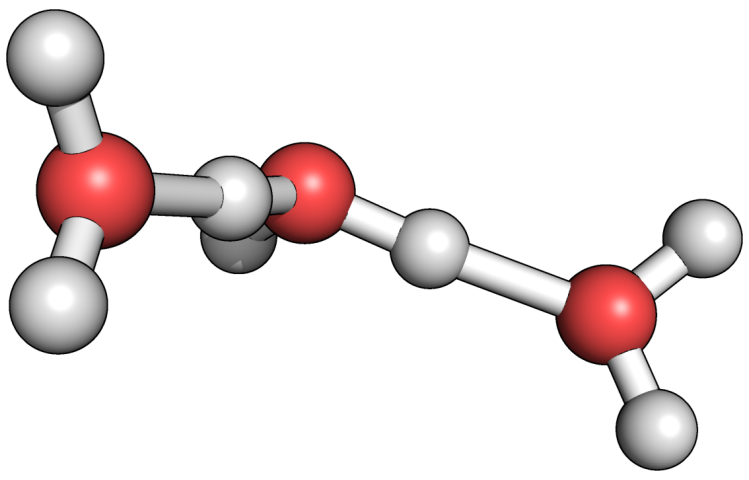

```
  10

O          -3.3464825968        3.3313553677        2.4986072452
H          -3.1294232785        4.2620714531        2.3834228523
H          -3.0480138923        3.0794274674        3.3777687516
O          -2.3622190435        0.0912093202       -0.2304965740
H          -1.8868828394        0.1687275592       -1.0639480322
H          -1.9313767498       -0.6041509528        0.2757610788
O          -3.8741282780        1.8998087396        0.5424755697
H          -3.2206470296        1.1436233814        0.2756008845
H          -4.7822541990        1.5754736966        0.6003035790
H          -3.6180350930        2.4529044676        1.3783525450
```

### CHW9/04: 3 Water, 1 Hydronium

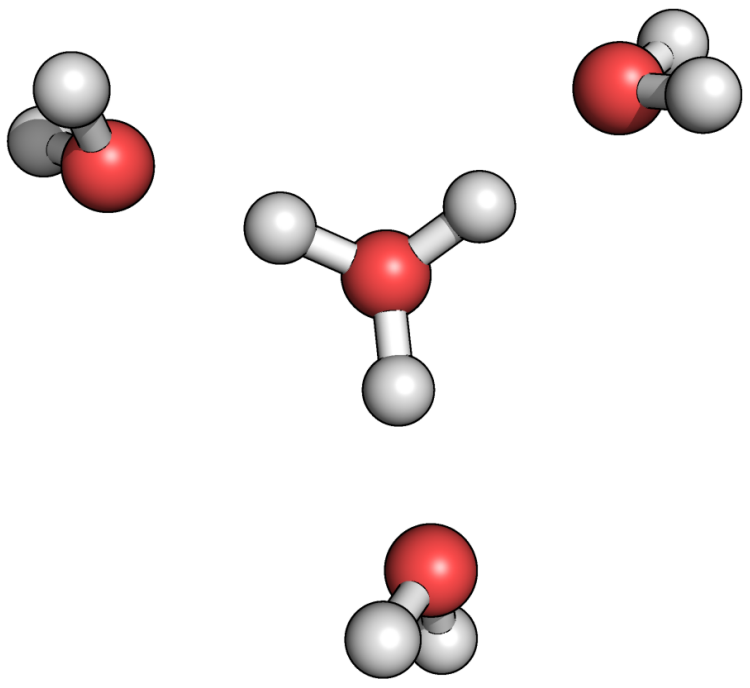

```
13

O      2.8748226451    -0.2572345142    -0.6795479135
H      3.2082739858    -1.1355973471    -0.4710768955
H      3.5915547668     0.3511201769    -0.4755761137
O     -0.2003528751     1.8829551664     1.3557887524
H     -0.6994014095     1.5869524947     2.1235804728
H     -0.4295275027     2.8093541201     1.2346734430
O     -0.7591761087     0.6889655650    -2.6990673014
H     -1.4165323020     0.0711028281    -3.0341785228
H     -0.3572384016     1.0913483560    -3.4749144238
O      0.3823289725     0.2138304976    -0.4747505796
H      0.1293253016     0.9067169620     0.2160096336
H     -0.0241344149     0.4063184293    -1.3797299621
H      1.3809255428     0.0670041654    -0.5247418895
```

**CHW9/05: 3 Water, 1 Hydronium**

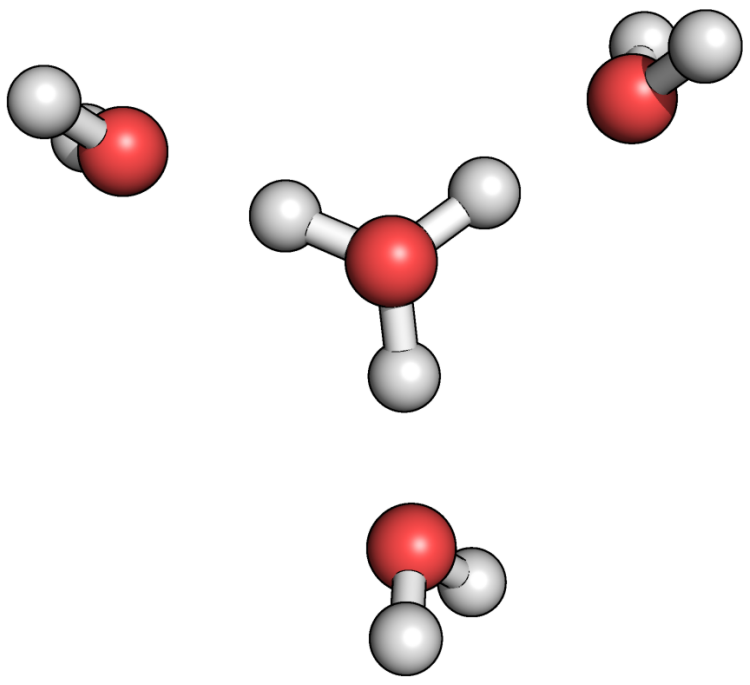

```
13

O        -0.0639189843     1.7390005297     0.6420588142
H        -0.9131637363     1.4411209241     0.3041629172
H         0.0705259436     1.2861147397     1.4786773337
O         3.5329984427     2.1555018468    -1.7960610284
H         3.5944127384     2.0774958222    -2.7530464208
H         4.1317552246     1.4943368873    -1.4363193644
O         1.9619039550     5.4040659438     0.4995464184
H         2.7964270246     5.6774574718     0.8910548905
H         1.5273734683     6.2131287244     0.2115134384
O         1.4873545672     3.2245176198    -0.7186627650
H         2.2926038157     2.7610266404    -1.1096107630
H         0.9141474568     2.6162308896    -0.1547986024
H         1.7245930836     4.0824846603    -0.2386116682
```

**CHW9/06: 3 Water, 1 Hydronium**

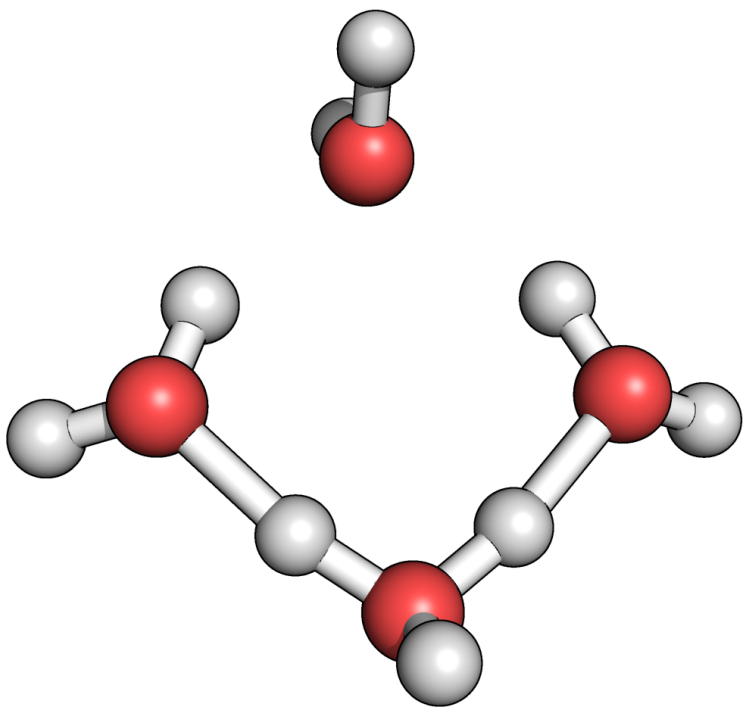

```
  13


O           0.0008745896       0.2552148196      -0.6190886732
H           0.1363742932      -0.2768320600       0.1742754767
H           0.1426801567      -0.3639458369      -1.3458634289
O          -2.2714351623       1.9079576936      -0.9616827206
H          -3.0787156897       1.8072939971      -0.4477307576
H          -1.6880040268       1.1584824826      -0.7530614314
O           1.2675403985       2.7583486779      -0.9961859678
H           1.0922619929       1.8265438169      -0.7802109151
H           2.0417402369       3.0377032703      -0.4976307524
O          -0.8922431476       3.9541546591      -1.0588194956
H          -1.0205842522       4.4551727309      -1.8753176537
H          -1.5607948650       3.1486077823      -1.0124574539
H           0.0700234758       3.5404744666      -1.0283439266
```

## CHW9/07: 3 Water, 1 Hydronium

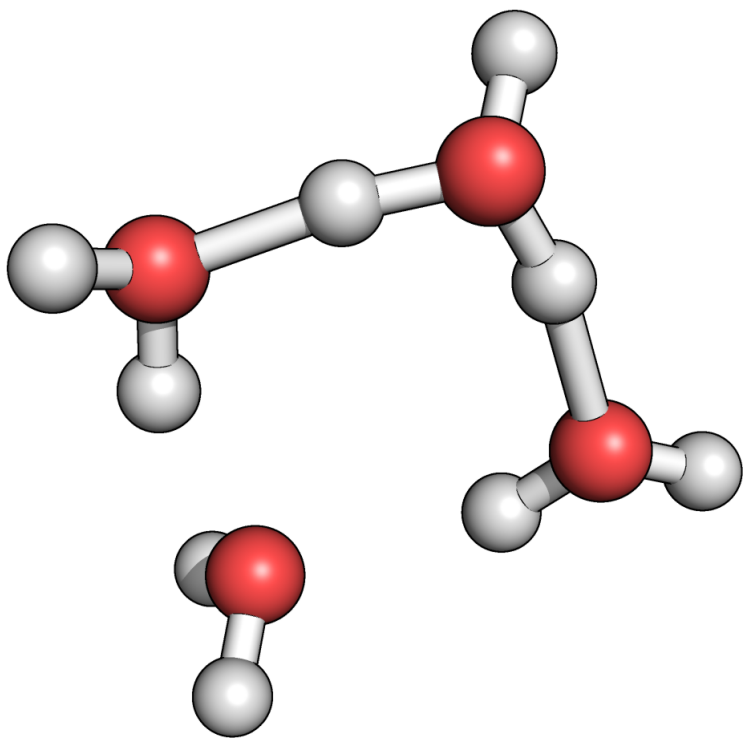

```
  13

O          -3.4884154492        3.0851398384        -2.0801865315
H          -4.4192120446        3.1274547557        -1.8293663446
H          -3.2579946641        3.9980856174        -2.2913085215
O          -2.8434945504        0.9773999385        -3.8205121640
H          -3.1531005206        1.8161365563        -3.4336150125
H          -2.6700604582        1.1273630694        -4.7553223622
O          -1.4777093138        1.7583568994        -0.5752587778
H          -2.2352299225        2.3261475216        -0.7857785469
H          -1.3269491999        1.7880008400         0.3734200839
O          -1.2063414442       -0.0185151316        -2.2799954233
H          -1.4352338763       -0.9205150768        -2.0206092098
H          -1.8798707749        0.3465706759        -3.0068542936
H          -1.2727922812        0.6284588960        -1.4725746962
```

## CHW9/08: 3 Water, 1 Hydronium

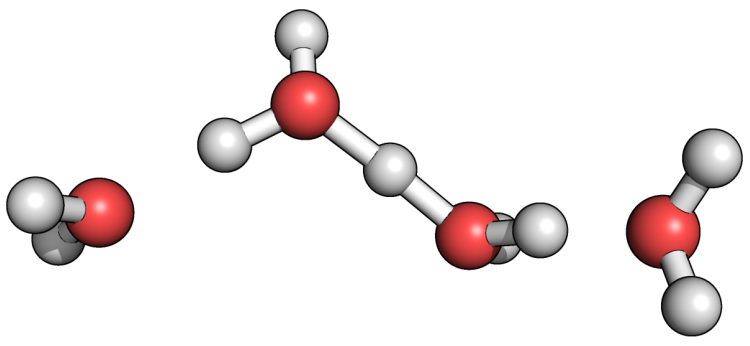

```
  13

O        2.1715892215     0.0668682826     1.0276979419
H        2.4300810064     0.3645249888     1.9044305156
H        1.9902778673     0.8646509967     0.5223571476
O        1.2709013232    -2.2321374320     0.2926696613
H        0.4749722602    -2.4693456668     0.7789932588
H        1.6015237968    -1.3459607549     0.6206131530
O        3.0598365465    -5.0631715010    -2.3207116496
H        3.8006353738    -4.9372190975    -2.9207845265
H        2.3464857399    -5.4270224446    -2.8523664639
O        2.9580698508    -3.8811424345    -0.0492524948
H        3.0859364642    -4.5496650483     0.6320392507
H        2.9520830828    -4.3292507795    -0.9481455571
H        2.1038402666    -3.1019529089     0.1596971630
```

**CHW9/09: 1 Water**

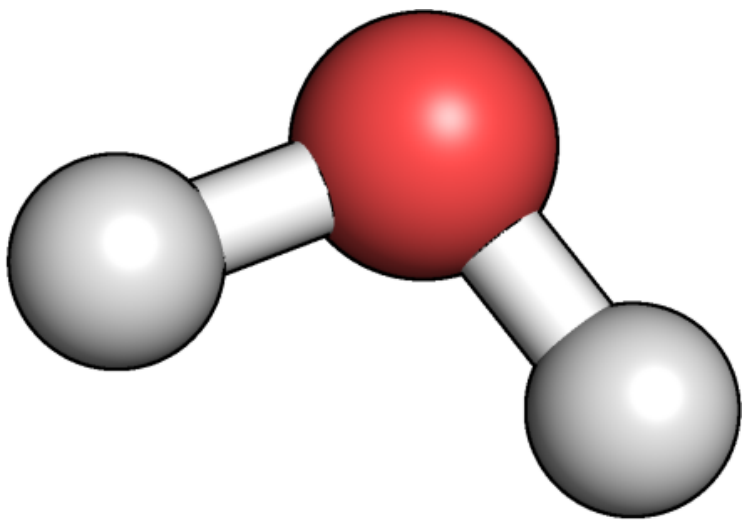

```
  3


O           1.5983771345        1.0106524941        3.8250480251
H           1.8865611993        1.5638492471        4.5535887940
H           1.5803917662        1.6135116588        3.0793364809
```

**CHW9/10: 1 Hydronium**

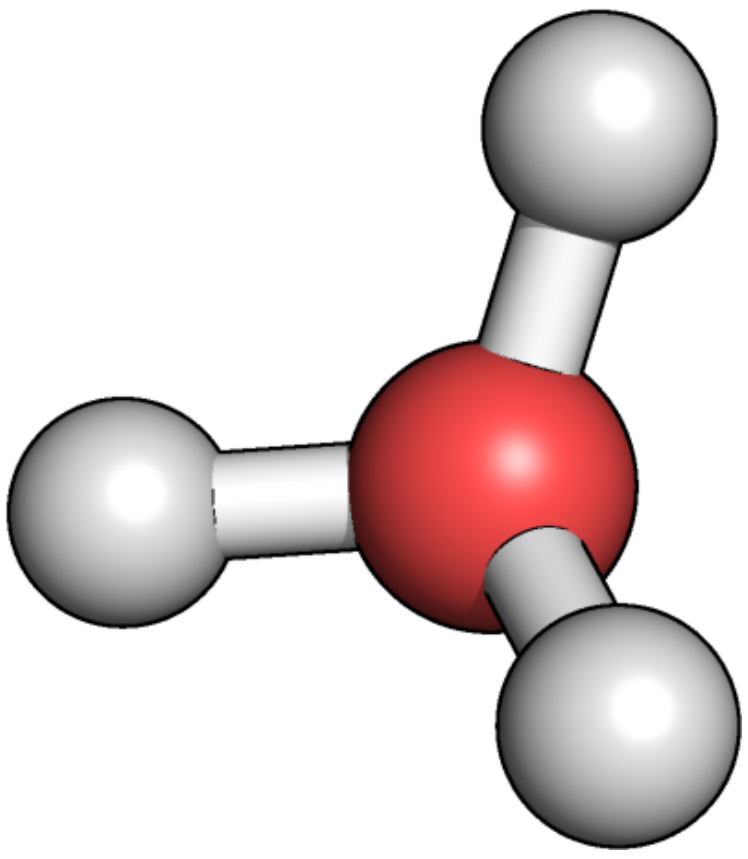

```
   4


O            2.4781953058        -1.2825558401          3.4277764063
H            3.3573087822        -1.4257622798          3.8321417449
H            1.8345452337        -1.9603189285          3.7162407697
H            2.1476428784        -0.3774304516          3.5961302791
```

## S2.3: Water ions - (W2)

Table S3: Interaction energies in kcal/mol for the W2×8 dataset as a function of displacement.

| ID | Description | E(0.90) | E(0.95) | E(1.00) | E(1.05) | E(1.10) | E(1.25) | E(1.50) | E(2.00) |
|---|---|---|---|---|---|---|---|---|---|
| 01 | Hydronium (H3O+) ... Water (H2O) | -35.95 | -49.13 | -52.13 | -50.36 | -46.48 | -32.74 | -17.34 | -6.82 |
| 02 | Hydroxide (OH-) ... Water (H2O) | -35.47 | -46.21 | -48.84 | -47.67 | -44.76 | -33.73 | -19.74 | -7.97 |

### W2/01: 1 Water ... 1 Hydronium

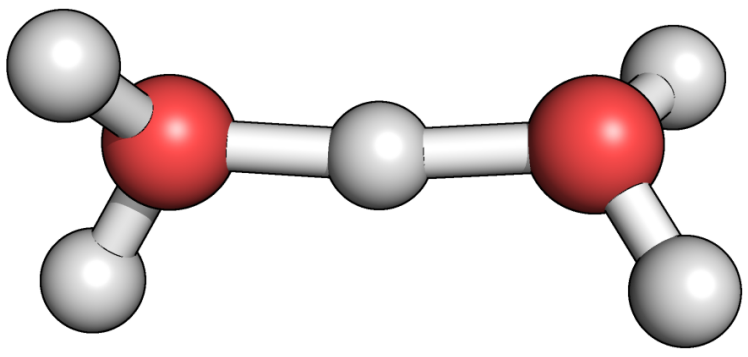

```
7


 O         2.463749936600      -1.248474117700        3.472682704300
 H         3.361306187200      -1.487701840400        3.745796353200
 H         1.858758547600      -1.960988341400        3.727563261100
 H         2.084815376100      -0.129844211300        3.682921776900
 O         1.610283063300       0.959261683100        3.803889099200
 H         1.864030609400       1.482539350500        4.578171657000
 H         1.640078579900       1.527153377300        3.019237648300
```

### W2/02: 1 Water ... 1 Hydroxide

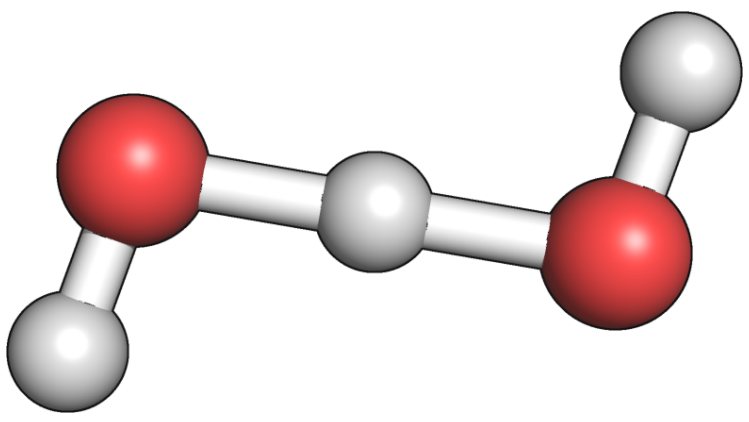

```
5


 O        -1.154095576700       0.207473003400        0.000000000000
 H        -1.483012527500      -0.694311688900        0.000000000000
 O         1.250806942700      -0.204302672300        0.000000000000
 H         1.579546804700       0.697542978000        0.000000000000
 H         0.048444426800       0.001519499700        0.000000000000
```

# S3: RMSD values

Table S4: S22 error in interaction energy for various methods, compared to a CCSD(T) reference. All values are in kcal/mol, except $r$ and $r^2$ which are unitless. *Reference energies are taken from Takatani *et al.* in Ref. S5.

| ID | Complex | $E$ Ref.* | Error PM6 | PM6-D3H4 | PBE/6-31G(d) | PBE-D3/6-31G(d) | PBE/def2-QZVP | PBE-D3/def2-QZVP |
|---|---|---|---|---|---|---|---|---|
| 1 | Ammonia dimer | -3.17 | 0.86 | -0.61 | -2.17 | -2.76 | 0.18 | -0.40 |
| 2 | Water dimer | -5.02 | 1.08 | 0.16 | -2.89 | -3.30 | -0.28 | -0.70 |
| 3 | Formic acid dimer | -18.61 | 7.48 | 0.44 | -3.90 | -5.44 | 0.13 | -1.41 |
| 4 | Formamide dimer | -15.96 | 3.41 | -0.88 | -3.60 | -5.24 | 0.94 | -0.71 |
| 5 | Uracil dimer h-bonded | -20.47 | 7.15 | 2.00 | -1.73 | -3.97 | 1.75 | -0.48 |
| 6 | 2-pyridoxine 2-aminopyridine | -16.71 | 6.73 | 0.43 | -2.91 | -5.59 | 1.19 | -1.48 |
| 7 | Adenine thymine Watson-Crick | -16.37 | 7.31 | 0.88 | -2.60 | -5.47 | 1.83 | -1.04 |
| 8 | Methane dimer | -0.53 | 0.47 | -0.13 | 0.28 | -0.32 | 0.44 | -0.16 |
| 9 | Ethene dimer | -1.51 | 1.11 | 0.51 | 0.27 | -1.04 | 1.15 | -0.17 |
| 10 | Benzene - Methane complex | -1.50 | 1.02 | -0.02 | 1.02 | -0.47 | 1.42 | -0.06 |
| 11 | Benzene dimer parallel displaced | -2.73 | 2.85 | -0.21 | 3.50 | -0.95 | 4.46 | 0.02 |
| 12 | Pyrazine dimer | -4.42 | 2.61 | -0.42 | 3.68 | -0.76 | 5.04 | 0.59 |
| 13 | Uracil dimer stack | -9.88 | 5.42 | 0.74 | 4.57 | -1.49 | 6.98 | 0.92 |
| 14 | Indole benzene complex stack | -5.22 | 5.29 | 1.14 | 6.02 | -0.23 | 7.26 | 1.01 |
| 15 | Adenine thymine complex stack | -12.23 | 7.28 | 1.48 | 7.45 | -1.04 | 10.53 | 2.05 |
| 16 | Ethene ethyne complex | -1.53 | 0.98 | 0.51 | -0.54 | -1.22 | 0.33 | -0.35 |
| 17 | Benzene water complex | -3.28 | 1.00 | -0.05 | -0.03 | -1.45 | 0.88 | -0.54 |
| 18 | Benzene ammonia complex | -2.35 | 0.82 | -0.26 | 0.58 | -0.91 | 1.25 | -0.24 |
| 19 | Benzene HCN complex | -4.46 | 2.47 | 1.63 | 0.63 | -1.19 | 1.56 | -0.26 |
| 20 | Benzene dimer T-shaped | -2.74 | 1.98 | 0.34 | 1.78 | -0.76 | 2.54 | 0.01 |
| 21 | Indole benzene T-shape complex | -5.73 | 3.32 | 1.02 | 1.92 | -1.49 | 3.52 | 0.11 |
| 22 | Phenol dimer | -7.05 | 3.67 | -0.40 | -0.01 | -2.76 | 2.97 | 0.22 |
| | RMSD | | 4.18 | 0.83 | 3.07 | 2.82 | 3.71 | 0.79 |
| | Mean deviation | | 3.38 | 0.38 | 0.51 | -2.18 | 2.55 | -0.14 |
| | Median deviation | | 2.73 | 0.38 | 0.28 | -1.34 | 1.49 | -0.20 |
| | Mean unsigned deviation | | 3.38 | 0.65 | 2.37 | 2.18 | 2.57 | 0.59 |
| | Median unsigned deviation | | 2.73 | 0.48 | 2.05 | 1.34 | 1.49 | 0.44 |
| | $r$ | | 0.96 | 0.99 | 0.93 | 0.99 | 0.91 | 0.99 |
| | $r^2$ | | 0.92 | 0.99 | 0.87 | 0.98 | 0.83 | 0.99 |
| | Max absolute deviation | | 7.48 | 2.00 | 7.45 | 5.59 | 10.53 | 2.05 |
| | lowest negative deviation | | N/A | -0.88 | -3.90 | -5.59 | -0.28 | -1.48 |
| | highest positive deviation | | 7.48 | 2.00 | 7.45 | N/A | 10.53 | 2.05 |

| ID | Complex | $E$ Ref.* | Error DFTB3 | DFTB3-D3 | DFTB3-D3H4 | DFTB3/CPE($U$)-D3* | DFTB3/CPE($U$)-D3 | DFTB3/CPE($\zeta$)-D3 | DFTB3/CPE($q$)-D3 |
|---|---|---|---|---|---|---|---|---|---|
| 1 | Ammonia dimer | -3.17 | 2.03 | 1.38 | 0.49 | 0.04 | 0.14 | -0.57 | -0.03 |
| 2 | Water dimer | -5.02 | 0.44 | -0.04 | -1.00 | -0.59 | -0.57 | -0.38 | -0.08 |
| 3 | Formic acid dimer | -18.61 | 1.23 | -0.43 | -2.61 | -0.39 | -0.36 | -0.05 | -0.17 |
| 4 | Formamide dimer | -15.96 | 2.39 | 0.65 | -1.38 | 0.50 | 0.69 | 0.73 | 2.13 |
| 5 | Uracil dimer h-bonded | -20.47 | 4.40 | 2.08 | -0.29 | 1.22 | 1.37 | 0.25 | -1.14 |
| 6 | 2-pyridoxine 2-aminopyridine complex | -16.71 | 6.62 | 3.85 | 0.42 | 3.19 | 3.24 | 3.07 | 2.62 |
| 7 | Adenine thymine Watson-Crick complex | -16.37 | 7.42 | 4.46 | 1.08 | 3.72 | 3.68 | 4.02 | 1.59 |
| 8 | Methane dimer | -0.53 | 0.43 | -0.19 | -0.26 | -0.27 | -0.20 | -0.17 | -0.18 |
| 9 | Ethene dimer | -1.51 | 1.29 | -0.08 | 0.63 | -0.19 | -0.08 | 0.05 | -0.20 |
| 10 | Benzene - Methane complex | -1.50 | 1.31 | -0.18 | 0.21 | -0.26 | -0.22 | -0.07 | -0.13 |
| 11 | Benzene dimer parallel displaced | -2.73 | 3.34 | -0.97 | -0.02 | -0.90 | -0.66 | -0.26 | -0.91 |
| 12 | Pyrazine dimer | -4.42 | 4.30 | -0.04 | 0.96 | -0.28 | -0.02 | 0.20 | -0.26 |
| 13 | Uracil dimer stack | -9.88 | 6.69 | 0.70 | 1.69 | -0.16 | 0.15 | 0.37 | 1.74 |
| 14 | Indole benzene complex stack | -5.22 | 5.84 | -0.22 | 1.32 | -0.19 | 0.12 | 0.68 | -0.12 |
| 15 | Adenine thymine complex stack | -12.23 | 8.78 | 0.41 | 2.52 | -0.79 | -0.43 | -0.17 | 1.95 |
| 16 | Ethene ethyne complex | -1.53 | 0.89 | 0.19 | 0.36 | 0.10 | 0.13 | 0.21 | 0.08 |
| 17 | Benzene water complex | -3.28 | 1.54 | 0.08 | 0.39 | -0.12 | -0.10 | -0.34 | 0.23 |
| 18 | Benzene ammonia complex | -2.35 | 1.64 | 0.13 | 0.48 | -0.70 | -0.55 | -0.89 | 0.11 |
| 19 | Benzene HCN complex | -4.46 | 3.02 | 1.18 | 2.19 | 0.99 | 1.03 | 1.27 | 1.56 |
| 20 | Benzene dimer T-shaped | -2.74 | 2.72 | 0.19 | 0.94 | 0.20 | 0.38 | 0.57 | 0.34 |
| 21 | Indole benzene T-shape complex | -5.73 | 4.44 | 1.01 | 1.86 | 0.76 | 1.00 | 1.15 | 1.38 |
| 22 | Phenol dimer | -7.05 | 3.66 | 0.86 | 0.62 | 0.64 | 0.83 | 0.81 | 0.63 |
| | RMSD | | 4.12 | 1.45 | 1.24 | 1.18 | 1.19 | 1.21 | 1.13 |
| | Mean deviation | | 3.38 | 0.68 | 0.48 | 0.30 | 0.44 | 0.48 | 0.51 |
| | Media deviation | | 2.87 | 0.19 | 0.49 | -0.14 | 0.13 | 0.20 | 0.10 |
| | Mean unsigned deviation | | 3.38 | 0.88 | 0.99 | 0.74 | 0.73 | 0.74 | 0.80 |
| | Median unsigned deviation | | 2.87 | 0.42 | 0.79 | 0.44 | 0.40 | 0.38 | 0.30 |
| | $r$ | | 0.93 | 0.98 | 0.99 | 0.99 | 0.99 | 0.99 | 0.99 |
| | $r^2$ | | 0.86 | 0.96 | 0.97 | 0.97 | 0.97 | 0.97 | 0.98 |
| | Max absolute deviation | | 8.78 | 4.46 | 2.61 | 3.72 | 3.68 | 4.02 | 2.62 |
| | lowest negative deviation | | N/A | -0.97 | -2.61 | -0.90 | -0.66 | -0.89 | -1.14 |
| | highest positive deviation | | 8.78 | 4.46 | 2.52 | 3.72 | 3.68 | 4.02 | 2.62 |

| $E$ | Error |
|---|---|

| ID | Complex | Ref.* | DFTB3/ CPE($q$) -D3 (original) 3OB | DFTB3/ CPE($q$) -D3 (original) MIO | DFTB3/ CPE($\zeta$) -D3 (pol) | DFTB3/ CPE($q$) -D3 (pol) |
|---|---|---|---|---|---|---|
| 1 | Ammonia dimer | -3.17 | -0.20 | 0.06 | 0.59 | 0.39 |
| 2 | Water dimer | -5.02 | -0.64 | -0.67 | -0.11 | 0.08 |
| 3 | Formic acid dimer | -18.61 | -1.63 | -2.66 | -1.04 | -0.26 |
| 4 | Formamide dimer | -15.96 | 0.34 | -0.58 | 0.39 | 0.96 |
| 5 | Uracil dimer h-bonded | -20.47 | 1.33 | -0.21 | 1.20 | -0.65 |
| 6 | 2-pyridoxine 2-aminopyridine complex | -16.71 | 3.19 | 1.78 | 1.31 | 2.88 |
| 7 | Adenine thymine Watson-Crick complex | -16.37 | 3.24 | 2.26 | 3.79 | 3.21 |
| 8 | Methane dimer | -0.53 | -0.19 | -0.20 | -0.19 | -0.19 |
| 9 | Ethene dimer | -1.51 | -0.15 | -0.27 | -0.03 | -0.16 |
| 10 | Benzene - Methane complex | -1.50 | -0.15 | -0.29 | -0.14 | -0.17 |
| 11 | Benzene dimer parallel displaced | -2.73 | -0.91 | -1.16 | -1.04 | -0.98 |
| 12 | Pyrazine dimer | -4.42 | 0.04 | -0.17 | -0.39 | -0.11 |
| 13 | Uracil dimer stack | -9.88 | -1.92 | -1.85 | -1.48 | 1.44 |
| 14 | Indole benzene complex stack | -5.22 | -0.08 | -0.53 | -0.15 | -0.19 |
| 15 | Adenine thymine complex stack | -12.23 | -0.37 | -0.86 | -1.67 | 1.37 |
| 16 | Ethene ethyne complex | -1.53 | 0.03 | -0.03 | 0.30 | 0.06 |
| 17 | Benzene water complex | -3.28 | 0.05 | -0.09 | 0.30 | 0.16 |
| 18 | Benzene ammonia complex | -2.35 | 0.16 | 0.04 | 0.14 | 0.09 |
| 19 | Benzene HCN complex | -4.46 | 1.17 | 0.60 | 1.49 | 1.38 |
| 20 | Benzene dimer T-shaped | -2.74 | 0.24 | -0.17 | 0.32 | 0.27 |
| 21 | Indole benzene T-shape complex | -5.73 | 1.05 | 0.34 | 1.41 | 1.30 |
| 22 | Phenol dimer | -7.05 | 0.40 | 0.08 | 0.20 | 0.61 |
| | RMSD | | 1.23 | 1.02 | 1.17 | 1.15 |
| | Mean deviation | | 0.23 | -0.21 | 0.24 | 0.52 |
| | Media deviation | | 0.03 | -0.18 | 0.17 | 0.12 |
| | Mean unsigned deviation | | 0.79 | 0.68 | 0.80 | 0.77 |
| | Median unsigned deviation | | 0.35 | 0.32 | 0.39 | 0.33 |
| | $r$ | | 0.98 | 0.99 | 0.98 | 0.99 |
| | $r^2$ | | 0.96 | 0.98 | 0.97 | 0.97 |
| | Max absolute deviation | | 3.24 | 2.66 | 3.79 | 3.21 |
| | lowest negative deviation | | -1.92 | -2.66 | -1.67 | -0.98 |
| | highest positive deviation | | 3.24 | 2.26 | 3.79 | 3.21 |

Table S5: S66 error in interaction energy for various methods, compared to a CCSD(T) reference. All values are in kcal/mol, except $r$ and $r^2$ which are unitless. Reference geometries and energies are taken from Ref. S4.

| | | $E$ | Error | | | | | |
|---|---|---|---|---|---|---|---|---|
| ID | Complex | Ref.* | PM6 | PM6-D3H4 | PBE/ 6-31G(d) | PBE-D3/ 6-31G(d) | PBE/ def2-QZVP | PBE-D3/ def2-QZVP |
| 1 | Water ... Water | -4.89 | 1.03 | 0.01 | -2.65 | -3.04 | -0.31 | -0.70 |
| 2 | Water ... MeOH | -5.57 | 1.35 | 0.09 | -1.97 | -2.65 | 0.15 | -0.53 |
| 3 | Water ... MeNH2 | -6.88 | 2.69 | -0.65 | -2.23 | -2.94 | -0.45 | -1.16 |
| 4 | Water ... Peptide | -8.08 | 1.81 | 0.23 | -1.67 | -2.68 | 0.59 | -0.41 |
| 5 | MeOH ... MeOH | -5.75 | 2.24 | -0.56 | -2.04 | -2.94 | 0.34 | -0.56 |
| 6 | MeOH ... MeNH2 | -7.54 | 4.31 | 0.49 | -2.35 | -3.62 | 0.10 | -1.16 |
| 7 | MeOH ... Peptide | -8.22 | 3.27 | 0.05 | -2.55 | -3.97 | 0.74 | -0.67 |
| 8 | MeOH ... Water | -5.00 | 1.80 | -0.68 | -2.71 | -3.24 | -0.14 | -0.67 |
| 9 | MeNH2 ... MeOH | -3.04 | 0.85 | -1.01 | -1.30 | -2.33 | 0.60 | -0.43 |
| 10 | MeNH2 ... MeNH2 | -4.15 | 2.19 | -0.43 | -1.08 | -2.43 | 0.88 | -0.47 |
| 11 | MeNH2 ... Peptide | -5.41 | 1.57 | -0.66 | -0.84 | -2.80 | 1.82 | -0.14 |
| 12 | MeNH2 ... Water | -7.25 | 3.21 | -0.17 | -2.05 | -2.90 | -0.08 | -0.94 |
| 13 | Peptide ... MeOH | -6.18 | 1.95 | -0.27 | -1.26 | -2.76 | 1.36 | -0.14 |
| 14 | Peptide ... MeNH2 | -7.45 | 3.20 | -0.04 | -1.62 | -3.46 | 1.07 | -0.77 |
| 15 | Peptide ... Peptide | -8.62 | 2.72 | -0.12 | -0.80 | -3.08 | 2.04 | -0.24 |
| 16 | Peptide ... Water | -5.12 | 1.30 | -0.34 | -2.09 | -2.85 | 0.42 | -0.34 |
| 17 | Uracil ... Uracil (BP) | -17.18 | 5.81 | 0.76 | -2.20 | -4.40 | 1.47 | -0.72 |
| 18 | Water ... Pyridine | -6.83 | 3.51 | -0.13 | -1.10 | -1.97 | -0.20 | -1.08 |
| 19 | MeOH ... Pyridine | -7.40 | 5.08 | 1.13 | -1.18 | -2.49 | 0.24 | -1.06 |
| 20 | AcOH ... AcOH | -19.09 | 7.95 | 0.63 | -3.96 | -5.62 | 0.17 | -1.48 |
| 21 | AcNH2 ... AcNH2 | -16.26 | 3.89 | -0.68 | -3.27 | -5.02 | 0.97 | -0.77 |
| 22 | AcOH ... Uracil | -19.49 | 7.43 | 1.08 | -2.93 | -4.83 | 0.90 | -1.00 |
| 23 | AcNH2 ... Uracil | -19.19 | 5.13 | 0.25 | -2.62 | -4.62 | 1.28 | -0.72 |
| 24 | Benzene ... Benzene (pi-pi) | -2.74 | 2.87 | -0.19 | 3.49 | -0.92 | 4.45 | 0.04 |
| 25 | Pyridine ... Pyridine (pi-pi) | -3.83 | 2.85 | -0.27 | 3.53 | -1.02 | 4.69 | 0.14 |
| 26 | Uracil ... Uracil (pi-pi) | -9.82 | 5.36 | 0.67 | 4.44 | -1.62 | 6.83 | 0.77 |
| 27 | Benzene ... Pyridine (pi-pi) | -3.37 | 2.87 | -0.22 | 3.57 | -0.91 | 4.63 | 0.15 |
| 28 | Benzene ... Uracil (pi-pi) | -5.71 | 4.11 | 0.27 | 4.43 | -0.94 | 6.04 | 0.67 |
| 29 | Pyridine ... Uracil (pi-pi) | -6.81 | 3.59 | -0.21 | 4.50 | -0.75 | 5.99 | 0.74 |
| 30 | Benzene ... Ethene | -1.41 | 1.56 | -0.22 | 1.73 | -0.73 | 2.31 | -0.16 |
| 31 | Uracil ... Ethene | -3.38 | 2.33 | 0.29 | 1.86 | -0.90 | 2.85 | 0.10 |
| 32 | Uracil ... Ethyne | -3.74 | 2.65 | 1.01 | 1.71 | -0.69 | 2.57 | 0.16 |
| 33 | Pyridine ... Ethene | -1.86 | 1.63 | -0.17 | 1.68 | -0.80 | 2.37 | -0.11 |
| 34 | Pentane ... Pentane | -3.77 | 3.11 | 0.59 | 3.06 | -1.19 | 4.03 | -0.21 |
| 35 | Neopentane ... Pentane | -2.61 | 1.91 | 0.25 | 1.77 | -1.18 | 2.66 | -0.29 |
| 36 | Neopentane ... Neopentane | -1.77 | 1.23 | -0.10 | 0.76 | -1.36 | 1.68 | -0.43 |
| 37 | Cyclopentane ... Neopentane | -2.41 | 1.71 | 0.12 | 1.66 | -1.24 | 2.54 | -0.35 |
| 38 | Cyclopentane ... Cyclopentane | -3.00 | 2.60 | 0.92 | 2.13 | -1.17 | 3.06 | -0.24 |
| 39 | Benzene ... Cyclopentane | -3.57 | 3.02 | 0.20 | 2.68 | -1.21 | 3.93 | 0.03 |
| 40 | Benzene ... Neopentane | -2.89 | 2.16 | -0.22 | 1.82 | -1.27 | 2.93 | -0.16 |
| 41 | Uracil ... Pentane | -4.84 | 3.06 | -0.68 | 2.92 | -1.85 | 5.05 | 0.29 |
| 42 | Uracil ... Cyclopentane | -4.13 | 2.89 | -0.36 | 2.70 | -1.48 | 4.41 | 0.23 |
| 43 | Uracil ... Neopentane | -3.70 | 2.66 | -0.09 | 2.05 | -1.36 | 3.45 | 0.04 |
| 44 | Ethene ... Pentane | -1.99 | 1.52 | 0.21 | 1.02 | -1.07 | 1.82 | -0.27 |
| 45 | Ethyne ... Pentane | -1.75 | 1.45 | -0.03 | 0.60 | -1.27 | 1.61 | -0.27 |
| 46 | Peptide ... Pentane | -4.24 | 3.00 | 0.24 | 2.08 | -1.82 | 3.94 | 0.04 |
| 47 | Benzene ... Benzene (TS) | -2.87 | 2.06 | 0.28 | 1.84 | -0.86 | 2.73 | 0.02 |
| 48 | Pyridine ... Pyridine (TS) | -3.53 | 2.32 | 0.60 | 1.55 | -1.10 | 2.73 | 0.08 |
| 49 | Benzene ... Pyridine (TS) | -3.32 | 2.16 | 0.40 | 1.81 | -0.90 | 2.73 | 0.01 |
| 50 | Benzene ... Ethyne (CH-pi) | -2.86 | 1.86 | 0.94 | 0.75 | -0.97 | 1.56 | -0.15 |
| 51 | Ethyne ... Ethyne (TS) | -1.52 | 1.06 | 0.71 | -0.54 | -1.15 | 0.26 | -0.35 |
| 52 | Benzene ... AcOH (OH-pi) | -4.70 | 2.11 | 0.52 | 0.53 | -1.70 | 2.14 | -0.09 |
| 53 | Benzene ... AcNH2 (NH-pi) | -4.36 | 1.98 | 0.47 | -0.20 | -2.20 | 1.88 | -0.13 |
| 54 | Benzene ... Water (OH-pi) | -3.27 | 0.98 | -0.06 | -0.04 | -1.46 | 0.87 | -0.55 |
| 55 | Benzene ... MeOH (OH-pi) | -4.19 | 2.27 | 0.55 | 0.74 | -1.64 | 2.04 | -0.34 |
| 56 | Benzene ... MeNH2 (NH-pi) | -3.23 | 1.80 | 0.00 | 1.25 | -1.22 | 2.27 | -0.20 |
| 57 | Benzene ... Peptide (NH-pi) | -5.28 | 3.00 | 0.46 | 1.91 | -1.63 | 3.49 | -0.05 |
| 58 | Pyridine ... Pyridine (CH-N) | -4.15 | 1.63 | 0.93 | -0.03 | -1.52 | 1.54 | 0.04 |
| 59 | Ethyne ... Water (CH-O) | -2.85 | 1.12 | 0.87 | -1.50 | -1.88 | -0.01 | -0.38 |
| 60 | Ethyne ... AcOH (OH-pi) | -4.86 | 2.97 | 2.36 | -1.49 | -2.67 | 0.62 | -0.56 |
| 61 | Pentane ... AcOH | -2.88 | 1.66 | -0.54 | 0.85 | -1.91 | 2.58 | -0.17 |
| 62 | Pentane ... AcNH2 | -3.51 | 2.07 | -0.27 | 0.71 | -2.35 | 2.87 | -0.18 |
| 63 | Benzene ... AcOH | -3.80 | 2.20 | 0.09 | 1.91 | -1.11 | 3.13 | 0.12 |
| 64 | Peptide ... Ethene | -2.99 | 1.77 | 0.27 | 0.28 | -1.70 | 1.86 | -0.12 |
| 65 | Pyridine ... Ethyne | -3.99 | 2.70 | 2.12 | -1.01 | -1.88 | 0.27 | -0.59 |
| 66 | MeNH2 ... Pyridine | -3.97 | 2.62 | -0.50 | 0.76 | -1.47 | 2.10 | -0.12 |
| | RMSD | | 2.99 | 0.64 | 2.14 | 2.34 | 2.65 | 0.52 |
| | Mean deviation | | 2.65 | 0.17 | 0.30 | -2.04 | 2.05 | -0.29 |
| | Median deviation | | 2.29 | 0.09 | 0.65 | -1.67 | 1.87 | -0.23 |
| | Mean unsigned deviation | | 2.65 | 0.47 | 1.85 | 2.04 | 2.09 | 0.40 |
| | Median unsigned deviation | | 2.29 | 0.31 | 1.79 | 1.67 | 1.87 | 0.28 |
| | $r$ | | 0.97 | 0.99 | 0.95 | 0.99 | 0.94 | 1.00 |
| | $r^2$ | | 0.94 | 0.98 | 0.89 | 0.99 | 0.88 | 0.99 |
| | Max absolute deviation | | 7.95 | 2.36 | 4.50 | 5.62 | 6.83 | 1.48 |
| | lowest negative deviation | | N/A | -1.01 | -3.96 | -5.62 | -0.45 | -1.48 |
| | highest positive deviation | | 7.95 | 2.36 | 4.50 | N/A | 6.83 | 0.77 |

| ID | Complex | $E$ Ref.* | Error DFTB3 | DFTB3-D3 | DFTB3-D3H4 | DFTB3/CPE($U$)-D3* | DFTB3/CPE($U$)-D3 | DFTB3/CPE($\zeta$)-D3 | DFTB3/CPE($q$)-D3 |
|---|---|---|---|---|---|---|---|---|---|
| 1 | Water ... Water | -4.89 | 0.29 | -0.17 | -1.20 | -0.76 | -0.74 | -0.56 | -0.21 |
| 2 | Water ... MeOH | -5.57 | 1.05 | 0.29 | -0.72 | -0.10 | -0.07 | -0.06 | -0.17 |
| 3 | Water ... MeNH2 | -6.88 | 3.10 | 2.29 | 0.26 | 1.77 | 1.77 | 1.67 | 0.06 |
| 4 | Water ... Peptide | -8.08 | 0.49 | -0.63 | -1.58 | -0.86 | -0.69 | -0.93 | 0.07 |
| 5 | MeOH ... MeOH | -5.75 | 1.73 | 0.75 | 0.06 | 0.41 | 0.45 | 0.49 | 0.11 |
| 6 | MeOH ... MeNH2 | -7.54 | 4.30 | 2.92 | 0.89 | 1.97 | 2.08 | 1.72 | 1.25 |
| 7 | MeOH ... Peptide | -8.22 | 1.73 | 0.21 | -0.42 | -0.12 | 0.13 | -0.15 | 0.54 |
| 8 | MeOH ... Water | -5.00 | 1.01 | 0.42 | -0.27 | -0.25 | -0.30 | -0.34 | -0.03 |
| 9 | MeNH2 ... MeOH | -3.04 | 1.60 | 0.46 | -0.15 | 0.04 | 0.16 | -0.09 | -0.53 |
| 10 | MeNH2 ... MeNH2 | -4.15 | 3.18 | 1.70 | 0.48 | 0.43 | 0.59 | -0.02 | -0.03 |
| 11 | MeNH2 ... Peptide | -5.41 | 3.09 | 1.00 | 0.98 | 0.80 | 0.97 | 0.72 | -0.26 |
| 12 | MeNH2 ... Water | -7.25 | 3.94 | 2.97 | 1.10 | 2.29 | 2.33 | 1.97 | -1.40 |
| 13 | Peptide ... MeOH | -6.18 | 2.60 | 1.01 | 0.52 | 0.32 | 0.29 | -0.04 | -0.08 |
| 14 | Peptide ... MeNH2 | -7.45 | 4.57 | 2.62 | 1.16 | 0.47 | 0.56 | -0.27 | 0.19 |
| 15 | Peptide ... Peptide | -8.62 | 2.99 | 0.61 | 0.20 | 0.22 | 0.51 | 0.24 | 0.80 |
| 16 | Peptide ... Water | -5.12 | 1.35 | 0.53 | -0.08 | -0.29 | -0.45 | -0.99 | -0.27 |
| 17 | Uracil ... Uracil (BP) | -17.18 | 4.13 | 1.84 | -0.46 | 0.96 | 0.94 | 1.17 | -0.70 |
| 18 | Water ... Pyridine | -6.83 | 3.46 | 2.50 | 0.31 | 2.32 | 2.37 | 2.13 | 0.30 |
| 19 | MeOH ... Pyridine | -7.40 | 4.39 | 3.01 | 0.87 | 2.74 | 2.86 | 2.58 | 1.40 |
| 20 | AcOH ... AcOH | -19.09 | 1.32 | -0.45 | -2.72 | -0.35 | -0.32 | 0.02 | -0.47 |
| 21 | AcNH2 ... AcNH2 | -16.26 | 2.36 | 0.51 | -1.68 | -0.01 | 0.19 | 0.13 | 1.12 |
| 22 | AcOH ... Uracil | -19.49 | 2.78 | 0.77 | -1.91 | 0.49 | 0.60 | 0.31 | -0.65 |
| 23 | AcNH2 ... Uracil | -19.19 | 3.56 | 1.46 | -0.82 | 0.70 | 0.90 | 0.26 | 0.50 |
| 24 | Benzene ... Benzene (pi-pi) | -2.74 | 3.26 | -0.56 | -0.01 | -0.48 | -0.23 | 0.09 | -0.44 |
| 25 | Pyridine ... Pyridine (pi-pi) | -3.83 | 3.98 | -0.07 | 0.63 | -0.13 | 0.12 | 0.44 | 0.51 |
| 26 | Uracil ... Uracil (pi-pi) | -9.82 | 6.66 | 0.56 | 1.65 | -0.31 | -0.01 | 0.25 | 1.59 |
| 27 | Benzene ... Pyridine (pi-pi) | -3.37 | 3.69 | -0.27 | 0.39 | -0.26 | -0.01 | 0.30 | -0.09 |
| 28 | Benzene ... Uracil (pi-pi) | -5.71 | 4.75 | -0.37 | 0.50 | -0.59 | -0.44 | -0.24 | -0.05 |
| 29 | Pyridine ... Uracil (pi-pi) | -6.81 | 4.89 | -0.16 | 0.67 | -0.43 | -0.22 | -0.18 | 0.55 |
| 30 | Benzene ... Ethene | -1.41 | 1.77 | -0.46 | -0.16 | -0.50 | -0.42 | -0.18 | -0.44 |
| 31 | Uracil ... Ethene | -3.38 | 2.51 | -0.26 | 0.25 | -0.56 | -0.59 | -0.51 | -0.24 |
| 32 | Uracil ... Ethyne | -3.74 | 2.24 | -0.14 | 0.45 | -0.35 | -0.35 | -0.32 | -0.09 |
| 33 | Pyridine ... Ethene | -1.86 | 2.12 | -0.19 | 0.15 | -0.29 | -0.17 | 0.06 | -0.20 |
| 34 | Pentane ... Pentane | -3.77 | 3.48 | -0.99 | 0.46 | -1.05 | -0.54 | -0.25 | -0.98 |
| 35 | Neopentane ... Pentane | -2.61 | 2.19 | -0.93 | 0.20 | -0.97 | -0.60 | -0.43 | -0.99 |
| 36 | Neopentane ... Neopentane | -1.77 | 1.28 | -1.00 | -0.35 | -1.02 | -0.74 | -0.64 | -0.96 |
| 37 | Cyclopentane ... Neopentane | -2.41 | 1.96 | -1.01 | -0.06 | -1.03 | -0.68 | -0.50 | -0.98 |
| 38 | Cyclopentane ... Cyclopentane | -3.00 | 2.73 | -0.70 | 0.61 | -0.64 | -0.23 | -0.03 | -0.62 |
| 39 | Benzene ... Cyclopentane | -3.57 | 3.64 | -0.16 | 0.53 | -0.08 | 0.24 | 0.55 | -0.10 |
| 40 | Benzene ... Neopentane | -2.89 | 2.68 | -0.37 | 0.06 | -0.31 | -0.06 | 0.17 | -0.42 |
| 41 | Uracil ... Pentane | -4.84 | 4.17 | -0.79 | 0.10 | -1.20 | -1.06 | -1.02 | -0.70 |
| 42 | Uracil ... Cyclopentane | -4.13 | 3.75 | -0.58 | 0.15 | -0.73 | -0.54 | -0.38 | -0.35 |
| 43 | Uracil ... Neopentane | -3.70 | 3.50 | -0.09 | 0.43 | -0.18 | 0.04 | 0.13 | 0.07 |
| 44 | Ethene ... Pentane | -1.99 | 1.80 | -0.48 | 0.40 | -0.59 | -0.38 | -0.20 | -0.56 |
| 45 | Ethyne ... Pentane | -1.75 | 1.67 | -0.21 | 0.02 | -0.30 | -0.18 | -0.02 | -0.37 |
| 46 | Peptide ... Pentane | -4.24 | 3.70 | -0.49 | 0.73 | -0.78 | -0.51 | -0.77 | -0.94 |
| 47 | Benzene ... Benzene (TS) | -2.87 | 2.74 | 0.15 | 0.82 | 0.16 | 0.36 | 0.56 | 0.31 |
| 48 | Pyridine ... Pyridine (TS) | -3.53 | 3.12 | 0.51 | 1.31 | 0.43 | 0.66 | 0.78 | 0.80 |
| 49 | Benzene ... Pyridine (TS) | -3.32 | 3.08 | 0.45 | 1.19 | 0.45 | 0.65 | 0.86 | 0.53 |
| 50 | Benzene ... Ethyne (CH-pi) | -2.86 | 1.98 | 0.29 | 1.02 | 0.22 | 0.32 | 0.51 | 0.42 |
| 51 | Ethyne ... Ethyne (TS) | -1.52 | 0.84 | 0.20 | 0.46 | 0.15 | 0.17 | 0.26 | 0.13 |
| 52 | Benzene ... AcOH (OH-pi) | -4.70 | 2.33 | 0.02 | 0.51 | -0.22 | 0.02 | -0.32 | 0.32 |
| 53 | Benzene ... AcNH2 (NH-pi) | -4.36 | 2.60 | 0.52 | 0.89 | 0.14 | 0.27 | -0.11 | 0.62 |
| 54 | Benzene ... Water (OH-pi) | -3.27 | 1.50 | -0.02 | 0.32 | -0.24 | -0.22 | -0.49 | 0.13 |
| 55 | Benzene ... MeOH (OH-pi) | -4.19 | 2.76 | 0.32 | 0.83 | 0.23 | 0.40 | 0.45 | 0.64 |
| 56 | Benzene ... MeNH2 (NH-pi) | -3.23 | 2.75 | 0.24 | 0.76 | -0.26 | -0.02 | 0.02 | 0.22 |
| 57 | Benzene ... Peptide (NH-pi) | -5.28 | 3.96 | 0.43 | 1.15 | -0.10 | 0.16 | 0.32 | 0.53 |
| 58 | Pyridine ... Pyridine (CH-N) | -4.15 | 3.63 | 2.09 | 2.86 | 2.01 | 2.13 | 2.20 | 1.16 |
| 59 | Ethyne ... Water (CH-O) | -2.85 | 0.42 | 0.01 | 0.15 | -0.29 | -0.33 | -0.60 | -0.51 |
| 60 | Ethyne ... AcOH (OH-pi) | -4.86 | 1.77 | 0.51 | 1.15 | 0.40 | 0.44 | 0.57 | -0.10 |
| 61 | Pentane ... AcOH | -2.88 | 2.39 | -0.63 | -0.09 | -0.86 | -0.64 | -0.80 | -0.88 |
| 62 | Pentane ... AcNH2 | -3.51 | 2.68 | -0.65 | 0.08 | -1.23 | -1.02 | -1.59 | -1.03 |
| 63 | Benzene ... AcOH | -3.80 | 3.01 | -0.01 | 0.75 | -0.22 | -0.07 | -0.16 | -0.03 |
| 64 | Peptide ... Ethene | -2.99 | 2.12 | 0.02 | 0.50 | -0.13 | -0.04 | -0.09 | -0.42 |
| 65 | Pyridine ... Ethyne | -3.99 | 2.09 | 1.19 | 1.45 | 1.11 | 1.19 | 1.23 | 0.85 |
| 66 | MeNH2 ... Pyridine | -3.97 | 3.46 | 1.16 | 0.21 | 0.61 | 0.76 | 0.55 | 0.39 |
| | RMSD | | 2.99 | 1.07 | 0.89 | 0.85 | 0.84 | 0.80 | 0.63 |
| | Mean deviation | | 2.74 | 0.36 | 0.29 | 0.04 | 0.19 | 0.16 | -0.02 |
| | Media deviation | | 2.74 | 0.18 | 0.39 | -0.16 | 0.01 | 0.00 | -0.06 |
| | Mean unsigned deviation | | 2.74 | 0.75 | 0.67 | 0.62 | 0.58 | 0.56 | 0.51 |
| | Median unsigned deviation | | 2.74 | 0.51 | 0.50 | 0.43 | 0.43 | 0.36 | 0.44 |
| | $r$ | | 0.96 | 0.97 | 0.99 | 0.98 | 0.98 | 0.98 | 0.99 |
| | $r^2$ | | 0.92 | 0.94 | 0.98 | 0.96 | 0.96 | 0.96 | 0.98 |
| | Max absolute deviation | | 6.66 | 3.01 | 2.86 | 2.74 | 2.86 | 2.58 | 1.59 |
| | lowest negative deviation | | N/A | -1.01 | -2.72 | -1.23 | -1.06 | -1.59 | -1.40 |
| | highest positive deviation | | 6.66 | 3.01 | 2.86 | 2.74 | 2.86 | 2.58 | 1.59 |

| | | $E$ | Error | | | |
|---|---|---|---|---|---|---|
| ID | Complex | Ref.* | DFTB3/ CPE($q$) -D3 (original) 3OB | DFTB3/ CPE($q$) -D3 (original) MIO | DFTB3/ CPE($\zeta$) -D3 (pol) | DFTB3/ CPE($q$) -D3 (pol) |
| 1 | Water ... Water | -4.89 | -0.73 | -0.75 | -0.23 | -0.05 |
| 2 | Water ... MeOH | -5.57 | -0.50 | -0.55 | 0.15 | 0.22 |
| 3 | Water ... MeNH2 | -6.88 | 1.73 | 1.07 | 1.55 | 0.15 |
| 4 | Water ... Peptide | -8.08 | -1.28 | -1.23 | -0.98 | -0.31 |
| 5 | MeOH ... MeOH | -5.75 | 0.35 | 0.11 | 0.34 | 0.44 |
| 6 | MeOH ... MeNH2 | -7.54 | 2.74 | 1.98 | 1.59 | 0.78 |
| 7 | MeOH ... Peptide | -8.22 | 0.14 | -0.12 | -0.07 | 0.26 |
| 8 | MeOH ... Water | -5.00 | 0.22 | 0.02 | -0.10 | 0.16 |
| 9 | MeNH2 ... MeOH | -3.04 | 0.32 | 0.23 | 0.36 | -0.17 |
| 10 | MeNH2 ... MeNH2 | -4.15 | 0.78 | 0.38 | 0.78 | 0.82 |
| 11 | MeNH2 ... Peptide | -5.41 | 0.59 | 0.41 | 1.15 | -0.12 |
| 12 | MeNH2 ... Water | -7.25 | 1.51 | 0.95 | 2.02 | -1.24 |
| 13 | Peptide ... MeOH | -6.18 | 0.82 | 0.50 | 0.43 | 0.28 |
| 14 | Peptide ... MeNH2 | -7.45 | 2.12 | 0.89 | 0.86 | 1.44 |
| 15 | Peptide ... Peptide | -8.62 | 0.57 | 0.21 | 0.61 | 0.46 |
| 16 | Peptide ... Water | -5.12 | 0.46 | 0.27 | -0.09 | -0.06 |
| 17 | Uracil ... Uracil (BP) | -17.18 | 1.40 | -0.09 | 1.98 | -0.30 |
| 18 | Water ... Pyridine | -6.83 | 2.40 | 1.65 | 2.24 | -0.05 |
| 19 | MeOH ... Pyridine | -7.40 | 3.09 | 2.31 | 2.50 | 1.07 |
| 20 | AcOH ... AcOH | -19.09 | -1.41 | -2.32 | -0.93 | -0.20 |
| 21 | AcNH2 ... AcNH2 | -16.26 | 0.42 | -0.46 | -0.25 | 0.04 |
| 22 | AcOH ... Uracil | -19.49 | -0.19 | -1.43 | -0.43 | -0.21 |
| 23 | AcNH2 ... Uracil | -19.19 | 1.03 | -0.14 | 0.64 | 0.00 |
| 24 | Benzene ... Benzene (pi-pi) | -2.74 | -0.49 | -0.59 | -0.60 | -0.55 |
| 25 | Pyridine ... Pyridine (pi-pi) | -3.83 | 0.04 | -0.09 | -0.07 | 0.12 |
| 26 | Uracil ... Uracil (pi-pi) | -9.82 | -2.18 | -2.17 | -1.73 | 1.30 |
| 27 | Benzene ... Pyridine (pi-pi) | -3.37 | -0.16 | -0.29 | -0.28 | -0.22 |
| 28 | Benzene ... Uracil (pi-pi) | -5.71 | -0.13 | -0.40 | 0.08 | -0.21 |
| 29 | Pyridine ... Uracil (pi-pi) | -6.81 | -0.14 | -0.32 | -0.68 | 0.17 |
| 30 | Benzene ... Ethene | -1.41 | -0.39 | -0.42 | -0.59 | -0.45 |
| 31 | Uracil ... Ethene | -3.38 | -0.24 | -0.31 | -0.29 | -0.33 |
| 32 | Uracil ... Ethyne | -3.74 | -0.06 | -0.16 | 0.07 | -0.24 |
| 33 | Pyridine ... Ethene | -1.86 | -0.08 | -0.12 | -0.23 | -0.21 |
| 34 | Pentane ... Pentane | -3.77 | -0.85 | -1.11 | -0.93 | -0.94 |
| 35 | Neopentane ... Pentane | -2.61 | -0.91 | -1.04 | -0.82 | -0.94 |
| 36 | Neopentane ... Neopentane | -1.77 | -0.93 | -0.97 | -0.90 | -0.97 |
| 37 | Cyclopentane ... Neopentane | -2.41 | -0.96 | -1.09 | -0.91 | -0.97 |
| 38 | Cyclopentane ... Cyclopentane | -3.00 | -0.58 | -0.78 | -0.65 | -0.62 |
| 39 | Benzene ... Cyclopentane | -3.57 | -0.06 | -0.32 | -0.16 | -0.12 |
| 40 | Benzene ... Neopentane | -2.89 | -0.34 | -0.49 | -0.29 | -0.37 |
| 41 | Uracil ... Pentane | -4.84 | -0.85 | -1.16 | -1.73 | -0.77 |
| 42 | Uracil ... Cyclopentane | -4.13 | -0.38 | -0.68 | -1.40 | -0.45 |
| 43 | Uracil ... Neopentane | -3.70 | -0.04 | -0.28 | -1.07 | -0.15 |
| 44 | Ethene ... Pentane | -1.99 | -0.51 | -0.66 | -0.42 | -0.51 |
| 45 | Ethyne ... Pentane | -1.75 | -0.24 | -0.33 | -0.23 | -0.28 |
| 46 | Peptide ... Pentane | -4.24 | -0.60 | -0.86 | -0.91 | -0.73 |
| 47 | Benzene ... Benzene (TS) | -2.87 | 0.20 | -0.09 | 0.27 | 0.23 |
| 48 | Pyridine ... Pyridine (TS) | -3.53 | 0.49 | 0.15 | 0.50 | 0.62 |
| 49 | Benzene ... Pyridine (TS) | -3.32 | 0.50 | 0.16 | 0.57 | 0.51 |
| 50 | Benzene ... Ethyne (CH-pi) | -2.86 | 0.24 | -0.08 | 0.56 | 0.32 |
| 51 | Ethyne ... Ethyne (TS) | -1.52 | 0.07 | -0.05 | 0.33 | 0.09 |
| 52 | Benzene ... AcOH (OH-pi) | -4.70 | -0.19 | -0.60 | 0.18 | 0.18 |
| 53 | Benzene ... AcNH2 (NH-pi) | -4.36 | 0.44 | 0.11 | -0.16 | 0.51 |
| 54 | Benzene ... Water (OH-pi) | -3.27 | -0.08 | -0.27 | 0.21 | 0.07 |
| 55 | Benzene ... MeOH (OH-pi) | -4.19 | 0.39 | 0.06 | 0.70 | 0.52 |
| 56 | Benzene ... MeNH2 (NH-pi) | -3.23 | 0.35 | 0.10 | 0.43 | 0.21 |
| 57 | Benzene ... Peptide (NH-pi) | -5.28 | 0.59 | 0.18 | 0.85 | 0.46 |
| 58 | Pyridine ... Pyridine (CH-N) | -4.15 | 2.09 | 1.55 | 1.56 | 1.86 |
| 59 | Ethyne ... Water (CH-O) | -2.85 | 0.12 | -0.07 | -1.13 | -0.45 |
| 60 | Ethyne ... AcOH (OH-pi) | -4.86 | -0.94 | -1.28 | 0.31 | 0.00 |
| 61 | Pentane ... AcOH | -2.88 | -0.66 | -0.81 | -0.74 | -0.79 |
| 62 | Pentane ... AcNH2 | -3.51 | -0.76 | -0.91 | -1.75 | -0.93 |
| 63 | Benzene ... AcOH | -3.80 | 0.03 | -0.20 | -0.02 | 0.01 |
| 64 | Peptide ... Ethene | -2.99 | -0.04 | -0.17 | -0.46 | -0.36 |
| 65 | Pyridine ... Ethyne | -3.99 | 1.07 | 0.41 | 0.52 | 0.96 |
| 66 | MeNH2 ... Pyridine | -3.97 | 0.88 | 0.47 | 0.58 | 0.83 |
| | RMSD | | 0.98 | 0.84 | 0.92 | 0.60 |
| | Mean deviation | | 0.16 | -0.18 | 0.04 | 0.00 |
| | Media deviation | | -0.01 | -0.17 | -0.07 | -0.05 |
| | Mean unsigned deviation | | 0.70 | 0.61 | 0.71 | 0.46 |
| | Median unsigned deviation | | 0.50 | 0.41 | 0.58 | 0.33 |
| | $r$ | | 0.97 | 0.98 | 0.97 | 0.99 |
| | $r^2$ | | 0.95 | 0.96 | 0.95 | 0.98 |
| | Max absolute deviation | | 3.09 | 2.32 | 2.50 | 1.86 |
| | lowest negative deviation | | -2.18 | -2.32 | -1.75 | -1.24 |
| | highest positive deviation | | 3.09 | 2.31 | 2.50 | 1.86 |

Table S6: C15 error in interaction energy for various methods, compared to a CCSD(T) reference. All values are in kcal/mol, except $r$ and $r^2$ which are unitless. Reference energies are taken from Ref. S6. Note that the statistics here differ from the ones found in the main text, as the Imidazolium ... Methylamine is left out in Table 2 of the main text.

| | | $E$ | Error | | | | | |
|---|---|---|---|---|---|---|---|---|
| ID | Complex | Ref.* | PM6 | PM6-D3H4 | PBE/6-31G(d) | PBE-D3/6-31G(d) | PBE/def2-QZVP | PBE-D3/def2-QZVP |
| 1 | acetate ... methanol | -19.75 | 6.20 | 0.72 | -6.24 | -7.44 | -0.08 | -1.27 |
| 2 | acetate ... water | -21.06 | 2.05 | -0.78 | -7.34 | -8.31 | 0.08 | -0.88 |
| 3 | acetate ... methylamine | -11.46 | 2.63 | -0.58 | -5.48 | -6.69 | 0.35 | -0.86 |
| 4 | methylammonium ... formaldehyde | -19.10 | 2.95 | -1.30 | -1.32 | -2.24 | 0.29 | -0.63 |
| 5 | methylammonium ... methylamine | -28.56 | 6.89 | -0.51 | -6.43 | -8.10 | -1.47 | -3.14 |
| 6 | methylammonium ... methanol | -21.23 | 5.98 | 1.76 | -3.40 | -4.60 | -0.03 | -1.23 |
| 7 | methylammonium ... water | -18.51 | 4.47 | 0.58 | -5.27 | -6.03 | -0.74 | -1.50 |
| 8 | guanidinium ... formaldehyde | -18.09 | 1.94 | -0.81 | -1.64 | -2.74 | 1.21 | 0.10 |
| 9 | guanidinium ... methylamine | -20.20 | 4.82 | 1.28 | -4.09 | -5.66 | 0.01 | -1.55 |
| 10 | guanidinium ... methanol | -19.79 | 3.96 | 1.00 | -3.98 | -5.45 | 1.09 | -0.38 |
| 11 | guanidinium ... water | -17.47 | 3.35 | 0.82 | -5.53 | -6.49 | 0.25 | -0.72 |
| 12 | imidazolium ... formaldehyde | -16.41 | 2.46 | 0.34 | -0.94 | -1.81 | 0.60 | -0.27 |
| 13 | imidazolium ... methylamine | -25.98 | 6.36 | 3.25 | -5.44 | -7.03 | -0.85 | -2.44 |
| 14 | imidazolium ... methanol | -18.91 | 5.71 | 3.65 | -2.81 | -3.94 | 0.38 | -0.75 |
| 15 | imidazolium ... water | -16.49 | 4.29 | 2.59 | -4.58 | -5.32 | -0.33 | -1.07 |
| | RMSD | | 4.57 | 1.66 | 4.69 | 5.80 | 0.68 | 1.37 |
| | Mean deviation | | 4.27 | 0.80 | -4.30 | -5.46 | 0.05 | -1.11 |
| | Median deviation | | 4.29 | 0.72 | -4.58 | -5.66 | 0.08 | -0.88 |
| | Mean unsigned deviation | | 4.27 | 1.33 | 4.30 | 5.46 | 0.52 | 1.12 |
| | Median unsigned deviation | | 4.29 | 0.82 | 4.58 | 5.66 | 0.35 | 0.88 |
| | $r$ | | 0.92 | 0.93 | 0.93 | 0.93 | 0.99 | 0.99 |
| | $r^2$ | | 0.84 | 0.86 | 0.86 | 0.86 | 0.98 | 0.99 |
| | Max absolute deviation | | 6.89 | 3.65 | 7.34 | 8.31 | 1.47 | 3.14 |
| | lowest negative deviation | | N/A | -1.30 | -7.34 | -8.31 | -1.47 | -3.14 |
| | highest positive deviation | | 6.89 | 3.65 | N/A | N/A | 1.21 | 0.10 |

| | | $E$ | Error | | | | | | |
|---|---|---|---|---|---|---|---|---|---|
| ID | Complex | Ref.* | DFTB3 | DFTB3-D3 | DFTB3-D3H4 | DFTB3/CPE($U$)-D3* | DFTB3/CPE($U$)-D3 | DFTB3/CPE($\zeta$)-D3 | DFTB3/CPE($q$)-D3 |
| 1 | acetate ... methanol | -19.75 | 2.60 | 1.35 | -0.31 | 0.21 | 0.42 | -0.22 | 1.29 |
| 2 | acetate ... water | -21.06 | -1.44 | -2.51 | -4.96 | -2.23 | -2.25 | -1.68 | -1.92 |
| 3 | acetate ... methylamine | -11.46 | 3.83 | 2.58 | 1.30 | 0.41 | 0.93 | -1.67 | -1.20 |
| 4 | methylammonium ... formaldehyde | -19.10 | 3.32 | 2.35 | 0.89 | 1.38 | 1.56 | 1.69 | 1.31 |
| 5 | methylammonium ... methylamine | -28.56 | 9.85 | 8.06 | 4.60 | 4.17 | 4.49 | 1.62 | -0.54 |
| 6 | methylammonium ... methanol | -21.23 | 2.77 | 1.50 | 0.25 | -0.35 | -0.15 | -1.55 | -0.42 |
| 7 | methylammonium ... water | -18.51 | 1.03 | 0.20 | -1.06 | -0.63 | -0.57 | -0.85 | -0.15 |
| 8 | guanidinium ... formaldehyde | -18.09 | 4.67 | 3.51 | 2.46 | 2.09 | 2.46 | 2.34 | 2.09 |
| 9 | guanidinium ... methylamine | -20.20 | 9.57 | 7.90 | 5.91 | 4.33 | 4.44 | 1.88 | -0.24 |
| 10 | guanidinium ... methanol | -19.79 | 4.79 | 3.22 | 2.31 | 0.25 | 0.32 | -2.02 | 0.47 |
| 11 | guanidinium ... water | -17.47 | 3.02 | 1.96 | 1.03 | 0.33 | 0.20 | -0.69 | 1.48 |
| 12 | imidazolium ... formaldehyde | -16.41 | 4.93 | 4.03 | 3.32 | 3.55 | 3.68 | 3.53 | 2.91 |
| 13 | imidazolium ... methylamine | -25.98 | 13.99 | 12.30 | 11.02 | 10.23 | 10.30 | 8.88 | 4.55 |
| 14 | imidazolium ... methanol | -18.91 | 5.72 | 4.53 | 4.07 | 3.35 | 3.38 | 2.19 | 2.35 |
| 15 | imidazolium ... water | -16.49 | 3.69 | 2.90 | 2.45 | 2.26 | 2.15 | 1.62 | 1.87 |
| | RMSD | | 6.07 | 4.99 | 4.10 | 3.48 | 3.57 | 2.90 | 1.90 |
| | Mean deviation | | 4.82 | 3.59 | 2.22 | 1.96 | 2.09 | 1.00 | 0.92 |

| | | | | | | | |
|---|---|---|---|---|---|---|---|
| Media deviation | 3.83 | 2.90 | 2.31 | 1.38 | 1.56 | 1.62 | 1.29 |
| Mean unsigned deviation | 5.02 | 3.93 | 3.06 | 2.38 | 2.49 | 2.16 | 1.52 |
| Median unsigned deviation | 3.83 | 2.90 | 2.45 | 2.09 | 2.15 | 1.68 | 1.31 |
| $r$ | 0.52 | 0.57 | 0.62 | 0.70 | 0.70 | 0.74 | 0.91 |
| $r^2$ | 0.27 | 0.33 | 0.39 | 0.49 | 0.49 | 0.55 | 0.83 |
| Max absolute deviation | 13.99 | 12.30 | 11.02 | 10.23 | 10.30 | 8.88 | 4.55 |
| lowest negative deviation | -1.44 | -2.51 | -4.96 | -2.23 | -2.25 | -2.02 | -1.92 |
| highest positive deviation | 13.99 | 12.30 | 11.02 | 10.23 | 10.30 | 8.88 | 4.55 |

| | | $E$ | Error | | | |
|---|---|---|---|---|---|---|
| ID | Complex | Ref.* | DFTB3/ CPE($q$) -D3 (original) 3OB | DFTB3/ CPE($q$) -D3 (original) MIO | DFTB3/ CPE($\zeta$) -D3 (pol) | DFTB3/ CPE($q$) -D3 (pol) |
| 1 | acetate ... methanol | -19.75 | 0.87 | 0.41 | 1.34 | 1.47 |
| 2 | acetate ... water | -21.06 | -2.73 | -2.52 | -2.56 | -1.86 |
| 3 | acetate ... methylamine | -11.46 | 1.61 | 1.50 | 1.54 | -0.19 |
| 4 | methylammonium ... formaldehyde | -19.10 | 1.39 | 0.35 | 2.35 | 2.33 |
| 5 | methylammonium ... methylamine | -28.56 | 6.52 | 6.24 | 2.09 | 3.16 |
| 6 | methylammonium ... methanol | -21.23 | -0.44 | -1.53 | -0.34 | 0.84 |
| 7 | methylammonium ... water | -18.51 | -0.40 | -1.63 | -0.43 | 0.05 |
| 8 | guanidinium ... formaldehyde | -18.09 | 3.16 | 2.65 | 4.15 | 2.65 |
| 9 | guanidinium ... methylamine | -20.20 | 6.46 | 6.20 | 3.29 | 3.40 |
| 10 | guanidinium ... methanol | -19.79 | 1.64 | 1.14 | 0.99 | 1.12 |
| 11 | guanidinium ... water | -17.47 | 1.20 | 0.82 | 0.86 | 1.65 |
| 12 | imidazolium ... formaldehyde | -16.41 | 3.84 | 2.69 | 3.65 | 3.72 |
| 13 | imidazolium ... methylamine | -25.98 | 11.05 | 10.71 | 8.93 | 8.25 |
| 14 | imidazolium ... methanol | -18.91 | 3.45 | 2.14 | 2.71 | 3.49 |
| 15 | imidazolium ... water | -16.49 | 2.78 | 1.37 | 2.00 | 2.09 |
| | RMSD | | 4.23 | 3.91 | 3.21 | 3.09 |
| | Mean deviation | | 2.69 | 2.04 | 2.04 | 2.15 |
| | Media deviation | | 1.64 | 1.37 | 2.00 | 2.09 |
| | Mean unsigned deviation | | 3.17 | 2.79 | 2.48 | 2.42 |
| | Median unsigned deviation | | 2.73 | 1.63 | 2.09 | 2.09 |
| | $r$ | | 0.63 | 0.61 | 0.80 | 0.82 |
| | $r^2$ | | 0.39 | 0.37 | 0.64 | 0.67 |
| | Max absolute deviation | | 11.05 | 10.71 | 8.93 | 8.25 |
| | lowest negative deviation | | -2.73 | -2.52 | -2.56 | -1.86 |
| | highest positive deviation | | 11.05 | 10.71 | 8.93 | 8.25 |

Table S7: I9 error in interaction energy for various methods, compared to a CCSD(T) reference. All values are in kcal/mol, except $r$ and $r^2$ which are unitless.

| | | $E$ | Error | | | | | |
|---|---|---|---|---|---|---|---|---|
| ID | Complex | Ref.* | PM6 | PM6-D3H4 | PBE/ 6-31G(d) | PBE-D3/ 6-31G(d) | PBE/ def2-QZVP | PBE-D3/ def2-QZVP |
| 1 | Guanidinium ... Methyl acetate | -134.31 | 10.94 | 5.59 | -11.89 | -13.95 | -0.69 | -2.75 |
| 2 | Guanidinium ... Thiometoxide | -125.30 | 9.76 | 8.92 | -8.00 | -10.66 | -4.39 | -7.06 |
| 3 | Guanidinium ... Methoxide | -181.95 | 10.54 | 8.11 | -20.43 | -22.34 | -1.69 | -3.60 |
| 4 | Imidazolium ... Methyl acetate | -120.39 | 8.09 | 5.41 | -10.70 | -12.57 | 0.89 | -0.98 |
| 5 | Imidazolium ... Thiometoxide | -103.97 | 3.56 | 2.78 | -5.75 | -7.91 | -2.39 | -4.55 |
| 6 | Imidazolium ... Methoxide | -120.35 | 3.88 | 2.21 | -15.17 | -17.05 | -1.02 | -2.90 |
| 7 | Methyl ammonium ... Methyl acetate | -145.10 | 14.71 | 6.22 | -14.86 | -16.85 | -1.75 | -3.74 |
| 8 | Methyl ammonium ... Thiometoxide | -117.60 | 8.53 | 7.66 | -6.78 | -8.97 | -3.72 | -5.91 |
| 9 | Methyl ammonium ... Methoxide | -132.43 | 6.45 | 3.64 | -15.64 | -17.72 | -2.12 | -4.20 |
| | RMSD | | 9.13 | 6.05 | 12.96 | 14.90 | 2.39 | 4.31 |
| | Mean deviation | | 8.50 | 5.61 | -12.13 | -14.22 | -1.88 | -3.97 |
| | Median deviation | | 8.53 | 5.59 | -11.89 | -13.95 | -1.75 | -3.74 |
| | Mean unsigned deviation | | 8.50 | 5.61 | 12.13 | 14.22 | 2.07 | 3.97 |
| | Median unsigned deviation | | 8.53 | 5.59 | 11.89 | 13.95 | 1.75 | 3.74 |
| | $r$ | | 0.99 | 1.00 | 0.99 | 1.00 | 1.00 | 1.00 |
| | $r^2$ | | 0.98 | 0.99 | 0.99 | 0.99 | 1.00 | 0.99 |
| | Max absolute deviation | | 14.71 | 8.92 | 20.43 | 22.34 | 4.39 | 7.06 |
| | lowest negative deviation | | N/A | N/A | -20.43 | -22.34 | -4.39 | -7.06 |
| | highest positive deviation | | 14.71 | 8.92 | N/A | N/A | 0.89 | N/A |

| | | $E$ | Error | | | | | | |
|---|---|---|---|---|---|---|---|---|---|
| ID | Complex | Ref.* | DFTB3 | DFTB3-D3 | DFTB3-D3H4 | DFTB3/ CPE($U$) -D3* | DFTB3/ CPE($U$) -D3 | DFTB3/ CPE($\zeta$) -D3 | DFTB3/ CPE($q$) -D3 |
| 1 | Guanidinium ... Methyl acetate | -134.31 | 4.96 | 2.78 | 0.68 | 1.81 | 2.25 | 0.12 | -0.16 |
| 2 | Guanidinium ... Thiometoxide | -125.30 | 9.19 | 6.54 | 8.31 | 4.47 | 4.65 | 2.69 | 4.09 |
| 3 | Guanidinium ... Methoxide | -181.95 | 6.42 | 4.36 | 4.37 | 2.30 | 2.24 | 0.68 | 0.78 |
| 4 | Imidazolium ... Methyl acetate | -120.39 | 5.61 | 3.66 | 3.60 | 4.04 | 4.27 | 3.75 | 0.54 |
| 5 | Imidazolium ... Thiometoxide | -103.97 | 7.14 | 5.02 | 6.30 | 3.82 | 3.97 | 2.32 | 1.98 |
| 6 | Imidazolium ... Methoxide | -120.35 | 5.06 | 3.06 | 3.56 | 2.07 | 2.28 | 1.00 | -2.28 |
| 7 | Methyl ammonium ... Methyl acetate | -145.10 | -0.52 | -2.66 | -4.86 | -1.77 | -1.34 | -1.76 | -0.23 |
| 8 | Methyl ammonium ... Thiometoxide | -117.60 | 4.89 | 2.69 | 3.91 | 0.11 | 0.65 | -2.07 | 0.12 |
| 9 | Methyl ammonium ... Methoxide | -132.43 | -0.15 | -2.45 | -1.67 | -3.90 | -3.61 | -4.93 | -0.23 |

| | | | | | | | |
|---|---|---|---|---|---|---|---|
| RMSD | 5.60 | 3.91 | 4.66 | 3.02 | 3.09 | 2.58 | 1.73 |
| Mean deviation | 4.74 | 2.56 | 2.69 | 1.44 | 1.71 | 0.20 | 0.51 |
| Media deviation | 5.06 | 3.06 | 3.60 | 2.07 | 2.25 | 0.68 | 0.12 |
| Mean unsigned deviation | 4.88 | 3.69 | 4.14 | 2.70 | 2.81 | 2.15 | 1.16 |
| Median unsigned deviation | 5.06 | 3.06 | 3.91 | 2.30 | 2.28 | 2.07 | 0.54 |
| $r$ | 0.99 | 0.99 | 0.99 | 0.99 | 0.99 | 0.99 | 1.00 |
| $r^2$ | 0.98 | 0.98 | 0.97 | 0.99 | 0.99 | 0.99 | 0.99 |
| Max absolute deviation | 9.19 | 6.54 | 8.31 | 4.47 | 4.65 | 4.93 | 4.09 |
| lowest negative deviation | -0.52 | -2.66 | -4.86 | -3.90 | -3.61 | -4.93 | -2.28 |
| highest positive deviation | 9.19 | 6.54 | 8.31 | 4.47 | 4.65 | 3.75 | 4.09 |

| ID | Complex | $E$ Ref.* | Error DFTB3/ CPE($q$) -D3 (original) 3OB | DFTB3/ CPE($q$) -D3 (original) MIO | DFTB3/ CPE($\zeta$) -D3 (pol) | DFTB3/ CPE($q$) -D3 (pol) |
|---|---|---|---|---|---|---|
| 1 | Guanidinium ... Methyl acetate | -134.31 | 0.90 | -2.37 | 0.90 | 0.91 |
| 2 | Guanidinium ... Thiometoxide | -125.30 | 5.57 | 4.37 | -0.70 | 5.12 |
| 3 | Guanidinium ... Methoxide | -181.95 | 2.98 | -1.13 | -0.83 | 1.35 |
| 4 | Imidazolium ... Methyl acetate | -120.39 | 3.40 | 0.34 | 2.86 | 2.64 |
| 5 | Imidazolium ... Thiometoxide | -103.97 | 4.03 | 2.22 | 3.80 | 3.51 |
| 6 | Imidazolium ... Methoxide | -120.35 | 2.47 | -0.94 | 1.68 | 1.16 |
| 7 | Methyl ammonium ... Methyl acetate | -145.10 | -3.15 | -7.91 | -3.56 | -0.02 |
| 8 | Methyl ammonium ... Thiometoxide | -117.60 | 1.98 | 0.28 | 1.08 | 1.71 |
| 9 | Methyl ammonium ... Methoxide | -132.43 | -4.63 | -8.42 | -3.02 | -0.14 |
| | RMSD | | 3.49 | 4.29 | 2.36 | 2.41 |
| | Mean deviation | | 1.51 | -1.51 | 0.24 | 1.80 |
| | Media deviation | | 2.47 | -0.94 | 0.90 | 1.35 |
| | Mean unsigned deviation | | 3.23 | 3.11 | 2.05 | 1.84 |
| | Median unsigned deviation | | 3.15 | 2.22 | 1.68 | 1.35 |
| | $r$ | | 0.99 | 0.99 | 1.00 | 1.00 |
| | $r^2$ | | 0.98 | 0.97 | 0.99 | 1.00 |
| | Max absolute deviation | | 5.57 | 8.42 | 3.80 | 5.12 |
| | lowest negative deviation | | -4.63 | -8.42 | -3.56 | -0.14 |
| | highest positive deviation | | 5.57 | 4.37 | 3.80 | 5.12 |

Table S8: S14 error in interaction energy for various methods, compared to a CCSD(T) reference. All values are in kcal/mol, except $r$ and $r^2$ which are unitless. Reference energies are taken from Ref. S7.

| ID | Complex | $E$ Ref.* | Error PM6 | PM6-D3H4 | PBE/ 6-31G(d) | PBE-D3/ 6-31G(d) | PBE/ def2-QZVP | PBE-D3/ def2-QZVP |
|---|---|---|---|---|---|---|---|---|
| 1 | ch3sh_benzene_cs | -3.50 | 1.95 | -0.06 | 1.72 | -1.21 | 2.84 | -0.09 |
| 2 | ch3sch3_benzene_c2v | -1.12 | 1.88 | 0.57 | 1.83 | -0.35 | 2.12 | -0.05 |
| 3 | h2o_h2s_cs | -2.90 | 1.19 | 0.95 | -0.59 | -1.16 | -0.26 | -0.82 |
| 4 | h2s_ch4 | -0.85 | 0.72 | 0.20 | 0.31 | -0.39 | 0.51 | -0.19 |
| 5 | ch3sh_h2co | -3.11 | 0.97 | -0.72 | -0.38 | -1.59 | 0.89 | -0.32 |
| 6 | ch3sh_nh3_cs | -2.95 | 0.10 | -0.34 | -2.13 | -2.82 | -0.10 | -0.80 |
| 7 | h2s_dimer_cs | -1.67 | 0.56 | 0.28 | -0.49 | -1.13 | 0.01 | -0.64 |
| 8 | h2o_dimer_cs | -4.96 | 1.10 | 0.08 | -2.55 | -2.94 | -0.23 | -0.62 |
| 9 | ch3sh_h2o_cs | -2.36 | -0.13 | -0.47 | -1.44 | -1.96 | 0.13 | -0.39 |
| 10 | h2s_h2o_cs | -2.66 | -0.38 | -0.63 | -1.78 | -2.22 | -0.03 | -0.48 |
| 11 | h2o_ch3sch3_cs | -5.35 | 2.56 | 2.11 | -1.01 | -2.22 | 0.31 | -0.90 |
| 12 | ch3sh_dimer | -3.15 | 2.34 | 1.49 | 0.60 | -1.17 | 1.28 | -0.50 |
| 13 | formamide_ch3sh | -6.33 | 2.51 | 1.55 | -1.13 | -2.72 | 0.73 | -0.86 |
| 14 | ch3oh_ch3sch3_cs | -5.69 | 3.46 | 2.83 | -0.70 | -2.18 | 0.59 | -0.90 |
| | RMSD | | 1.74 | 1.19 | 1.38 | 1.90 | 1.08 | 0.61 |
| | Mean deviation | | 1.35 | 0.56 | -0.55 | -1.72 | 0.63 | -0.54 |
| | Median deviation | | 1.14 | 0.24 | -0.65 | -1.78 | 0.41 | -0.56 |
| | Mean unsigned deviation | | 1.42 | 0.88 | 1.19 | 1.72 | 0.72 | 0.54 |
| | Median unsigned deviation | | 1.14 | 0.60 | 1.07 | 1.78 | 0.41 | 0.56 |
| | $r$ | | 0.74 | 0.77 | 0.88 | 0.97 | 0.89 | 0.99 |
| | $r^2$ | | 0.55 | 0.59 | 0.77 | 0.94 | 0.79 | 0.99 |
| | Max absolute deviation | | 3.46 | 2.83 | 2.55 | 2.94 | 2.84 | 0.90 |
| | lowest negative deviation | | -0.38 | -0.72 | -2.55 | -2.94 | -0.26 | -0.90 |
| | highest positive deviation | | 3.46 | 2.83 | 1.83 | N/A | 2.84 | N/A |

| ID | Complex | $E$ Ref.* | Error DFTB3 | DFTB3-D3 | DFTB3-D3H4 | DFTB3/ CPE($U$) -D3* | DFTB3/ CPE($U$) -D3 | DFTB3/ CPE($\zeta$) -D3 | DFTB3/ CPE($q$) -D3 |
|---|---|---|---|---|---|---|---|---|---|
| 1 | ch3sh_benzene_cs | -3.50 | 4.02 | 2.50 | 3.32 | 2.33 | 2.36 | 2.44 | 1.75 |
| 2 | ch3sch3_benzene_c2v | -1.12 | 1.18 | -0.82 | -0.19 | -0.81 | -0.76 | -0.57 | -0.67 |
| 3 | h2o_h2s_cs | -2.90 | 2.63 | -0.23 | 0.38 | -0.24 | -0.06 | 0.25 | -0.10 |
| 4 | h2s_ch4 | -0.85 | 1.92 | 0.25 | 1.05 | 0.18 | 0.15 | 0.36 | -0.11 |
| 5 | ch3sh_h2co | -3.11 | 1.81 | 0.59 | 1.20 | 0.36 | 0.48 | 0.72 | 0.15 |
| 6 | ch3sh_nh3_cs | -2.95 | 0.28 | -0.26 | -0.10 | -0.53 | -0.44 | -1.02 | -0.59 |
| 7 | h2s_dimer_cs | -1.67 | 0.92 | 0.21 | 0.42 | -0.95 | -0.05 | -1.40 | -0.89 |
| 8 | h2o_dimer_cs | -4.96 | 3.44 | 1.84 | 2.38 | 1.52 | 1.59 | 1.02 | 0.53 |
| 9 | ch3sh_h2o_cs | -2.36 | 3.39 | 2.13 | 2.90 | 1.73 | 1.70 | 1.39 | 1.74 |
| 10 | h2s_h2o_cs | -2.66 | 0.41 | -0.03 | -1.12 | -0.64 | -0.62 | -0.43 | -0.05 |
| 11 | h2o_ch3sch3_cs | -5.35 | 1.37 | 0.80 | 1.11 | 0.48 | 0.67 | 0.26 | 0.71 |
| 12 | ch3sh_dimer | -3.15 | 0.59 | -0.10 | 0.00 | -0.23 | -0.21 | -0.14 | -0.08 |
| 13 | formamide_ch3sh | -6.33 | 0.58 | -0.02 | 0.29 | -0.04 | -0.03 | 0.05 | -0.18 |
| 14 | ch3oh_ch3sch3_cs | -5.69 | -0.08 | -0.54 | -0.35 | -0.80 | -0.70 | -0.99 | -1.29 |
| | RMSD | | 2.04 | 1.08 | 1.48 | 1.00 | 0.98 | 1.01 | 0.85 |
| | Mean deviation | | 1.60 | 0.45 | 0.81 | 0.17 | 0.29 | 0.14 | 0.07 |
| | Media deviation | | 1.27 | 0.10 | 0.40 | -0.14 | -0.04 | 0.15 | -0.09 |
| | Mean unsigned deviation | | 1.62 | 0.74 | 1.06 | 0.78 | 0.70 | 0.79 | 0.63 |
| | Median unsigned deviation | | 1.27 | 0.40 | 0.74 | 0.59 | 0.55 | 0.65 | 0.56 |
| | $r$ | | 0.63 | 0.82 | 0.65 | 0.80 | 0.83 | 0.79 | 0.86 |
| | $r^2$ | | 0.40 | 0.67 | 0.43 | 0.64 | 0.69 | 0.62 | 0.73 |
| | Max absolute deviation | | 4.02 | 2.50 | 3.32 | 2.33 | 2.36 | 2.44 | 1.75 |
| | lowest negative deviation | | -0.08 | -0.82 | -1.12 | -0.95 | -0.76 | -1.40 | -1.29 |
| | highest positive deviation | | 4.02 | 2.50 | 3.32 | 2.33 | 2.36 | 2.44 | 1.75 |

| | | $E$ | Error | | | |
|---|---|---|---|---|---|---|
| ID | Complex | Ref.* | DFTB3/ CPE($q$) -D3 (original) 3OB | DFTB3/ CPE($q$) -D3 (original) MIO | DFTB3/ CPE($\zeta$) -D3 (pol) | DFTB3/ CPE($q$) -D3 (pol) |
| 1 | ch3sh_benzene_cs | -3.50 | 2.19 | 1.92 | 1.38 | 2.13 |
| 2 | ch3sch3_benzene_c2v | -1.12 | -0.76 | -1.32 | -1.07 | -0.74 |
| 3 | h2o_h2s_cs | -2.90 | -0.33 | -0.73 | -0.13 | -0.21 |
| 4 | h2s_ch4 | -0.85 | 0.13 | -0.51 | 0.23 | 0.04 |
| 5 | ch3sh_h2co | -3.11 | 0.54 | 0.26 | 0.43 | 0.41 |
| 6 | ch3sh_nh3_cs | -2.95 | -0.23 | -0.29 | -0.70 | -0.44 |
| 7 | h2s_dimer_cs | -1.67 | 0.05 | -0.24 | -1.18 | -0.56 |
| 8 | h2o_dimer_cs | -4.96 | 1.63 | 1.33 | 0.38 | 1.09 |
| 9 | ch3sh_h2o_cs | -2.36 | 1.94 | 1.77 | 0.44 | 1.85 |
| 10 | h2s_h2o_cs | -2.66 | -0.56 | -0.51 | -0.07 | 0.10 |
| 11 | h2o_ch3sch3_cs | -5.35 | 0.73 | 0.76 | 0.56 | 0.79 |
| 12 | ch3sh_dimer | -3.15 | -0.23 | -0.38 | -0.14 | -0.31 |
| 13 | formamide_ch3sh | -6.33 | -0.02 | -0.18 | -0.18 | -0.11 |
| 14 | ch3oh_ch3sch3_cs | -5.69 | -0.55 | -0.64 | -0.80 | -1.00 |
| | RMSD | | 0.98 | 0.95 | 0.68 | 0.93 |
| | Mean deviation | | 0.32 | 0.09 | -0.06 | 0.22 |
| | Media deviation | | 0.01 | -0.27 | -0.10 | -0.04 |
| | Mean unsigned deviation | | 0.71 | 0.77 | 0.55 | 0.70 |
| | Median unsigned deviation | | 0.55 | 0.58 | 0.43 | 0.50 |
| | $r$ | | 0.83 | 0.82 | 0.91 | 0.85 |
| | $r^2$ | | 0.69 | 0.67 | 0.83 | 0.72 |
| | Max absolute deviation | | 2.19 | 1.92 | 1.38 | 2.13 |
| | lowest negative deviation | | -0.76 | -1.32 | -1.18 | -1.00 |
| | highest positive deviation | | 2.19 | 1.92 | 1.38 | 2.13 |

Table S9: Large water error in interaction energy for various methods, compared to a CCSD(T) reference. All values are in kcal/mol, except $r$ and $r^2$ which are unitless. Reference energies are take from Ref. S8

| | | $E$ | Error | | | | | |
|---|---|---|---|---|---|---|---|---|
| ID | Complex | Ref.* | PM6 | PM6-D3H4 | PBE/ 6-31G(d) | PBE-D3/ 6-31G(d) | PBE/ def2-QZVP | PBE-D3/ def2-QZVP |
| 1 | 4444-a | -171.06 | 50.78 | 14.59 | -101.57 | -121.55 | -0.67 | -20.64 |
| 2 | 4444-b | -170.52 | 50.18 | 14.21 | -102.13 | -122.23 | -0.77 | -20.87 |
| 3 | 552-5 | -181.84 | 53.78 | 17.04 | -103.08 | -122.75 | -4.23 | -23.91 |
| 4 | antiboat | -170.55 | 50.44 | 16.79 | -96.95 | -115.08 | -4.32 | -22.45 |
| 5 | bag | -44.30 | 10.74 | 2.83 | -29.51 | -33.82 | -5.04 | -9.35 |
| 6 | boat-1 | -43.13 | 11.18 | 5.19 | -25.28 | -28.53 | -5.61 | -8.86 |
| 7 | boat-2 | -44.90 | 12.52 | 6.49 | -23.53 | -26.79 | -3.71 | -6.97 |
| 8 | boat-a | -170.80 | 49.99 | 16.67 | -98.17 | -116.31 | -4.64 | -22.78 |
| 9 | Boat-b | -170.64 | 49.91 | 16.70 | -98.00 | -116.12 | -4.58 | -22.70 |
| 10 | book-1 | -45.20 | 10.84 | 3.49 | -29.46 | -33.50 | -5.16 | -9.19 |
| 11 | book-2 | -44.90 | 10.76 | 3.24 | -29.74 | -33.86 | -5.21 | -9.34 |
| 12 | cage | -45.67 | 10.61 | 1.96 | -32.12 | -37.01 | -4.07 | -8.97 |
| 13 | chair | -44.12 | 10.71 | 4.98 | -25.75 | -28.94 | -5.76 | -8.95 |
| 14 | sphere | -182.54 | 53.91 | 16.68 | -103.88 | -125.30 | -1.53 | -22.95 |
| 15 | prism | -45.92 | 9.91 | 0.72 | -33.68 | -38.78 | -3.70 | -8.81 |
| | RMSD | | 34.81 | 11.04 | 69.57 | 82.69 | 4.11 | 16.05 |
| | Mean deviation | | 27.89 | 8.85 | -58.30 | -68.79 | -3.69 | -14.17 |
| | Median deviation | | 11.85 | 5.84 | -32.90 | -37.90 | -4.28 | -9.34 |
| | Mean unsigned deviation | | 27.89 | 8.85 | 58.30 | 68.79 | 3.69 | 14.17 |
| | Median unsigned deviation | | 11.85 | 5.84 | 32.90 | 37.90 | 4.28 | 9.34 |
| | $r$ | | 1.00 | 1.00 | 1.00 | 1.00 | 1.00 | 1.00 |
| | $r^2$ | | 1.00 | 1.00 | 1.00 | 1.00 | 1.00 | 1.00 |
| | Max absolute deviation | | 53.91 | 17.04 | 103.88 | 125.30 | 5.76 | 23.91 |
| | lowest negative deviation | | N/A | N/A | -103.88 | -125.30 | -5.76 | -23.91 |
| | highest positive deviation | | 53.91 | 17.04 | N/A | N/A | N/A | N/A |

| | | *E* | Error | | | | | | |
|---|---|---|---|---|---|---|---|---|---|
| ID | Complex | Ref.* | DFTB3 | DFTB3-D3 | DFTB3-D3H4 | DFTB3/ CPE($U$) -D3* | DFTB3/ CPE($U$) -D3 | DFTB3/ CPE($\zeta$) -D3 | DFTB3/ CPE($q$) -D3 |
| 1 | 4444-a | -171.06 | 18.94 | -3.33 | -36.75 | -1.30 | -4.15 | -2.86 | -3.96 |
| 2 | 4444-b | -170.52 | 18.53 | -3.89 | -36.95 | -5.54 | -4.68 | -3.47 | -5.50 |
| 3 | 552-5 | -181.84 | 21.95 | 0.08 | -34.22 | -2.84 | -1.58 | -0.34 | -0.04 |
| 4 | antiboat | -170.55 | 20.43 | 0.27 | -31.24 | -1.78 | -1.13 | 0.98 | 1.04 |
| 5 | bag | -44.30 | 3.30 | -1.53 | -9.10 | -2.51 | -2.60 | -1.84 | -1.62 |
| 6 | boat-1 | -43.13 | 2.35 | -1.32 | -7.24 | -2.74 | -2.53 | -1.68 | -0.98 |
| 7 | boat-2 | -44.90 | 3.50 | -0.18 | -6.14 | -1.62 | -1.41 | -0.51 | 0.20 |
| 8 | boat-a | -170.80 | 20.96 | 0.77 | -30.40 | -1.25 | -0.39 | 2.00 | 1.30 |
| 9 | Boat-b | -170.64 | 20.94 | 0.78 | -30.24 | -1.30 | -0.41 | 2.05 | 0.99 |
| 10 | book-1 | -45.20 | 2.86 | -1.71 | -8.82 | -2.79 | -2.79 | -2.30 | -1.68 |
| 11 | book-2 | -44.90 | 3.26 | -1.39 | -8.68 | -2.46 | -2.47 | -1.98 | -1.43 |
| 12 | cage | -45.67 | 2.95 | -2.59 | -10.61 | -5.60 | -5.70 | -4.03 | -4.30 |
| 13 | chair | -44.12 | 1.70 | -1.91 | -7.54 | -3.41 | -3.22 | -2.48 | -1.57 |
| 14 | sphere | -182.54 | 21.53 | -2.24 | -34.95 | -7.07 | -5.92 | -2.46 | -4.53 |
| 15 | prism | -45.92 | 2.56 | -3.21 | -11.76 | -4.86 | -5.21 | -4.58 | -6.22 |
| | RMSD | | 14.16 | 2.04 | 23.88 | 3.59 | 3.44 | 2.51 | 3.04 |
| | Mean deviation | | 11.05 | -1.43 | -20.31 | -3.14 | -2.95 | -1.57 | -1.89 |
| | Media deviation | | 3.50 | -1.53 | -11.76 | -2.74 | -2.60 | -1.98 | -1.57 |
| | Mean unsigned deviation | | 11.05 | 1.68 | 20.31 | 3.14 | 2.95 | 2.24 | 2.36 |
| | Median unsigned deviation | | 3.50 | 1.53 | 11.76 | 2.74 | 2.60 | 2.05 | 1.57 |
| | $r$ | | 1.00 | 1.00 | 1.00 | 1.00 | 1.00 | 1.00 | 1.00 |
| | $r^2$ | | 1.00 | 1.00 | 1.00 | 1.00 | 1.00 | 1.00 | 1.00 |
| | Max absolute deviation | | 21.95 | 3.89 | 36.95 | 7.07 | 5.92 | 4.58 | 6.22 |
| | lowest negative deviation | | N/A | -3.89 | -36.95 | -7.07 | -5.92 | -4.58 | -6.22 |
| | highest positive deviation | | 21.95 | 0.78 | N/A | N/A | N/A | 2.05 | 1.30 |

| | | *E* | Error | | | |
|---|---|---|---|---|---|---|
| ID | Complex | Ref.* | DFTB3/ CPE($q$) -D3 (original) 3OB | DFTB3/ CPE($q$) -D3 (original) MIO | DFTB3/ CPE($\zeta$) -D3 (pol) | DFTB3/ CPE($q$) -D3 (pol) |
| 1 | 4444-a | -171.06 | -418.24 | -271.84 | -3.38 | -0.01 |
| 2 | 4444-b | -170.52 | -422.10 | -277.28 | -4.05 | -0.68 |
| 3 | 552-5 | -181.84 | -222.80 | -165.86 | 1.82 | 4.46 |
| 4 | antiboat | -170.55 | -194.18 | -154.00 | 2.15 | 4.50 |
| 5 | bag | -44.30 | -41.55 | -34.63 | -1.04 | -0.43 |
| 6 | boat-1 | -43.13 | -8.09 | -9.25 | -0.60 | -0.14 |
| 7 | boat-2 | -44.90 | -7.77 | -8.72 | 0.58 | 1.04 |
| 8 | boat-a | -170.80 | -216.83 | -149.45 | 2.63 | 5.08 |
| 9 | Boat-b | -170.64 | -211.44 | -146.44 | 2.66 | 5.02 |
| 10 | book-1 | -45.20 | -19.78 | -18.75 | -1.19 | -0.55 |
| 11 | book-2 | -44.90 | -26.01 | -23.68 | -0.87 | -0.26 |
| 12 | cage | -45.67 | -45.08 | -37.62 | -2.87 | -1.73 |
| 13 | chair | -44.12 | -5.00 | -6.67 | -1.31 | -0.72 |
| 14 | sphere | -182.54 | -458.60 | -351.77 | -1.26 | 1.90 |
| 15 | prism | -45.92 | -55.74 | -44.41 | -5.05 | -3.62 |
| | RMSD | | 223.69 | 158.16 | 2.46 | 2.75 |
| | Mean deviation | | -156.88 | -113.36 | -0.78 | 0.92 |
| | Media deviation | | -55.74 | -44.41 | -1.04 | -0.14 |
| | Mean unsigned deviation | | 156.88 | 113.36 | 2.10 | 2.01 |
| | Median unsigned deviation | | 55.74 | 44.41 | 1.82 | 1.04 |
| | $r$ | | 0.94 | 0.95 | 1.00 | 1.00 |
| | $r^2$ | | 0.88 | 0.91 | 1.00 | 1.00 |
| | Max absolute deviation | | 458.60 | 351.77 | 5.05 | 5.08 |
| | lowest negative deviation | | -458.60 | -351.77 | -5.05 | -3.62 |
| | highest positive deviation | | N/A | N/A | 2.66 | 5.08 |

Table S10: Charged water error in interaction energy for various methods, compared to a CCSD(T) reference. All values are in kcal/mol, except $r$ and $r^2$ which are unitless.

| | | *E* | Error | | | | | |
|---|---|---|---|---|---|---|---|---|
| ID | Complex | Ref.* | PM6 | PM6-D3H4 | PBE/ 6-31G(d) | PBE-D3/ 6-31G(d) | PBE/ def2-QZVP | PBE-D3/ def2-QZVP |
| 1 | 1 Water, 1 Hydronium | -33.93 | 7.70 | 8.01 | -9.16 | -9.76 | -3.52 | -4.12 |
| 2 | 2 Water, 1 Hydronium | -57.42 | 14.19 | 10.32 | -14.14 | -15.34 | -4.03 | -5.22 |
| 3 | 2 Water, 1 Hydronium | -57.37 | 14.99 | 11.17 | -13.63 | -14.81 | -3.83 | -5.01 |
| 4 | 3 Water, 1 Hydronium | -77.14 | 18.55 | 12.03 | -18.53 | -20.36 | -4.07 | -5.90 |
| 5 | 3 Water, 1 Hydronium | -77.08 | 19.40 | 12.87 | -17.86 | -19.67 | -3.86 | -5.67 |
| 6 | 3 Water, 1 Hydronium | -73.06 | 14.81 | 7.55 | -22.56 | -24.98 | -4.85 | -7.26 |
| 7 | 3 Water, 1 Hydronium | -72.69 | 15.57 | 8.38 | -21.94 | -24.35 | -4.68 | -7.09 |
| 8 | 3 Water, 1 Hydronium | -73.13 | 17.20 | 11.82 | -19.82 | -21.61 | -5.69 | -7.48 |
| | RMSD | | 14.01 | 9.34 | 15.86 | 17.42 | 3.90 | 5.43 |

| | | | | | | | | | |
|---|---|---|---|---|---|---|---|---|---|
| | Mean deviation | | 12.24 | 8.22 | -13.76 | -15.09 | -3.45 | -4.78 | |
| | Median deviation | | 14.90 | 9.35 | -16.00 | -17.50 | -3.94 | -5.45 | |
| | Mean unsigned deviation | | 12.24 | 8.22 | 13.76 | 15.09 | 3.45 | 4.78 | |
| | Median unsigned deviation | | 14.90 | 9.35 | 16.00 | 17.50 | 3.94 | 5.45 | |
| | $r$ | | 1.00 | 1.00 | 1.00 | 1.00 | 1.00 | 1.00 | |
| | $r^2$ | | 1.00 | 0.99 | 1.00 | 1.00 | 1.00 | 1.00 | |
| | Max absolute deviation | | 19.40 | 12.87 | 22.56 | 24.98 | 5.69 | 7.48 | |
| | lowest negative deviation | | N/A | N/A | -22.56 | -24.98 | -5.69 | -7.48 | |
| | highest positive deviation | | 19.40 | 12.87 | N/A | N/A | N/A | N/A | |

| | | $E$ | Error | | | | | | |
|---|---|---|---|---|---|---|---|---|---|
| ID | Complex | Ref.* | DFTB3 | DFTB3-D3 | DFTB3-D3H4 | DFTB3/CPE($U$)-D3* | DFTB3/CPE($U$)-D3 | DFTB3/CPE($\zeta$)-D3 | DFTB3/CPE($q$)-D3 |
| 1 | 1 water 1hydroxide | -33.93 | 3.31 | 2.61 | 3.41 | 1.61 | 1.57 | 0.75 | 1.96 |
| 2 | 2 water 1hydroxide | -57.42 | 5.13 | 3.75 | 2.95 | 2.32 | 2.29 | 1.29 | 3.09 |
| 3 | 2 water 1hydroxide | -57.37 | 4.56 | 3.21 | 2.44 | 1.70 | 1.66 | 0.65 | 2.54 |
| 4 | 3 water 1hydroxide | -77.14 | 6.13 | 4.05 | 2.07 | 2.78 | 2.79 | 1.98 | 3.58 |
| 5 | 3 water 1hydroxide | -77.08 | 5.52 | 3.46 | 1.48 | 2.23 | 2.22 | 1.46 | 3.03 |
| 6 | 3 water 1hydroxide | -73.06 | 7.33 | 4.57 | 2.40 | 1.82 | 1.69 | 0.94 | 3.13 |
| 7 | 3 water 1hydroxide | -72.69 | 7.29 | 4.54 | 2.39 | 1.75 | 1.61 | 1.05 | 3.11 |
| 8 | 3 water 1hydroxide | -73.13 | 5.60 | 3.56 | 2.53 | 1.86 | 1.90 | 1.03 | 3.05 |
| | RMSD | | 5.75 | 3.77 | 2.51 | 2.04 | 2.01 | 1.21 | 2.97 |
| | Mean deviation | | 5.61 | 3.72 | 2.46 | 2.01 | 1.97 | 1.14 | 2.94 |
| | Media deviation | | 5.56 | 3.65 | 2.42 | 1.84 | 1.79 | 1.04 | 3.07 |
| | Mean unsigned deviation | | 5.61 | 3.72 | 2.46 | 2.01 | 1.97 | 1.14 | 2.94 |
| | Median unsigned deviation | | 5.56 | 3.65 | 2.42 | 1.84 | 1.79 | 1.04 | 3.07 |
| | $r$ | | 1.00 | 1.00 | 1.00 | 1.00 | 1.00 | 1.00 | 1.00 |
| | $r^2$ | | 1.00 | 1.00 | 1.00 | 1.00 | 1.00 | 1.00 | 1.00 |
| | Max absolute deviation | | 7.33 | 4.57 | 3.41 | 2.78 | 2.79 | 1.98 | 3.58 |
| | lowest negative deviation | | N/A | N/A | N/A | N/A | N/A | N/A | N/A |
| | highest positive deviation | | 7.33 | 4.57 | 3.41 | 2.78 | 2.79 | 1.98 | 3.58 |

| | | $E$ | Error | | | |
|---|---|---|---|---|---|---|
| ID | Complex | Ref.* | DFTB3/CPE($q$)-D3 (original) 3OB | DFTB3/CPE($q$)-D3 (original) MIO | DFTB3/CPE($\zeta$)-D3 (pol) | DFTB3/CPE($q$)-D3 (pol) |
| 1 | 1 water 1hydroxide | -33.93 | 1.69 | -0.25 | 0.99 | 2.50 |
| 2 | 2 water 1hydroxide | -57.42 | 0.56 | -0.23 | 2.33 | 3.86 |
| 3 | 2 water 1hydroxide | -57.37 | -0.01 | -0.63 | 1.75 | 3.32 |
| 4 | 3 water 1hydroxide | -77.14 | -3.26 | -3.40 | 3.21 | 4.42 |
| 5 | 3 water 1hydroxide | -77.08 | -3.85 | -3.87 | 2.70 | 3.87 |
| 6 | 3 water 1hydroxide | -73.06 | -5.42 | -5.93 | 2.85 | 4.91 |
| 7 | 3 water 1hydroxide | -72.69 | -5.43 | -5.92 | 2.94 | 4.89 |
| 8 | 3 water 1hydroxide | -73.13 | -0.79 | -2.90 | 2.50 | 4.01 |
| | RMSD | | 3.32 | 3.63 | 2.50 | 4.04 |
| | Mean deviation | | -2.06 | -2.89 | 2.41 | 3.97 |
| | Media deviation | | -2.02 | -3.15 | 2.60 | 3.94 |
| | Mean unsigned deviation | | 2.63 | 2.89 | 2.41 | 3.97 |
| | Median unsigned deviation | | 2.47 | 3.15 | 2.60 | 3.94 |
| | $r$ | | 1.00 | 1.00 | 1.00 | 1.00 |
| | $r^2$ | | 0.99 | 0.99 | 1.00 | 1.00 |
| | Max absolute deviation | | 5.43 | 5.93 | 3.21 | 4.91 |
| | lowest negative deviation | | -5.43 | -5.93 | N/A | N/A |
| | highest positive deviation | | 1.69 | N/A | 3.21 | 4.91 |

Table S11: L7 error in interaction energy for various methods, compared to a CCSD(T) reference. All values are in kcal/mol, except $r$ and $r^2$ which are unitless. Geometries are taken from Ref. S9, while DPLNO-CCSD(T)/CBS interaction energies were supplied by Stephan Grimme (unpublished).

| | | $E$ | Error | | | | | |
|---|---|---|---|---|---|---|---|---|
| ID | Complex | Ref.* | PM6 | PM6-D3H4 | PBE/6-31G(d) | PBE-D3/6-31G(d) | PBE/def2-QZVP | PBE-D3/def2-QZVP |
| 1 | octadecanedimer | -9.80 | 8.59 | 0.12 | 6.11 | -6.60 | 9.18 | -3.52 |
| 2 | guaninetrimer | -1.60 | 3.29 | -1.64 | 3.54 | -2.24 | 5.62 | -0.16 |
| 3 | circumcoroneneadenine | -15.80 | 12.43 | 0.42 | 14.83 | -1.67 | 17.73 | 1.23 |
| 4 | circumcoroneneGCbasepair | -26.70 | 22.20 | 1.05 | 25.58 | -3.36 | 30.82 | 1.88 |
| 5 | phenylalanineresiduestrimer | -23.60 | 6.33 | -1.78 | -0.58 | -8.20 | 6.48 | -1.14 |
| 6 | coronenedimer | -19.60 | 18.94 | 2.65 | 21.56 | -1.86 | 24.59 | 1.17 |
| 7 | GCGCbasepairstack | -11.20 | 4.65 | -8.22 | 9.41 | -6.56 | 14.50 | -1.47 |
| | RMSD | | 12.83 | 3.42 | 14.53 | 5.02 | 17.88 | 1.79 |
| | Mean deviation | | 10.92 | -1.06 | 11.49 | -4.35 | 15.56 | -0.29 |
| | Median deviation | | 8.59 | 0.12 | 9.41 | -3.36 | 14.50 | -0.16 |
| | Mean unsigned deviation | | 10.92 | 2.27 | 11.66 | 4.35 | 15.56 | 1.51 |
| | Median unsigned deviation | | 8.59 | 1.64 | 9.41 | 3.36 | 14.50 | 1.23 |

| | | | | | | | |
|---|---|---|---|---|---|---|---|
| | $r$ | 0.57 | 0.91 | 0.42 | 0.96 | 0.34 | 0.98 |
| | $r^2$ | 0.32 | 0.84 | 0.18 | 0.92 | 0.12 | 0.96 |
| | Max absolute deviation | 22.20 | 8.22 | 25.58 | 8.20 | 30.82 | 3.52 |
| | lowest negative deviation | N/A | -8.22 | -0.58 | -8.20 | N/A | -3.52 |
| | highest positive deviation | 22.20 | 2.65 | 25.58 | N/A | 30.82 | 1.88 |

| | | $E$ | Error | | | | | | |
|---|---|---|---|---|---|---|---|---|---|
| ID | Complex | Ref.* | DFTB3 | DFTB3-D3 | DFTB3-D3H4 | DFTB3/CPE($U$)-D3* | DFTB3/CPE($U$)-D3 | DFTB3/CPE($\zeta$)-D3 | DFTB3/CPE($q$)-D3 |
| 1 | octadecanedimer | -9.80 | 8.53 | -4.35 | -1.81 | -3.73 | -2.14 | -1.53 | -3.41 |
| 2 | guaninetrimer | -1.60 | 6.89 | 1.24 | 1.35 | -3.23 | -2.48 | -3.62 | -0.48 |
| 3 | circumcoroneneadenine | -15.80 | 16.20 | 0.25 | 2.88 | -0.04 | 0.83 | 1.41 | 2.00 |
| 4 | circumcoroneneGCbasepair | -26.70 | 27.61 | -0.39 | 3.99 | -1.40 | -0.62 | -0.57 | 1.14 |
| 5 | phenylalanineresiduestrimer | -23.60 | 6.34 | -1.32 | -1.87 | -2.40 | -1.56 | -2.90 | -0.68 |
| 6 | coronenedimer | -19.60 | 21.09 | -1.32 | 3.06 | 0.16 | 1.70 | 3.37 | 1.06 |
| 7 | GCGCbasepairstack | -11.20 | 12.02 | -3.61 | -2.38 | -9.64 | -8.94 | -8.72 | -3.52 |
| | RMSD | | 15.92 | 2.31 | 2.61 | 4.23 | 3.72 | 4.03 | 2.11 |
| | Mean deviation | | 14.10 | -1.36 | 0.75 | -2.90 | -1.89 | -1.79 | -0.55 |
| | Media deviation | | 12.02 | -1.32 | 1.35 | -2.40 | -1.56 | -1.53 | -0.48 |
| | Mean unsigned deviation | | 14.10 | 1.78 | 2.48 | 2.94 | 2.61 | 3.16 | 1.76 |
| | Median unsigned deviation | | 12.02 | 1.32 | 2.38 | 2.40 | 1.70 | 2.90 | 1.14 |
| | $r$ | | 0.51 | 0.97 | 0.95 | 0.92 | 0.92 | 0.89 | 0.97 |
| | $r^2$ | | 0.26 | 0.95 | 0.90 | 0.85 | 0.84 | 0.80 | 0.94 |
| | Max absolute deviation | | 27.61 | 4.35 | 3.99 | 9.64 | 8.94 | 8.72 | 3.52 |
| | lowest negative deviation | | N/A | -4.35 | -2.38 | -9.64 | -8.94 | -8.72 | -3.52 |
| | highest positive deviation | | 27.61 | 1.24 | 3.99 | 0.16 | 1.70 | 3.37 | 2.00 |

| | | $E$ | Error | | | |
|---|---|---|---|---|---|---|
| ID | Complex | Ref.* | DFTB3/CPE($q$)-D3 (original) 3OB | DFTB3/CPE($q$)-D3 (original) MIO | DFTB3/CPE($\zeta$)-D3 (pol) | DFTB3/CPE($q$)-D3 (pol) |
| 1 | octadecanedimer | -9.80 | -3.53 | -3.85 | -3.83 | -3.75 |
| 2 | guaninetrimer | -1.60 | -0.05 | -0.49 | -1.11 | 0.31 |
| 3 | circumcoroneneadenine | -15.80 | 1.36 | 0.55 | 0.33 | 1.44 |
| 4 | circumcoroneneGCbasepair | -26.70 | 1.05 | -0.22 | -0.36 | 0.80 |
| 5 | phenylalanineresiduestrimer | -23.60 | -1.65 | -2.00 | -0.84 | -1.01 |
| 6 | coronenedimer | -19.60 | 0.20 | -0.52 | 0.09 | 0.06 |
| 7 | GCGCbasepairstack | -11.20 | -3.90 | -4.87 | -4.84 | -3.52 |
| | RMSD | | 2.18 | 2.49 | 2.40 | 2.08 |
| | Mean deviation | | -0.93 | -1.63 | -1.51 | -0.81 |
| | Media deviation | | -0.05 | -0.52 | -0.84 | 0.06 |
| | Mean unsigned deviation | | 1.68 | 1.79 | 1.63 | 1.56 |
| | Median unsigned deviation | | 1.36 | 0.55 | 0.84 | 1.01 |
| | $r$ | | 0.97 | 0.97 | 0.97 | 0.97 |
| | $r^2$ | | 0.94 | 0.95 | 0.95 | 0.94 |
| | Max absolute deviation | | 3.90 | 4.87 | 4.84 | 3.75 |
| | lowest negative deviation | | -3.90 | -4.87 | -4.84 | -3.75 |
| | highest positive deviation | | 1.36 | 0.55 | 0.33 | 1.44 |

Table S12: W2 error in interaction energy for various methods, compared to a CCSD(T) reference. All values are in kcal/mol, except $r$ and $r^2$ which are unitless.

| | | $E$ | Error | | | | | |
|---|---|---|---|---|---|---|---|---|
| ID | Complex | Ref.* | PM6 | PM6-D3H4 | PBE/6-31G(d) | PBE-D3/6-31G(d) | PBE/def2-QZVP | PBE-D3/def2-QZVP |
| 1 | Water Hydronium | -52.14 | 16.03 | 16.39 | -6.45 | -7.04 | -1.70 | -2.29 |
| 2 | Water Hydroxide | -48.84 | -2.77 | -2.55 | -19.59 | -19.98 | -6.00 | -6.40 |
| | RMSD | | 11.50 | 11.73 | 14.58 | 14.98 | 4.41 | 4.81 |
| | Mean deviation | | 6.63 | 6.92 | -13.02 | -13.51 | -3.85 | -4.35 |
| | Median deviation | | 6.63 | 6.92 | -13.02 | -13.51 | -3.85 | -4.35 |
| | Mean unsigned deviation | | 9.40 | 9.47 | 13.02 | 13.51 | 3.85 | 4.35 |
| | Median unsigned deviation | | 9.40 | 9.47 | 13.02 | 13.51 | 3.85 | 4.35 |
| | $r$ | | -1.00 | -1.00 | -1.00 | -1.00 | -1.00 | -1.00 |
| | $r^2$ | | 1.00 | 1.00 | 1.00 | 1.00 | 1.00 | 1.00 |
| | Max absolute deviation | | 16.03 | 16.39 | 19.59 | 19.98 | 6.00 | 6.40 |
| | lowest negative deviation | | -2.77 | -2.55 | -19.59 | -19.98 | -6.00 | -6.40 |
| | highest positive deviation | | 16.03 | 16.39 | N/A | N/A | N/A | N/A |

| | | $E$ | Error | | | | | | |
|---|---|---|---|---|---|---|---|---|---|
| ID | Complex | Ref.* | DFTB3 | DFTB3-D3 | DFTB3-D3H4 | DFTB3/CPE($U$)-D3* | DFTB3/CPE($U$)-D3 | DFTB3/CPE($\zeta$)-D3 | DFTB3/CPE($q$)-D3 |
| 1 | Water Hydronium | -52.14 | 5.18 | 4.49 | 5.30 | 3.59 | 3.57 | 2.78 | 3.55 |
| 2 | Water Hydroxide | -48.84 | -6.80 | -7.27 | -6.68 | -7.35 | -7.31 | -7.24 | -7.13 |
| | RMSD | | 6.04 | 6.04 | 6.03 | 5.78 | 5.75 | 5.49 | 5.63 |
| | Mean deviation | | -0.81 | -1.39 | -0.69 | -1.88 | -1.87 | -2.23 | -1.79 |

| | | | | | | | |
|---|---|---|---|---|---|---|---|
| Media deviation | -0.81 | -1.39 | -0.69 | -1.88 | -1.87 | -2.23 | -1.79 |
| Mean unsigned deviation | 5.99 | 5.88 | 5.99 | 5.47 | 5.44 | 5.01 | 5.34 |
| Median unsigned deviation | 5.99 | 5.88 | 5.99 | 5.47 | 5.44 | 5.01 | 5.34 |
| $r$ | -1.00 | -1.00 | -1.00 | -1.00 | -1.00 | -1.00 | -1.00 |
| $r^2$ | 1.00 | 1.00 | 1.00 | 1.00 | 1.00 | 1.00 | 1.00 |
| Max absolute deviation | 6.80 | 7.27 | 6.68 | 7.35 | 7.31 | 7.24 | 7.13 |
| lowest negative deviation | -6.80 | -7.27 | -6.68 | -7.35 | -7.31 | -7.24 | -7.13 |
| highest positive deviation | 5.18 | 4.49 | 5.30 | 3.59 | 3.57 | 2.78 | 3.55 |

| | | $E$ | Error | | | |
|---|---|---|---|---|---|---|
| ID | Complex | Ref.* | DFTB3/ CPE($q$) -D3 (original) 3OB | DFTB3/ CPE($q$) -D3 (original) MIO | DFTB3/ CPE($\zeta$) -D3 (pol) | DFTB3/ CPE($q$) -D3 (pol) |
| 1 | Water Hydronium | -52.14 | 3.57 | 1.65 | 3.41 | 4.17 |
| 2 | Water Hydroxide | -48.84 | -8.27 | -9.47 | -6.96 | -6.97 |
| | RMSD | | 6.37 | 6.79 | 5.48 | 5.75 |
| | Mean deviation | | -2.35 | -3.91 | -1.78 | -1.40 |
| | Media deviation | | -2.35 | -3.91 | -1.78 | -1.40 |
| | Mean unsigned deviation | | 5.92 | 5.56 | 5.19 | 5.57 |
| | Median unsigned deviation | | 5.92 | 5.56 | 5.19 | 5.57 |
| | $r$ | | -1.00 | -1.00 | -1.00 | -1.00 |
| | $r^2$ | | 1.00 | 1.00 | 1.00 | 1.00 |
| | Max absolute deviation | | 8.27 | 9.47 | 6.96 | 6.97 |
| | lowest negative deviation | | -8.27 | -9.47 | -6.96 | -6.97 |
| | highest positive deviation | | 3.57 | 1.65 | 3.41 | 4.17 |

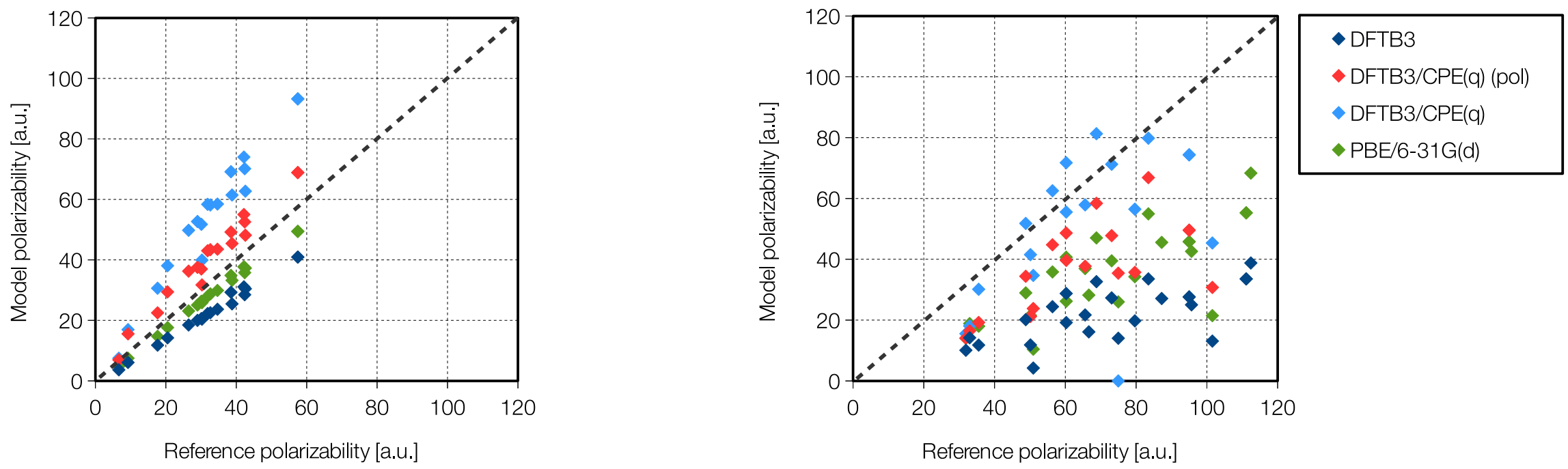

(a) Cations

(b) Anions

Figure S1: The polarizability of two data sets of 19 cations (a) and 27 anions (b) calculated with four different methods with respect to a B3LYP/aug-cc-pVTZ reference.

Table S13: Error for the data set of neutral molecular polarizability relative to a B3LYP/aug-cc-pVTZ reference in bohr$^3$.

| ID | Molecule | Ref | DFTB3 | DFTB3/ CPE($q$) (original) MIO | DFTB3/ CPE($q$) (original) 3OB | DFTB3/ CPE($\zeta$) (pol) | DFTB3/ CPE($q$) (pol) | DFTB3/ CPE($\zeta$) | DFTB3/ CPE($q$) | PBE/ 6-31G(d) |
|---|---|---|---|---|---|---|---|---|---|---|
| ID | 4-methylimidazole[0] | 63.12 | -25.09 | -3.35 | -2.81 | 0.54 | 6.77 | 124.73 | 31.88 | -12.64 |
| 1 | 5-methylimidazole[0] | 62.98 | -24.86 | -3.36 | -2.84 | 0.66 | 6.78 | 120.87 | 31.76 | -12.40 |
| 3 | C2H6[0] | 29.32 | -11.92 | -0.71 | -0.20 | 6.73 | 4.92 | 82.82 | 19.69 | -5.67 |
| 4 | C3H6[0] | 41.32 | -16.51 | -2.45 | -1.91 | 4.13 | 4.70 | 98.98 | 24.66 | -9.06 |
| 5 | C3H6-cyclo[0] | 37.12 | -13.85 | -0.02 | 0.42 | 5.69 | 7.75 | 97.42 | 27.26 | -6.82 |
| 6 | C3H6O2[0] | 45.30 | -17.78 | -5.01 | -4.43 | 3.92 | 5.31 | 97.61 | 27.12 | -7.68 |
| 7 | C3H8[0] | 41.77 | -16.15 | -1.37 | -0.73 | 8.54 | 6.66 | 109.73 | 26.80 | -7.19 |
| 8 | C4H4O[0] | 49.42 | -20.30 | -3.95 | -3.77 | -0.95 | 3.90 | 97.97 | 25.86 | -11.50 |
| 9 | C4H6-O[0] | 51.11 | -17.69 | -0.49 | 0.24 | 4.66 | 6.80 | 105.81 | 28.74 | -8.93 |
| 10 | C4H6O[0] | 51.48 | -18.71 | -2.62 | -2.13 | 4.33 | 7.30 | 112.97 | 30.77 | -9.52 |
| 11 | C5H5-CH2[0] | 75.24 | -26.59 | -5.32 | -4.77 | -0.49 | 4.73 | conv | 34.45 | -15.80 |
| 12 | C5H5N[0] | 64.58 | -24.54 | -5.09 | -4.68 | 0.03 | 5.16 | 119.33 | 29.39 | -13.02 |
| 13 | C5H6[0] | 58.90 | -21.28 | -4.22 | -3.73 | 2.79 | 5.63 | 122.85 | 28.71 | -11.85 |
| 14 | C6H4[0] | 67.93 | -25.90 | -7.74 | -7.36 | -4.77 | 0.91 | 111.83 | 24.72 | -15.21 |
| 15 | C6H6[0] | 70.11 | -26.03 | -7.04 | -6.62 | 0.52 | 4.64 | conv | 31.21 | -14.49 |
| 16 | CH2[0] | 18.46 | -14.03 | -6.71 | -6.46 | -5.46 | -4.85 | 29.58 | 2.70 | |
| 17 | CH2C(CH3)CH3[0] | 54.03 | -20.37 | -1.76 | -0.93 | 6.19 | 6.19 | 124.31 | 30.19 | -10.29 |
| 18 | Pyrazole[0] | 49.49 | -21.02 | -3.08 | -2.95 | -1.43 | 3.39 | 89.50 | 20.81 | -11.59 |
| 19 | CH2CH2[0] | 28.32 | -13.31 | -2.91 | -2.64 | 0.71 | 2.12 | 67.54 | 17.25 | -8.02 |
| 20 | CH2CH2CH(CCH)[0] | 59.94 | -21.41 | 0.80 | 1.58 | 3.64 | 10.76 | 137.08 | 42.23 | -12.33 |
| 21 | CH2CH2C-O[0] | 39.45 | -14.26 | 1.13 | 1.70 | 1.71 | 6.13 | 79.13 | 25.04 | -7.62 |
| 22 | CH2CHCH2CH3[0] | 54.25 | -20.56 | -2.09 | -1.34 | 6.21 | 7.28 | 128.56 | 34.67 | -10.54 |
| 23 | CH2CHCH2OH[0] | 45.88 | -19.05 | -4.97 | -4.34 | 2.65 | 2.47 | 106.44 | 23.23 | -9.98 |
| 24 | CH2CHCHCH2[0] | 58.08 | -21.81 | -4.43 | -3.90 | 0.39 | 3.72 | 115.16 | 30.29 | -13.01 |
| 25 | CH2CHCOOH[0] | 47.23 | -19.79 | -4.54 | -3.75 | -2.22 | -0.11 | 91.04 | 19.82 | -10.31 |
| 26 | CH2CHOCHCH2[0] | 56.98 | -22.30 | -2.13 | -1.54 | 1.03 | 5.02 | 112.07 | 31.99 | -12.54 |
| 27 | CH2CO[0] | 29.25 | -13.70 | 0.26 | 0.61 | -3.13 | 0.42 | 46.35 | 15.37 | -8.42 |
| 28 | CH2NH[0] | 23.30 | -11.77 | -3.41 | -3.12 | -0.22 | 2.80 | 49.34 | 11.97 | -6.25 |
| 29 | CH2OCH2[0] | 29.18 | -11.63 | -2.28 | -1.94 | 3.66 | 4.11 | 69.77 | 19.22 | -5.74 |
| 30 | Pyridazine[0] | 59.53 | -23.12 | -3.96 | -3.73 | -1.65 | 4.10 | 98.02 | 23.89 | -11.96 |
| 31 | (CH3)2CHOH[0] | 46.72 | -19.39 | -2.93 | -2.04 | 6.10 | 3.92 | 113.91 | 24.92 | -8.42 |
| 32 | (CH3)2CHSH[0] | 63.13 | -27.96 | -4.25 | -4.74 | 6.94 | 5.08 | 109.24 | 24.21 | -13.03 |
| 33 | CH3CH2CCH[0] | 50.40 | -20.03 | -1.03 | -0.27 | 4.18 | 7.53 | 120.24 | 34.14 | -11.29 |
| 34 | CH3CH2CHO[0] | 43.25 | -16.38 | -2.00 | -1.29 | 4.74 | 5.10 | 93.49 | 25.86 | -7.79 |
| 35 | CH3CH2COOH[0] | 46.90 | -18.42 | -2.03 | -1.01 | 3.37 | 3.07 | 105.86 | 24.01 | -8.68 |
| 36 | CH3CH2NH2[0] | 39.15 | -17.59 | -3.42 | -2.59 | 4.31 | 5.68 | 96.42 | 22.22 | -8.31 |
| 37 | CH3CH2OCH3[0] | 46.67 | -19.11 | -4.01 | -3.36 | 6.54 | 5.51 | 110.97 | 27.71 | -8.21 |
| 38 | CH3CH2OCHO[0] | 47.60 | -18.55 | -3.74 | -2.99 | 3.38 | 4.75 | 100.25 | 28.23 | -8.34 |
| 39 | CH3CH2OH[0] | 34.30 | -14.92 | -3.42 | -2.78 | 4.56 | 2.21 | 88.53 | 17.79 | -6.91 |
| 40 | CH3CH2SH[0] | 50.47 | -23.62 | -5.54 | -6.11 | 5.88 | 1.56 | 87.80 | 15.34 | -12.12 |
| 41 | CH3CHCHCOOH[0] | 61.73 | -22.68 | -3.42 | -2.23 | 2.06 | 2.98 | 125.89 | 29.42 | -11.09 |
| 42 | CH3CHOCH2[0] | 41.76 | -15.57 | -1.73 | -1.13 | 5.92 | 6.53 | 99.92 | 27.77 | -7.19 |
| 43 | CH3CHSHCH3[0] | 63.22 | -28.02 | -4.85 | -5.34 | 7.04 | 4.42 | 110.88 | 23.13 | -13.28 |
| 44 | CH3COCH3[0] | 43.16 | -16.32 | -0.09 | 0.76 | 4.86 | 5.44 | 97.41 | 25.93 | -8.12 |
| 45 | CH3C(O)C(O)CH3[0] | 56.19 | -19.38 | 0.26 | 1.26 | 4.60 | 7.66 | 121.15 | 35.40 | -9.78 |
| 46 | CH3COH[0] | 30.97 | -12.96 | -2.13 | -1.61 | 2.15 | 2.65 | 65.92 | 18.03 | -6.58 |
| 47 | CH3C(OH)CH2[0] | 46.24 | -19.23 | -1.75 | -0.92 | 2.27 | 2.52 | 103.14 | 23.30 | -10.32 |
| 48 | CH3CONH2[0] | 40.35 | -17.81 | -2.52 | -1.51 | 0.32 | 3.86 | 82.63 | 19.56 | -9.25 |
| 49 | CH3CONHCH3[0] | 52.45 | -20.57 | -2.15 | -1.13 | 3.25 | 6.74 | 105.63 | 28.65 | -9.85 |
| 50 | CH3C(O)OCH3[0] | 47.15 | -18.11 | -1.79 | -0.88 | 3.80 | 5.63 | 102.93 | 29.21 | -8.40 |
| 51 | CH3COOH[0] | 34.88 | -14.75 | -2.00 | -1.13 | 0.76 | 0.81 | 74.37 | 16.30 | -7.38 |
| 52 | CH3NC(NH2)2[0] | 56.82 | -26.40 | -5.55 | -4.15 | -2.12 | 7.60 | 109.98 | 27.65 | -12.66 |
| 53 | CH3NH2[0] | 26.55 | -13.21 | -3.34 | -2.75 | 2.34 | 3.93 | 67.43 | 14.75 | -6.89 |
| 54 | CH3NHCH3[0] | 39.43 | -17.83 | -4.47 | -3.89 | 3.77 | 4.98 | 91.34 | 22.49 | -8.21 |
| 55 | CH3NHC(NH2)(NH)[0] | 55.98 | -25.49 | -3.96 | -2.65 | -1.01 | 8.62 | 110.03 | 28.29 | -12.60 |
| 56 | CH3NO[0] | 28.70 | -13.19 | -4.40 | -4.12 | -2.97 | -0.17 | 38.73 | 9.03 | -6.80 |
| 57 | CH3NO2[0] | 32.91 | -13.01 | -2.31 | -2.05 | -3.16 | -1.23 | 37.97 | 8.02 | -7.33 |
| 58 | CH3OCH3[0] | 34.67 | -15.27 | -4.75 | -4.34 | 4.23 | 3.13 | 82.26 | 19.14 | -6.97 |
| 59 | CH3OCHO[0] | 34.84 | -14.71 | -4.25 | -3.73 | 0.84 | 2.20 | 68.91 | 18.33 | -7.08 |
| 60 | CH3OC(O)CHCH2[0] | 60.09 | -22.54 | -4.08 | -3.19 | 1.40 | 5.40 | 118.03 | 33.46 | -11.39 |
| 61 | CH3OH[0] | 21.78 | -10.67 | -3.49 | -3.09 | 2.60 | 0.11 | 57.89 | 9.00 | -5.49 |
| 62 | CH3OOH[0] | 28.42 | -13.65 | -6.61 | -6.20 | -0.31 | -2.62 | 56.09 | 6.68 | -6.80 |
| 63 | CH3SH[0] | 37.67 | -19.70 | -5.76 | -6.38 | 3.56 | -1.50 | 58.09 | 5.96 | -10.81 |
| 64 | CH4[0] | 17.16 | -8.20 | 0.20 | 0.68 | 4.10 | 2.48 | 53.09 | 10.62 | -4.37 |
| 65 | CHCOH[0] | 29.44 | -15.80 | -3.01 | -2.61 | -4.27 | -1.45 | 55.46 | 12.68 | -9.70 |
| 66 | CO2[0] | 17.38 | -9.10 | -4.25 | -4.05 | -5.65 | -2.45 | 16.44 | 6.79 | -4.69 |
| 67 | Diazine-1-3[0] | 58.79 | -22.79 | -2.91 | -2.39 | -1.38 | 6.43 | 100.91 | 28.61 | -11.66 |
| 68 | GLY[0] | 43.75 | -19.16 | -4.14 | -3.18 | -0.78 | 3.72 | 81.31 | 20.80 | -9.70 |
| 69 | H2CO[0] | 18.17 | -9.22 | -3.81 | -3.64 | -0.43 | -0.02 | 34.24 | 9.06 | -4.67 |
| 70 | H2NCH2CH2CN[0] | 52.04 | -21.27 | -2.12 | -1.21 | 4.20 | 10.45 | 102.24 | 31.68 | -10.23 |
| 71 | HCCH[0] | 23.86 | -13.85 | -3.46 | -3.30 | -3.35 | 0.23 | 49.73 | 13.35 | -9.05 |
| 72 | HCN[0] | 17.35 | -10.21 | -2.64 | -2.57 | -3.35 | 1.82 | 23.50 | 9.50 | -5.38 |
| 73 | HC(O)CHO[0] | 32.35 | -13.56 | -4.57 | -4.17 | -1.84 | 0.90 | 49.80 | 16.75 | -7.15 |
| 74 | HCOH[0] | 18.18 | -9.23 | -3.82 | -3.64 | -0.44 | -0.03 | 34.23 | 9.05 | -4.68 |
| 75 | pyridine[0] | 64.58 | -24.53 | -5.09 | -4.67 | 0.03 | 5.16 | 119.34 | 29.39 | -13.02 |
| 76 | HCOOH[0] | 22.90 | -11.49 | -4.81 | -4.35 | -2.23 | -2.64 | 40.56 | 6.51 | -5.93 |
| 77 | HO-CH2CH2-OH[0] | 38.93 | -17.64 | -5.69 | -4.94 | 2.78 | -0.26 | 92.81 | 15.93 | -8.29 |
| 78 | HO-CH2CH2-SH[0] | 55.51 | -26.55 | -7.35 | -8.06 | 3.78 | -1.41 | 92.01 | 12.46 | -12.92 |
| 79 | HOH[0] | 9.86 | -7.11 | -2.87 | -2.44 | -0.49 | -4.50 | 27.95 | -3.57 | -4.44 |
| 80 | HS-CH2CH2-SH[0] | 73.27 | -35.78 | -9.33 | -11.57 | 4.99 | -1.90 | 90.39 | 11.13 | -17.78 |
| 81 | HSH[0] | 25.09 | -17.27 | -8.59 | -8.58 | 2.65 | -3.62 | 26.82 | -4.21 | -10.18 |
| 82 | imidazole[0] | 49.73 | -21.78 | -4.53 | -4.31 | -2.76 | 3.39 | 88.32 | 21.34 | -11.66 |
| 83 | URA[0] | 70.88 | -24.54 | -2.95 | -1.87 | -4.43 | 5.59 | 103.74 | 28.53 | -13.59 |
| 84 | NH2CH2CH2NH2[0] | 48.53 | -22.73 | -5.46 | -4.34 | 2.26 | 7.75 | 112.04 | 28.01 | -10.90 |
| 85 | NH2CHO[0] | 28.50 | -14.44 | -4.97 | -4.28 | -2.92 | 0.88 | 50.10 | 11.03 | -7.70 |
| 86 | NH3[0] | 14.65 | -9.70 | -2.73 | -1.95 | -0.54 | 1.90 | 38.26 | 5.37 | -6.16 |
| 87 | Pyrazine[0] | 59.72 | -23.45 | -3.92 | -3.62 | -1.56 | 4.63 | 96.51 | 25.56 | -11.81 |
| | RMSD | | 19.08 | 4.02 | 3.76 | 3.64 | 4.75 | 91.38 | 23.27 | 9.88 |
| | Mean deviation | | -18.30 | -3.48 | -3.00 | 1.53 | 3.41 | 86.35 | 21.29 | -9.46 |
| | Media deviation | | -18.42 | -3.42 | -2.95 | 2.06 | 3.92 | 96.42 | 23.30 | -9.16 |
| | Mean unsigned deviation | | 18.30 | 3.55 | 3.16 | 3.05 | 4.07 | 86.35 | 21.47 | 9.46 |
| | Median unsigned deviation | | 18.42 | 3.42 | 2.95 | 2.97 | 4.10 | 96.42 | 23.30 | 9.16 |
| | $r$ | | 0.98 | 0.99 | 0.99 | 0.98 | 0.99 | 0.94 | 0.97 | 1.00 |
| | $r^2$ | | 0.97 | 0.98 | 0.98 | 0.96 | 0.98 | 0.89 | 0.94 | 0.99 |
| | Max absolute deviation | | 35.78 | 9.33 | 11.57 | 8.54 | 10.76 | 137.08 | 42.23 | 137.08 |
| | lowest negative deviation | | -35.78 | -9.33 | -11.57 | -5.65 | -4.85 | N/A | -4.21 | -17.78 |
| | highest positive deviation | | N/A | 1.13 | 1.70 | 8.54 | 10.76 | 137.08 | 42.23 | N/A |

Table S14: Error for the data set of cation molecular polarizability relative to a B3LYP/aug-cc-pVTZ reference in bohr$^3$.

| ID | Molecule | Ref | DFTB3 | DFTB3/ CPE($q$) (original) MIO | DFTB3/ CPE($q$) (original) 3OB | DFTB3/ CPE($\zeta$) (pol) | DFTB3/ CPE($q$) (pol) | DFTB3/ CPE($\zeta$) | DFTB3/ CPE($q$) | PBE/ 6-31G(d) |
|---|---|---|---|---|---|---|---|---|---|---|
| 1 | C5H5-NH[+1] | 57.47 | -16.56 | -1.02 | -0.68 | 9.42 | 12.76 | 138.32 | 35.76 | -8.04 |
| 2 | (CH3)2NH2[+1] | 42.17 | -11.08 | 3.68 | 4.56 | 15.29 | 13.71 | 128.70 | 31.83 | -4.46 |
| 3 | (CH3)2OH[+1] | 28.99 | -8.99 | 0.09 | 0.56 | 13.03 | 9.33 | 100.58 | 23.71 | -3.93 |
| 4 | (CH3)2SH[+1] | 42.58 | -12.13 | 2.37 | 0.47 | 17.51 | 6.32 | 106.19 | 20.14 | -5.33 |
| 5 | CH3CH2NH3[+1] | 32.65 | -10.14 | 1.09 | 1.83 | 13.96 | 11.30 | 111.71 | 25.51 | -3.89 |
| 6 | CH3CH2OH2[+1] | 30.09 | -9.50 | 0.05 | 0.65 | 12.57 | 7.61 | 104.08 | 21.68 | -3.77 |
| 7 | CH3CNH[+1] | 26.47 | -8.01 | 1.64 | 2.22 | 8.60 | 10.48 | 86.24 | 23.32 | -3.37 |
| 8 | CH3C(OH)CH3[+1] | 38.50 | -9.22 | 3.99 | 4.89 | 13.98 | 11.81 | 119.12 | 30.66 | -3.65 |
| 9 | CH3C(OH)NH2[+1] | 34.67 | -11.00 | 0.92 | 1.73 | 9.98 | 9.50 | 105.21 | 23.80 | -4.78 |
| 10 | CH3NH2CH3[+1] | 31.87 | -9.62 | 1.54 | 2.15 | 14.07 | 11.82 | 108.35 | 26.52 | -3.91 |
| 11 | CH3NH3[+1] | 20.50 | -6.29 | 0.84 | 1.40 | 11.50 | 9.19 | 83.69 | 17.56 | -2.89 |
| 12 | CH3OH2[+1] | 17.66 | -5.86 | -0.51 | -0.12 | 10.11 | 5.24 | 74.94 | 12.96 | -2.93 |
| 13 | CH3SH2[+1] | 30.26 | -9.87 | -1.35 | -2.32 | 15.77 | 1.83 | 78.67 | 9.67 | -4.41 |
| 14 | GLY[+1] | 38.80 | -13.34 | -1.88 | -1.16 | 7.39 | 7.38 | 100.40 | 22.65 | -5.57 |
| 15 | H3O[+1] | 6.65 | -3.00 | -2.01 | -1.80 | 6.56 | 0.29 | 45.11 | 0.83 | -1.83 |
| 16 | ImidazoleH[+1] | 42.40 | -13.93 | -0.27 | -0.09 | 6.94 | 10.98 | 109.57 | 27.75 | -6.62 |
| 17 | imidazolium[+1] | 42.40 | -13.93 | -0.27 | -0.09 | 6.94 | 10.99 | 109.58 | 27.76 | -6.62 |
| 18 | NH4[+1] | 9.26 | -3.18 | -0.36 | 0.10 | 8.48 | 6.13 | 55.48 | 7.67 | -1.79 |
| 19 | pyridinium[+1] | 57.48 | -16.56 | -1.02 | -0.68 | 9.42 | 12.75 | 138.32 | 35.75 | -8.05 |
| | RMSD | | 10.79 | 1.70 | 1.98 | 11.60 | 9.61 | 103.11 | 24.16 | 4.85 |
| | Mean deviation | | -10.12 | 0.39 | 0.72 | 11.13 | 8.92 | 100.22 | 22.40 | -4.52 |
| | Media deviation | | -9.87 | 0.05 | 0.47 | 10.11 | 9.50 | 105.21 | 23.71 | -3.93 |
| | Mean unsigned deviation | | 10.12 | 1.31 | 1.45 | 11.13 | 8.92 | 100.22 | 22.40 | 4.52 |
| | Median unsigned deviation | | 9.87 | 1.02 | 1.16 | 10.11 | 9.50 | 105.21 | 23.71 | 3.93 |
| | $r$ | | 1.00 | 0.99 | 0.99 | 0.97 | 0.99 | 0.97 | 0.98 | 1.00 |
| | $r^2$ | | 0.99 | 0.99 | 0.98 | 0.95 | 0.97 | 0.95 | 0.96 | 1.00 |
| | Max absolute deviation | | 16.56 | 3.99 | 4.89 | 17.51 | 13.71 | 138.32 | 35.76 | 138.32 |
| | lowest negative deviation | | -16.56 | -2.01 | -2.32 | N/A | N/A | N/A | N/A | -8.05 |
| | highest positive deviation | | N/A | 3.99 | 4.89 | 17.51 | 13.71 | 138.32 | 35.76 | N/A |

Table S15: Error for the data set of anionic molecular polarizability relative to a B3LYP/aug-cc-pVTZ reference in bohr$^3$. Some molecules displayed convergence problems and are left out of the statistics. These are marked "conv" instead of the polarizability error.

| ID | Molecule | Ref | DFTB3 | DFTB3/ CPE($q$) (original) MIO | DFTB3/ CPE($q$) (original) 3OB | DFTB3/ CPE($\zeta$) (pol) | DFTB3/ CPE($q$) (pol) | DFTB3/ CPE($\zeta$) | DFTB3/ CPE($q$) | PBE/ 6-31G(d) |
|---|---|---|---|---|---|---|---|---|---|---|
| 1 | C6H5-O[-1] | 95.29 | -43.66 | -15.21 | -14.29 | -19.31 | -13.66 | conv | 15.47 | -29.44 |
| 2 | C6H5-S[-1] | 126.85 | -64.45 | -8.31 | -10.83 | -29.18 | 141.82 | conv | 174.12 | -42.59 |
| 3 | CH2CN[-1] | 60.25 | -41.07 | -11.48 | -10.61 | -28.03 | -20.06 | 21.85 | -4.72 | -34.05 |
| 4 | CH2N[-1] | 101.55 | -88.46 | conv | conv | -79.35 | -70.65 | -43.94 | -56.20 | -80.09 |
| 5 | (CH3)2CHO[-1] | 95.00 | -67.39 | 0.89 | conv | -44.93 | -44.38 | 51.21 | -20.64 | -49.19 |
| 6 | (CH3)2CHS[-1] | 111.13 | -77.59 | -20.61 | -20.49 | -45.03 | 337.59 | 50.64 | 382.28 | -55.86 |
| 7 | CH3CH2COO[-1] | 60.24 | -31.50 | -6.98 | -5.50 | -12.94 | -10.41 | 70.74 | 11.51 | -19.56 |
| 8 | CH3CH2NH[-1] | 170.81 | -148.86 | conv | conv | -129.01 | -121.13 | -45.82 | conv | -130.25 |
| 9 | CH3CH2O[-1] | 79.64 | -59.89 | -19.38 | conv | -43.58 | -43.20 | 26.70 | -23.16 | -45.38 |
| 10 | CH3CH2S[-1] | 95.66 | -70.64 | -19.26 | -19.30 | -43.96 | 309.68 | 27.88 | 347.62 | -53.07 |
| 11 | CH3CH(NH2)COO[-1] | 68.77 | -36.13 | -8.39 | -6.68 | -14.41 | -9.37 | 83.45 | 12.52 | -21.73 |
| 12 | CH3CHOCH3[-1] | 95.00 | -67.39 | 0.88 | conv | -44.93 | -44.38 | 51.22 | -20.64 | -49.19 |
| 13 | CH3CHSCH3[-1] | 111.11 | -77.56 | -20.57 | -20.44 | -45.01 | 338.04 | 50.67 | 382.81 | -55.84 |
| 14 | CH3COCH2[-1] | 73.08 | -45.78 | -13.08 | -11.16 | -25.15 | -24.09 | 60.33 | -1.83 | -33.60 |
| 15 | CH3COO[-1] | 48.80 | -28.65 | -7.21 | -5.82 | -16.33 | -13.45 | 40.89 | 3.00 | -19.86 |
| 16 | CH3NH[-1] | 74.95 | -60.93 | conv | conv | -47.91 | -39.42 | 8.44 | conv | -48.97 |
| 17 | CH3O[-1] | 50.13 | -38.30 | -2.98 | 6.92 | -28.19 | -28.65 | 14.38 | -8.61 | -28.61 |
| 18 | CH3S[-1] | 66.63 | -50.50 | -5.37 | -5.66 | -30.37 | 295.03 | 15.79 | 326.43 | -38.42 |
| 19 | GLY[-1] | 56.36 | -31.95 | -8.22 | -6.69 | -16.14 | -10.85 | 55.25 | 6.18 | -20.53 |
| 20 | HCOO[-1] | 35.50 | -23.72 | -3.94 | -2.63 | -17.52 | -15.79 | 10.54 | -5.37 | -17.52 |
| 21 | HO-CH2CH2-O[-1] | 65.59 | -43.90 | -16.25 | -13.76 | -26.47 | -27.08 | 48.39 | -7.70 | -28.64 |
| 22 | HO-CH2CH2-S[-1] | 87.23 | -60.14 | -8.97 | -9.01 | -32.57 | 288.00 | 44.66 | 333.99 | -41.68 |
| 23 | HS-CH2CH2-O[-1] | 83.46 | -49.93 | -8.70 | -15.94 | -23.86 | -16.78 | 48.32 | -3.64 | -28.51 |
| 24 | HS-CH2CH2-S[-1] | 112.42 | -73.64 | -9.28 | -16.04 | -37.23 | 266.41 | 37.76 | 308.14 | -44.08 |
| 25 | NH2[-1] | 50.94 | -46.71 | -17.06 | -14.25 | -40.16 | -27.46 | -10.90 | -16.26 | -40.52 |
| 26 | NO2[-1] | 31.89 | -21.82 | -12.33 | -12.08 | -21.82 | -17.87 | -21.51 | -16.37 | -17.88 |
| 27 | NO3[-1] | 32.96 | -18.78 | -4.39 | -4.18 | -18.78 | -16.51 | -18.47 | -14.89 | -14.11 |
| | RMSD | | 60.38 | 12.01 | 12.26 | 42.55 | 151.21 | 43.07 | 174.70 | 46.53 |
| | Mean deviation | | -54.42 | -10.26 | -10.40 | -35.64 | 50.42 | 27.14 | 84.16 | -40.34 |
| | Media deviation | | -49.93 | -8.84 | -10.83 | -29.18 | -16.51 | 37.76 | -1.83 | -38.42 |
| | Mean unsigned deviation | | 54.42 | 10.41 | 11.06 | 35.64 | 95.99 | 38.39 | 100.16 | 40.34 |
| | Median unsigned deviation | | 49.93 | 8.84 | 10.83 | 29.18 | 28.65 | 43.94 | 16.26 | 38.42 |
| | $r$ | | 0.56 | 0.97 | 0.98 | 0.66 | 0.45 | 0.65 | 0.65 | 0.68 |
| | $r^2$ | | 0.32 | 0.94 | 0.96 | 0.44 | 0.20 | 0.42 | 0.42 | 0.46 |
| | Max absolute deviation | | 148.86 | 20.61 | 20.49 | 129.01 | 338.04 | 83.45 | 382.81 | 130.25 |
| | lowest negative deviation | | -148.86 | -20.61 | -20.49 | -129.01 | -121.13 | -45.82 | -56.20 | -130.25 |
| | highest positive deviation | | N/A | 0.89 | 6.93 | N/A | 338.04 | 83.45 | 382.81 | N/A |

# S4: Optimization with DFTB3/CPE($q$)-D3

Table S16: Interaction energies in kcal/mol for the various data sets using either reference geometries or DFTB3/CPE($q$)-D3 optimized geometries. All RMSD values are in kcal/mol. Energies are evaluated using the DFTB3/CPE($q$)-D3 model. In most cases, the change in geometry upon DFTB3/CPE($q$)-D3 optimization is very small. The few largest changes are illustrated in Figs.S2-S4

| Data set | RMSD (reference geometry) | RMSD (Optimized geometry) |
|---|---|---|
| S22 | 1.13 | 1.02 |
| S66 | 0.63 | 1.14 |
| C15 | 1.49 | 2.09 |
| Large water | 3.04 | 4.87 |
| Charged Water | 2.97 | 3.11 |
| L7 | 2.11 | 3.77 |

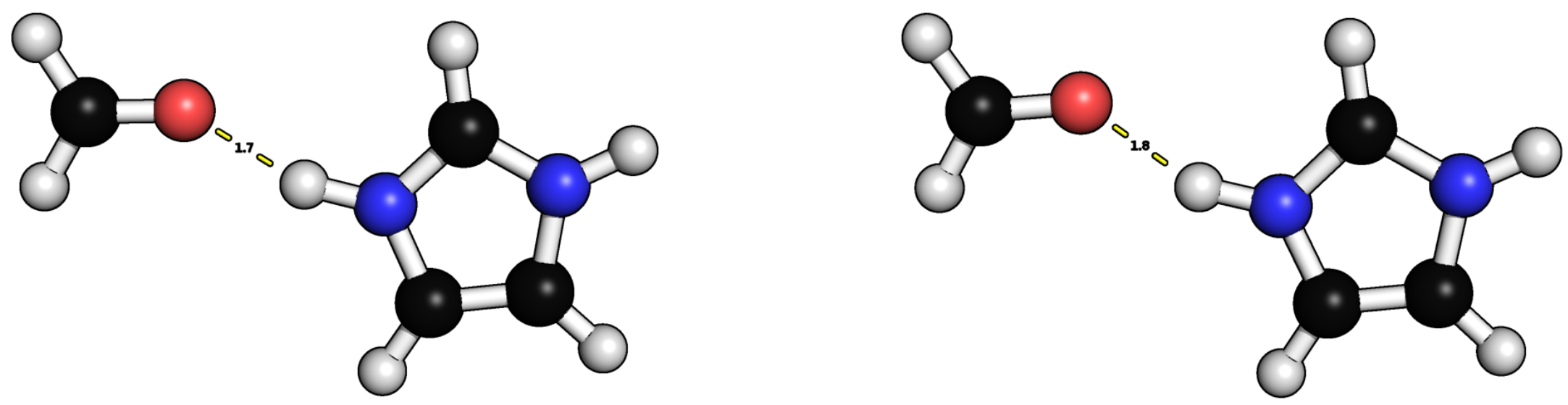

(a) MP2/cc-pVTZ geometry (b) DFTB3/CPE(q)-D3 geometry

Figure S2: Imidazolium ... formaldehyde complex from Ref. S6, (a) is the reference MP2/cc-pVTZ geometry, and (b) is the DFTB3/CPE($q$)-D3 minimized geometry.

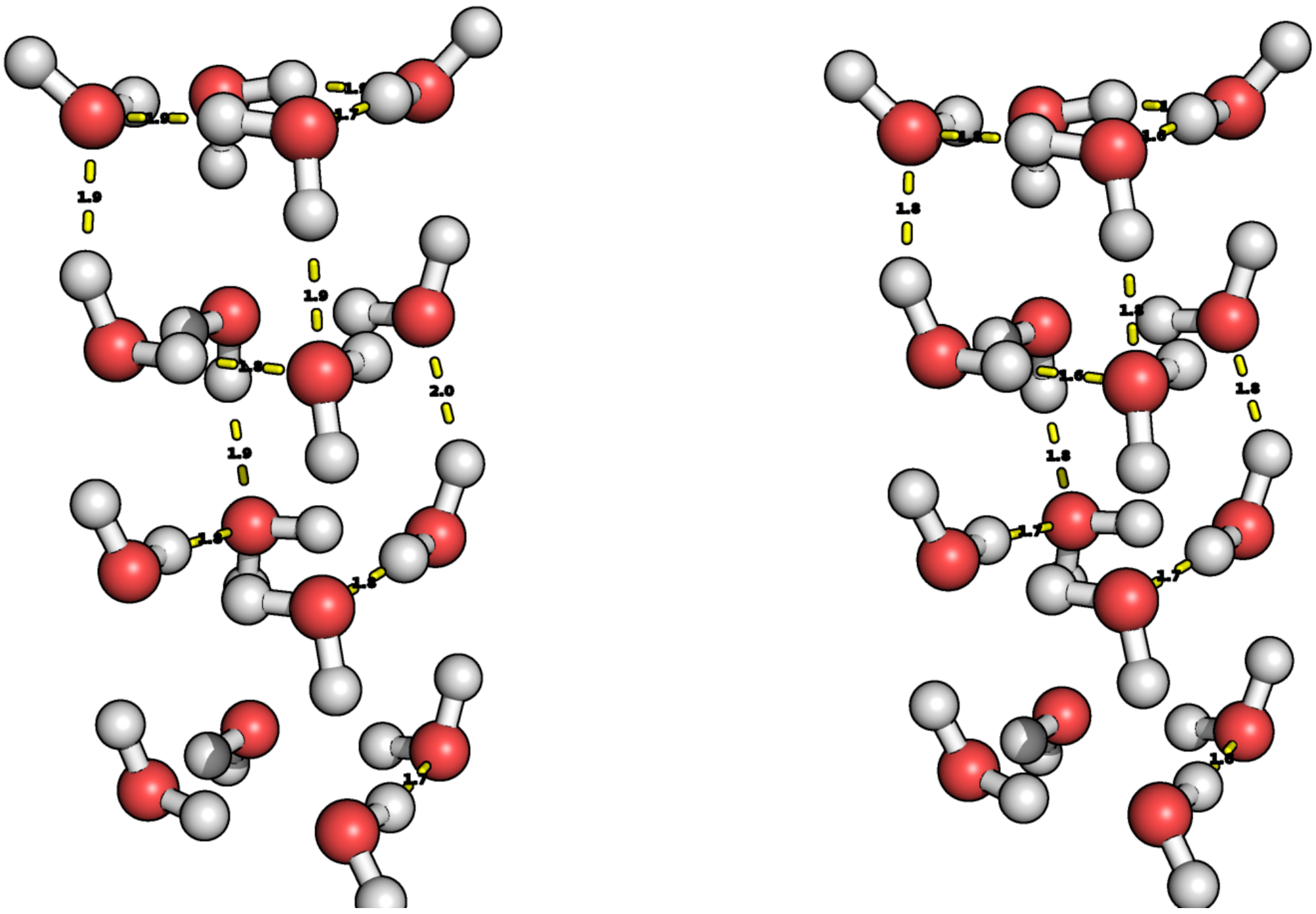

(a) MP2/cc-pVTZ geometry (b) DFTB3/CPE(q)-D3 geometry

Figure S3: 16-Water cluster (4444-b) from Ref. S8, (a) is the reference MP2/cc-pVTZ geometry, and (b) is the DFTB3/CPE($q$)-D3 minimized geometry.

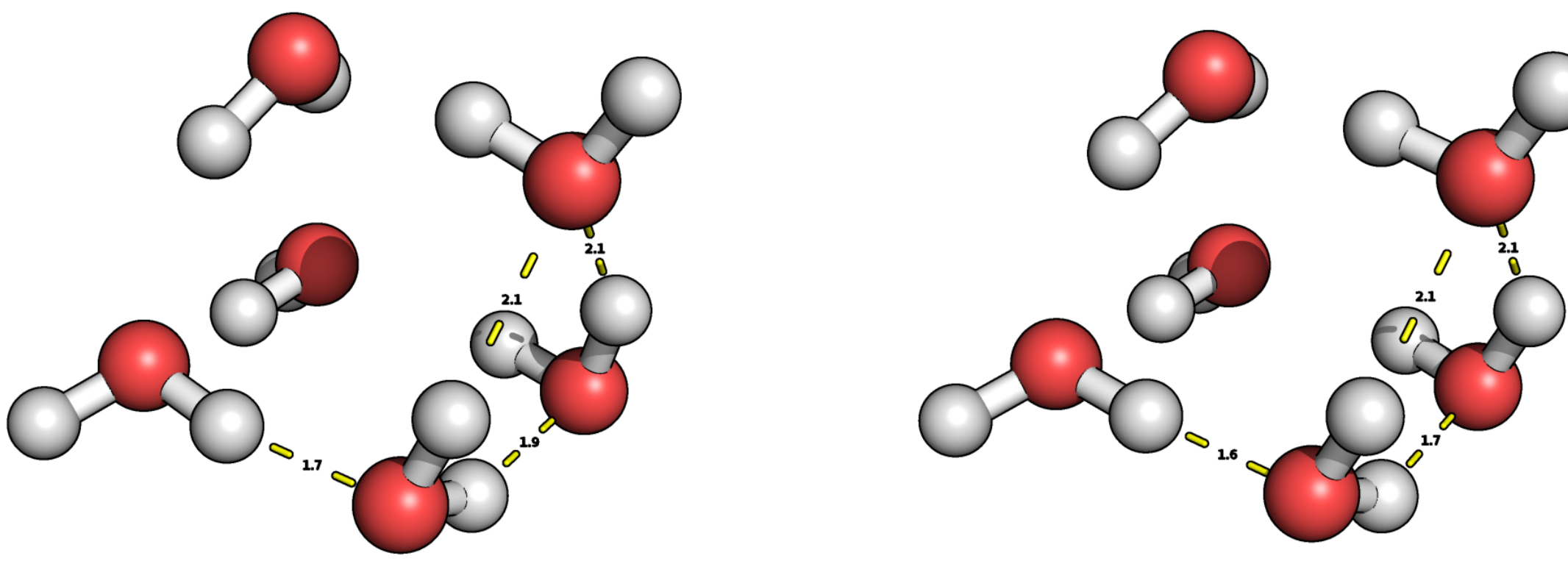

(a) MP2/cc-pVTZ geometry (b) DFTB3/CPE(q)-D3 geometry

Figure S4: 6-Water cluster (prism) from Ref. S8, (a) is the reference MP2/cc-pVTZ geometry, and (b) is the DFTB3/CPE($q$)-D3 minimized geometry.

## S5: Thiomethoxide polarizability in solution

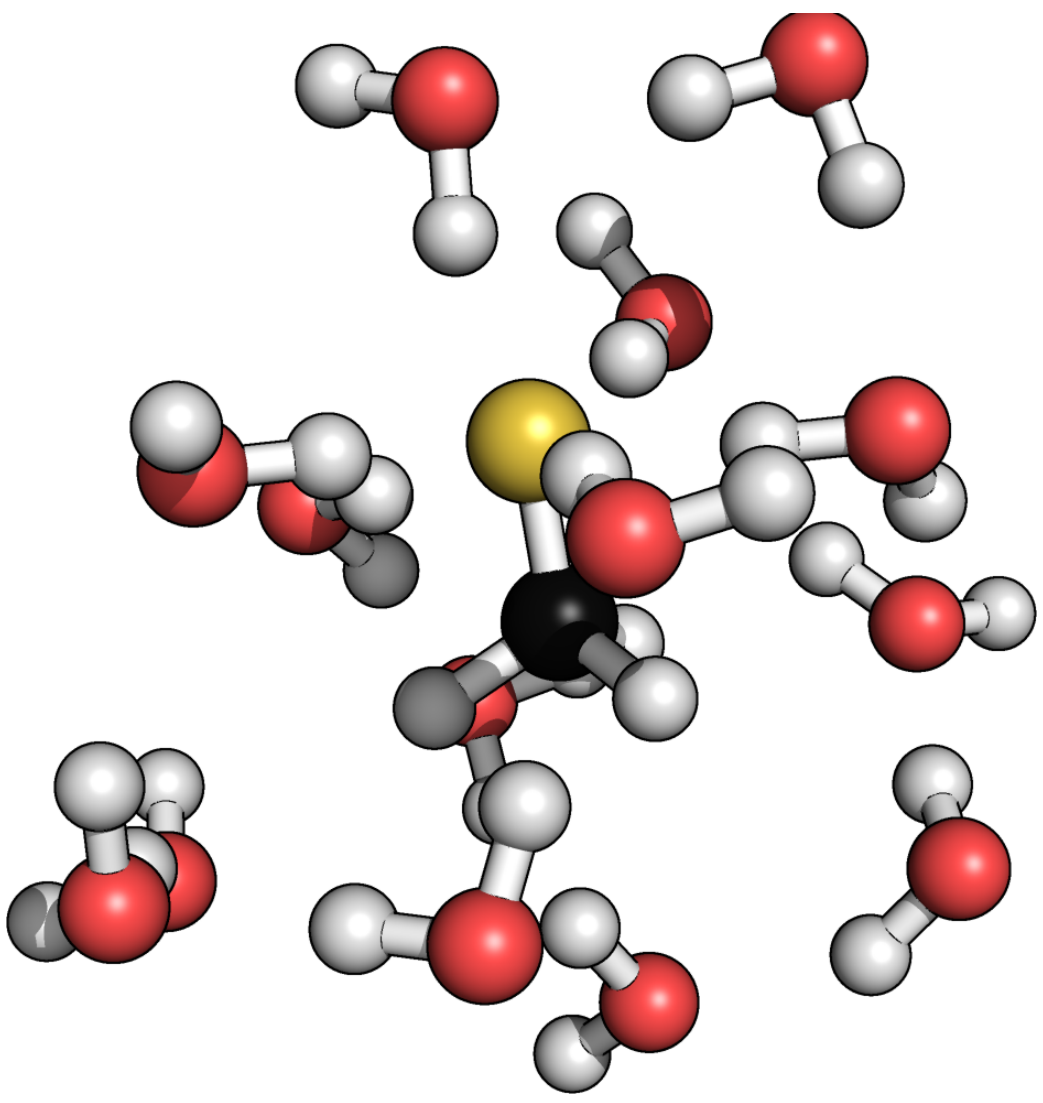

Table S17: Polarizability of a thiomethoxide anion in the gasp phase and in a droplet of 14 water molecules. Polarizabilities are given in bohr$^3$. The polarizability is greatly overestimated in the gas phase, but is very well described when the charge is stabilized by the surrounding water molecules.

| | Gas phase | Solution |
|---|---|---|
| DFTB/CPE($q$) | 414 bohr$^3$ | 192 bohr$^3$ |
| B3LYP/aug-cc-pVDZ | 66 bohr$^3$ | 177 bohr$^3$ |

# S6: DFTB2/CPE energy

## Notation

Summation over atom centers is denoted by summation over indices $a, b, c$, etc. Summation over AO-basis functions is denoted by summation over indices such as $\mu, \nu$, etc. The notation "$\mu \in a$" means that the basis function $\mu$ is centered on the atom $a$.

In the following sections we let $q_a$ be the *Mulliken population* (always a strictly positive quantity), defined as:

$$q_a = \sum_i^{\text{occ}} n_i \sum_{\mu \in a} \sum_b \sum_{\nu \in b} C_{\mu i} C_{\nu i} S_{\mu\nu} \tag{17}$$

Similarly the *partial Mulliken charge* is defined as

$$\Delta q_a = q_a^0 - q_a \tag{18}$$

All the equations are presented for DFTB2, but the presented derived terms are the same for DFTB3, since the extra DFTB3 energy terms are additive. Where the DFTB2 and DFTB3 CPE equations differ, this is explicitly noted.

## CPE energy

The CPE energy is given by:[S10]

$$E_{\text{cpe}} = \mathbf{c}^T \cdot \mathbf{M} \cdot \mathbf{q} + \frac{1}{2}\mathbf{c}^T \cdot \mathbf{N} \cdot \mathbf{c}, \tag{19}$$

where the first order CPE-DFTB2 Coulomb interaction matrix elements are given by:

$$M_{ij} = f(R_{ij}) \iint \frac{\phi_i^{\text{cpe}}(\mathbf{r})\,\phi_j^{\text{dftb2}}(\mathbf{r}')}{|\mathbf{r}-\mathbf{r}'|}\mathrm{d}\mathbf{r}\mathrm{d}\mathbf{r}' \tag{20}$$

and the second order CPE-CPE Coulomb interaction matrix elements are given by:

$$N_{ij} = \iint \frac{\phi_i^{\text{cpe}}(\mathbf{r})\,\phi_j^{\text{cpe}}(\mathbf{r}')}{|\mathbf{r}-\mathbf{r}'|}\mathrm{d}\mathbf{r}\mathrm{d}\mathbf{r}' \tag{21}$$

The CPE basis functions depend on the Mulliken population, while the DFTB basis functions only have a charge dependence in DFTB3.

The set of coefficients of the CPE response density basis that variationally minimizes the total CPE energy in Eqn. 19 is given (analytically) by:

$$\mathbf{c} = -\mathbf{N}^{-1} \cdot \mathbf{M} \cdot \mathbf{q} \tag{22}$$

Using the above relation, the CPE energy can be recast into:

$$\begin{aligned}
E_{\text{cpe}} &= \mathbf{c}^T \cdot \mathbf{M} \cdot \mathbf{q} + \frac{1}{2}\mathbf{c}^T \cdot \mathbf{N} \cdot \mathbf{c} && (23)\\
&= -(\mathbf{N}^{-1} \cdot \mathbf{M} \cdot \mathbf{q})^T \cdot \mathbf{M} \cdot \mathbf{q} + \frac{1}{2} + (\mathbf{N}^{-1} \cdot \mathbf{M} \cdot \mathbf{q})^T \cdot \mathbf{N} \cdot (\mathbf{N}^{-1} \cdot \mathbf{M} \cdot \mathbf{q}) && (24)\\
&= -(\mathbf{N}^{-1} \cdot \mathbf{M} \cdot \mathbf{q})^T \cdot \mathbf{M} \cdot \mathbf{q} + \frac{1}{2}(\mathbf{N}^{-1} \cdot \mathbf{M} \cdot \mathbf{q})^T \cdot \mathbf{M} \cdot \mathbf{q} && (25)\\
&= -\frac{1}{2}(\mathbf{N}^{-1} \cdot \mathbf{M} \cdot \mathbf{q})^T \cdot \mathbf{M} \cdot \mathbf{q} && (26)
\end{aligned}$$

## DFTB2 energy

The DFTB2 energy is given by:[S11]

$$E_{\text{dftb2}} = \sum_i^{\text{occ}} n_i \sum_\mu \sum_\nu C_{\mu i} C_{\nu i} H^0_{\mu\nu} + \frac{1}{2} \sum_{ab} \Delta q_a \Delta q_b \gamma_{ab} + \frac{1}{2} \sum_{ab} V^{\text{rep}}_{ab} \tag{27}$$

The DFTB2 Hamiltonian matrix elements are given by:[S11]

$$H^{(\text{dftb2})}_{\mu\nu} = H^0_{\mu\nu} + \frac{1}{2} S_{\mu\nu} \sum_c \left(\gamma_{ac} + \gamma_{bc}\right) \Delta q_c \tag{28}$$

## Combined DFTB2/CPE energy

The full DFTB2/CPE energy is given by:

$$\begin{aligned} E_{\text{dftb2/cpe}} &= \sum_i^{\text{occ}} n_i \sum_\mu \sum_\nu C_{\mu i} C_{\nu i} H^{(\text{dftb2})}_{\mu\nu} + \frac{1}{2} \sum_{ab} V^{\text{rep}}_{ab} + E_{\text{cpe}} \\ &= \underbrace{\sum_i^{\text{occ}} n_i \sum_\mu \sum_\nu C_{\mu i} C_{\nu i} H^0_{\mu\nu}}_{E_{\text{H0}}} + \underbrace{\frac{1}{2} \sum_{ab} \Delta q_a \Delta q_b \gamma_{ab}}_{E_\gamma} + \underbrace{\frac{1}{2} \sum_{ab} V^{\text{rep}}_{ab}}_{E_{\text{rep}}} + E_{\text{cpe}} \qquad (29) \\ &= E_{\text{H0}} + E_\gamma + E_{\text{rep}} + E_{\text{cpe}} \qquad (30) \end{aligned}$$

The CPE Hamiltonian shift is given by:[S12]

$$\Delta H^{(\text{cpe})}_{\mu\nu} = \frac{1}{2} S_{\mu\nu} \left( \frac{\partial E_{\text{cpe}}\left[\mathbf{q}, \mathbf{c}\right]}{\partial q_a} + \frac{\partial E_{\text{cpe}}\left[\mathbf{q}, \mathbf{c}\right]}{\partial q_b} \right) \qquad \mu \in a, \nu \in b \tag{31}$$

Note that here $q_a$ and $q_b$ are the Mulliken populations.

The occupied orbital energies is given in the terms of the (optimized) coefficients and the matrix elements mentioned previously:

$$\begin{aligned}
\sum_{i}^{\mathrm{occ}} n_i \varepsilon_i &= \sum_{i}^{\mathrm{occ}} n_i \sum_{\mu} \sum_{\nu} C_{\mu i} C_{\nu i} H_{\mu\nu} \\
&= \sum_{i}^{\mathrm{occ}} n_i \sum_{\mu} \sum_{\nu} C_{\mu i} C_{\nu i} \left( H_{\mu\nu}^{(\mathrm{dftb2})} + \Delta H_{\mu\nu}^{(\mathrm{cpe})} \right) \\
&= \sum_{i}^{\mathrm{occ}} n_i \sum_{\mu} \sum_{\nu} C_{\mu i} C_{\nu i} H_{\mu\nu}^{0} + \frac{1}{2} \sum_{i}^{\mathrm{occ}} n_i \sum_{a} \sum_{\mu \in a} \sum_{b} \sum_{\nu \in b} C_{\mu i} C_{\nu i} S_{\mu\nu} \sum_{c} \left( \gamma_{ac} + \gamma_{bc} \right) \Delta q_c \\
&\quad + \frac{1}{2} \sum_{i}^{\mathrm{occ}} n_i \sum_{a} \sum_{\mu \in a} \sum_{b} \sum_{\nu \in b} C_{\mu i} C_{\nu i} S_{\mu\nu} \left( \frac{\partial E_{\mathrm{cpe}} \left[ \mathbf{q}, \mathbf{c} \right]}{\partial q_a} + \frac{\partial E_{\mathrm{cpe}} \left[ \mathbf{q}, \mathbf{c} \right]}{\partial q_b} \right)
\end{aligned} \tag{32}$$

Using the relation above, the energy can be calculated in terms of the orbital energies (as implemented in CHARMM), by isolating $E_{\mathrm{H0}}$ in Eqn. 32 and inserting into Eqn. 30.

$$\begin{aligned}
E_{\mathrm{dftb2/cpe}} &= \sum_{i}^{\mathrm{occ}} n_i \varepsilon_i \underbrace{- \frac{1}{2} \sum_{i}^{\mathrm{occ}} n_i \sum_{a} \sum_{\mu \in a} \sum_{b} \sum_{\nu \in b} C_{\mu i} C_{\nu i} S_{\mu\nu} \sum_{c} \left( \gamma_{ac} + \gamma_{bc} \right) \Delta q_c - \frac{1}{2} \sum_{i}^{\mathrm{occ}} n_i \sum_{a} \sum_{\mu \in a} \sum_{b} \sum_{\nu \in b} C_{\mu i} C_{\nu i} S_{\mu\nu} \left( \frac{\partial E_{\mathrm{cpe}} \left[ \mathbf{q}, \mathbf{c} \right]}{\partial q_a} + \frac{\partial E_{\mathrm{cpe}} \left[ \mathbf{q}, \mathbf{c} \right]}{\partial q_b} \right)}_{E_{\mathrm{shift}}} \\
&\quad + \frac{1}{2} \sum_{ab} \Delta q_a \Delta q_b \gamma_{ab} + \frac{1}{2} \sum_{ab} V_{ab}^{\mathrm{rep}} + E_{\mathrm{cpe}}
\end{aligned} \tag{33}$$

The $E_{\mathrm{shift}}$ must be subtracted from the electronic energy to compensate for double counting when adding $E_{\mathrm{cpe}}$ to the electronic energy in terms of the occupied orbital energies.

The following relation is useful:

$$\begin{aligned}
E_{\text{shift}} &= -\frac{1}{2}\sum_{i}^{\text{occ}} n_i \sum_{a}\sum_{\mu\in a}\sum_{b}\sum_{\nu\in b} C_{\mu i}C_{\nu i}S_{\mu\nu}\left(\frac{\partial E_{\text{cpe}}\left[\mathbf{q},\mathbf{c}\right]}{\partial q_a}+\frac{\partial E_{\text{cpe}}\left[\mathbf{q},\mathbf{c}\right]}{\partial q_b}\right)\\
&= -\frac{1}{2}\sum_{i}^{\text{occ}} n_i \sum_{a}\sum_{\mu\in a}\sum_{b}\sum_{\nu\in b} C_{\mu i}C_{\nu i}S_{\mu\nu}\frac{\partial E_{\text{cpe}}\left[\mathbf{q},\mathbf{c}\right]}{\partial q_a}\\
&\quad -\frac{1}{2}\sum_{i}^{\text{occ}} n_i \sum_{a}\sum_{\mu\in a}\sum_{b}\sum_{\nu\in b} C_{\mu i}C_{\nu i}S_{\mu\nu}\frac{\partial E_{\text{cpe}}\left[\mathbf{q},\mathbf{c}\right]}{\partial q_b}\\
&= -\sum_{i}^{\text{occ}} n_i \sum_{a}\sum_{\mu\in a}\sum_{b}\sum_{\nu\in b} C_{\mu i}C_{\nu i}S_{\mu\nu}\frac{\partial E_{\text{cpe}}\left[\mathbf{q},\mathbf{c}\right]}{\partial q_a}\\
&= -\sum_{a}\frac{\partial E_{\text{cpe}}\left[\mathbf{q},\mathbf{c}\right]}{\partial q_a} q_a
\end{aligned} \tag{34}$$

Using the above relation, the DFTB2/CPE energy can be simplified to:

$$E_{\text{dftb2/cpe}} = \sum_{i}^{\text{occ}} n_i\varepsilon_i - \frac{1}{2}\sum_{ab}\left(q_a + q_a^0\right)\Delta q_b\gamma_{ab} - \sum_{a}\frac{\partial E_{\text{cpe}}\left[\mathbf{q},\mathbf{c}\right]}{\partial q_a} q_a + E_{\text{cpe}} \tag{35}$$

# DFTB2/CPE energy gradient

The energy gradient is the derivative of the energy with respect to the nuclear coordinates under the following constraint:[S13]

$$-F_{kx} = \frac{\partial}{\partial R_{kx}}\left[E_{\text{dftb2/cpe}} - \sum_{i}^{\text{occ}} n_i\varepsilon_i\left(\sum_{\mu}\sum_{\nu} C_{\mu i}C_{\nu i}S_{\mu\nu} - 1\right)\right] \tag{36}$$

Using the notation of Eq. 30, we can rewrite this as:

$$-F_{kx} = \frac{\partial}{\partial R_{kx}}\left[E_{\text{H0}} + E_{\gamma} + E_{\text{rep}} + E_{\text{cpe}} - \sum_{i}^{\text{occ}} n_i\varepsilon_i\left(\sum_{\mu}\sum_{\nu} C_{\mu i}C_{\nu i}S_{\mu\nu} - 1\right)\right] \tag{37}$$

Separating the terms that appear in the standard DFTB2 gradient, and the DFTB2/CPE gradient, we arrive at the CPE gradient correction:

$$
\begin{aligned}
-F_{kx} &= \frac{\partial}{\partial R_{kx}} \left[ E_{\mathrm{H0}} + E_{\gamma} + E_{\mathrm{rep}} + E_{\mathrm{cpe}} - \sum_{i}^{\mathrm{occ}} n_i \varepsilon_i \left( \sum_{\mu} \sum_{\nu} C_{\mu i} C_{\nu i} S_{\mu\nu} - 1 \right) \right] \\
&= \underbrace{\frac{\partial}{\partial R_{kx}} \left[ E_{\mathrm{H0}} + E_{\gamma} + E_{\mathrm{rep}} - \sum_{i}^{\mathrm{occ}} n_i \varepsilon_i \left( \sum_{\mu} \sum_{\nu} C_{\mu i} C_{\nu i} S_{\mu\nu} - 1 \right) \right]}_{\text{Same as the DFTB2 gradient}} + \frac{\partial}{\partial R_{kx}} E_{\mathrm{cpe}} \\
&= -F_{kx}^{(\mathrm{dftb2})} + \frac{\partial E_{\mathrm{cpe}}}{\partial R_{kx}}
\end{aligned} \tag{38}
$$

The last term is new, and is presented in the next sections.

## CPE gradient: $-F_{kx}^{(\mathrm{cpe})} = \frac{\partial E_{\mathrm{cpe}}}{\partial R_{kx}}$

The CPE energy depends explicitly on the coordinates via the Coulomb integrals, and implicitly on the coordinates via the CPE coefficients and the Mulliken population:

$$
\begin{aligned}
-F_{kx}^{(\mathrm{cpe})} &= \frac{\partial E_{\mathrm{cpe}}\left(R_{kx}\right)}{\partial R_{kx}} \\
&= \sum_{i} \frac{\partial E_{\mathrm{cpe}}\left(\mathbf{q}, c_i(R_{kx})\right)}{\partial c_i} \frac{\partial c_i(R_{kx})}{\partial R_{kx}} + \sum_{a} \frac{\partial E_{\mathrm{cpe}}\left(q_a(R_{kx}), \mathbf{c}\right)}{\partial q_a} \frac{\partial q_a(R_{kx})}{\partial R_{kx}} + \frac{\partial E_{\mathrm{cpe}}\left[\mathbf{q}, \mathbf{c}\right]}{\partial R_{kx}}
\end{aligned} \tag{39}
$$

The first derivative term is zero, since the CPE energy is variationally optimized with respect to the $\mathbf{c}$ coefficients.

### Dependence on $q_a$

The second term can be divided into two factors. The first factor can be calculated as described in the previous section (it is the same term as found in the Hamiltonian-shift.) - i.e. the derivatives with respect to the Mulliken populations. In the CPE charge-independent

case:

$$\frac{\partial E_{\text{cpe}}\left[\mathbf{q},\mathbf{c}\right]}{\partial q_a} = \left[\mathbf{c}^{\mathbf{T}} \cdot \mathbf{M}\right]_a \tag{40}$$

And in the CPE charge-dependent case:

$$\frac{\partial E_{\text{cpe}}\left[\mathbf{q},\mathbf{c}\right]}{\partial q_a} = \mathbf{c}^{\mathbf{T}} \cdot \left(\frac{\partial \mathbf{M}}{\partial q_a}\right) \cdot \mathbf{q} + \left[\mathbf{c}^{\mathbf{T}} \cdot \mathbf{M}\right]_a + \frac{1}{2}\mathbf{c}^T \cdot \left(\frac{\partial \mathbf{N}}{\partial q_a}\right) \cdot \mathbf{c}. \tag{41}$$

The second factor can be calculated for $a \neq k$ by:[S13]

$$\frac{\partial q_{a\neq k}}{\partial R_{kx}} = \sum_i^{\text{occ}} n_i \sum_{\mu\in a} \sum_{\nu\in k} C_{\mu i} C_{\nu i} \frac{\partial S_{\mu\nu}}{\partial R_{kx}} \tag{42}$$

or for $a = k$:

$$\frac{\partial q_k}{\partial R_{kx}} = \sum_i^{\text{occ}} n_i \sum_{\mu\in k} \sum_{\nu\notin k} C_{\mu i} C_{\nu i} \frac{\partial S_{\mu\nu}}{\partial R_{kx}} \tag{43}$$

This term must be calculated in the DFTB2 gradient code, where the derivative of the overlap matrix is already being calculated.

**(Explicit) dependence on $R_{kx}$**

The derivative with respect to $R_{kx}$ is written via the matrix derivatives.

$$\frac{\partial E_{\text{cpe}}\left[\mathbf{q},\mathbf{c}\right]}{\partial R_{kx}} = \mathbf{c}^{\mathbf{T}} \cdot \left(\frac{\partial \mathbf{M}}{\partial R_{kx}}\right) \cdot \mathbf{q} + \frac{1}{2}\mathbf{c}^T \cdot \left(\frac{\partial \mathbf{N}}{\partial R_{kx}}\right) \cdot \mathbf{c}. \tag{44}$$

## DFTB2 electric field contribution

The energy in an electric field $\vec{F}$ is given by the interaction of the field with the partial charges:

$$
\begin{aligned}
E_{\mathrm{EF/dftb2}} &= \sum_i^{\mathrm{occ}} n_i \sum_\mu \sum_\nu C_{\mu i} C_{\nu i} H^0_{\mu\nu} + \frac{1}{2}\sum_{ab} \Delta q_a \Delta q_b \gamma_{ab} + \frac{1}{2}\sum_{ab} V^{\mathrm{rep}}_{ab} - \sum_a \Delta q_a \ \vec{F}\cdot\vec{r}_a \\
&= \sum_i^{\mathrm{occ}} n_i \sum_\mu \sum_\nu C_{\mu i} C_{\nu i} H^0_{\mu\nu} + \frac{1}{2}\sum_{ab} \Delta q_a \Delta q_b \gamma_{ab} + \frac{1}{2}\sum_{ab} V^{\mathrm{rep}}_{ab} \qquad (45)\\
&+ \underbrace{\sum_a q_a \ \vec{F}\cdot\vec{r}_a}_{\Delta E_{\mathrm{EF/dftb2}}} - \sum_a q^0_a \ \vec{F}\cdot\vec{r}_a \qquad (46)
\end{aligned}
$$

where $q_a$ are the Mulliken populations, $\Delta q_a$ are the partial Mulliken charges, and $q^0_a$ are the charges of the nuclei. Before deriving the Hamiltonian we note the following relation:

$$
\frac{\partial \Delta E_{\mathrm{EF/dftb2}}}{\partial q_a} = \frac{\partial}{\partial q_a}\sum_{a'} q_{a'} \ \vec{F}\cdot\vec{r}_{a'} = \vec{F}\cdot\vec{r}_a \qquad (47)
$$

Combining the above with Eq. 67, the corresponding Hamiltonian element to be added to the DFTB2 Hamiltonian element is given by:

$$
\begin{aligned}
\Delta H^{(\mathrm{EF/dftb2})}_{\mu\nu} &\equiv \frac{\partial \Delta E_{\mathrm{EF/dftb2}}}{\partial \rho_{\mu\nu}} \\
&= \sum_a \frac{\partial \Delta E_{\mathrm{EF/dftb2}}}{\partial q_a}\frac{\partial q_a}{\partial \rho_{\mu\nu}} \\
&= \frac{1}{2} S_{\mu\nu}\left(\vec{r}_a + \vec{r}_b\right)\cdot\vec{F} \qquad \mu\in a, \nu\in b \qquad (48)
\end{aligned}
$$

So the matrix elements of the DFTB2 Hamiltonian matrix in the presence of an electric field are:

$$
H_{\mu\nu} = H^0_{\mu\nu} + \frac{1}{2} S_{\mu\nu}\sum_c \left(\gamma_{ac} + \gamma_{bc}\right)\Delta q_c + \frac{1}{2} S_{\mu\nu}\left(\vec{r}_a + \vec{r}_b\right)\cdot\vec{F} \qquad \mu\in a, \nu\in b \qquad (49)
$$

The DFTB2 energy in an electric field has the following orbital energies:

$$
\begin{aligned}
\sum_i^{\text{occ}} n_i \varepsilon_i &= \sum_i^{\text{occ}} n_i \sum_\mu \sum_\nu C_{\mu i} C_{\nu i} H_{\mu\nu} \\
&= \sum_i^{\text{occ}} n_i \sum_\mu \sum_\nu C_{\mu i} C_{\nu i} \left( H_{\mu\nu}^{(\text{dftb2})} + \Delta H_{\mu\nu}^{(\text{EF/dftb2})} \right) \\
&= \sum_i^{\text{occ}} n_i \sum_\mu \sum_\nu C_{\mu i} C_{\nu i} H_{\mu\nu}^0 + \frac{1}{2} \sum_i^{\text{occ}} n_i \sum_a \sum_{\mu \in a} \sum_b \sum_{\nu \in b} C_{\mu i} C_{\nu i} S_{\mu\nu} \sum_c (\gamma_{ac} + \gamma_{bc}) \Delta q_c \\
&\quad + \frac{1}{2} \sum_i^{\text{occ}} n_i \sum_a \sum_{\mu \in a} \sum_b \sum_{\nu \in b} C_{\mu i} C_{\nu i} S_{\mu\nu} (\vec{r}_a + \vec{r}_b) \cdot \vec{F}
\end{aligned}
\tag{50}
$$

Isolating $H_0$ in the above and inserting into Eq. 46 the DFTB2 energy in the presence of an external field is written in terms of the orbital energies:

$$
\begin{aligned}
E_{\text{EF/dftb2}} &= \sum_i^{\text{occ}} n_i \varepsilon_i - \frac{1}{2} \sum_i^{\text{occ}} n_i \sum_a \sum_{\mu \in a} \sum_b \sum_{\nu \in b} C_{\mu i} C_{\nu i} S_{\mu\nu} \sum_c (\gamma_{ac} + \gamma_{bc}) \Delta q_c \\
&\quad - \frac{1}{2} \sum_i^{\text{occ}} n_i \sum_a \sum_{\mu \in a} \sum_b \sum_{\nu \in b} C_{\mu i} C_{\nu i} S_{\mu\nu} (\vec{r}_a + \vec{r}_b) \cdot \vec{F} + \frac{1}{2} \sum_{ab} V_{ab}^{\text{rep}} + \sum_a q_a \ \vec{F} \cdot \vec{r}_a \\
&= \sum_i^{\text{occ}} n_i \varepsilon_i - \frac{1}{2} \sum_{ab} \left( q_a + q_a^0 \right) \Delta q_b \gamma_{ab} + \frac{1}{2} \sum_{ab} V_{ab}^{\text{rep}} - \sum_a q_a^0 \ \vec{F} \cdot \vec{r}_a
\end{aligned}
\tag{51}
$$

So only the nuclear charge term has to be added to the energy expressed in terms of the orbital energies.

## DFTB2 dipole moment

The dipole moment of a molecule is given by the DFTB2 Mulliken partial charges.

$$
\vec{\mu}^{(dftb2)} = \sum_a \Delta q_a \ \vec{r}_a = \sum_a q_a^0 \ \vec{r}_a - \sum_a q_a \ \vec{r}_a \tag{52}
$$

## DFTB2/CPE electric field contribution

The CPE-dipole functions interact directly with an external electric field and enter the energy CPE energy as:

$$E_{\text{EF/cpe}} = \mathbf{c}^T \cdot (\mathbf{M} \cdot \mathbf{q} - \mathbf{f}) + \frac{1}{2}\mathbf{c}^T \cdot \mathbf{N} \cdot \mathbf{c}, \tag{53}$$

where $\mathbf{f}$ is a $3N$ vector containing $N$ repeats of the components of the electric field, i.e.

$$\mathbf{f} = \begin{bmatrix} F_x \\ F_y \\ F_z \\ F_x \\ F_y \\ F_z \\ \vdots \end{bmatrix} \tag{54}$$

In the electric field, the variational analytical solution of the coefficients becomes:

$$\mathbf{c} = -\mathbf{N}^{-1} \cdot (\mathbf{M} \cdot \mathbf{q} - \mathbf{f}) \tag{55}$$

The DFTB2/CPE energy in the electric field is simply:

$$E_{\text{EF/dftb2/cpe}} = \sum_i^{\text{occ}} n_i \sum_\mu \sum_\nu C_{\mu i} C_{\nu i} H^0_{\mu\nu} + \frac{1}{2}\sum_{ab} \Delta q_a \Delta q_b \gamma_{ab} + \frac{1}{2}\sum_{ab} V^{\text{rep}}_{ab} - \sum_a \Delta q_a\, \vec{F} \cdot \vec{r}_a + E_{\text{EF/cpe}} \tag{56}$$

Since no additional terms from the CPE basis/electric field interaction enter the Hamiltonian matrix (since this interaction does not depend on $\mathbf{q}$), we can write the DFTB2/CPE energy in the presence of an external field in terms of the orbital energies as:

$$\begin{aligned} E_{\mathrm{EF/dftb2/cpe}} &= \sum_i^{\mathrm{occ}} n_i \varepsilon_i - \frac{1}{2} \sum_{ab} \left(q_a + q_a^0\right) \Delta q_b \gamma_{ab} + \frac{1}{2} \sum_{ab} V_{ab}^{\mathrm{rep}} - \sum_a \Delta q_a \ \vec{F} \cdot \vec{r}_a \\ & - \sum_a \frac{\partial E_{\mathrm{EF/cpe}} \left[\mathbf{q}, \mathbf{c}\right]}{\partial q_a} q_a + \ E_{\mathrm{EF/cpe}} \end{aligned} \tag{57}$$

Again, $q_a$ is a Mulliken *population* and $\Delta q_a$ is the partial Mulliken *charge*.

### DFTB2/CPE dipole moment

We define $\vec{\mu}_a^{(\mathrm{cpe})}$ as the total dipole due to the CPE-basis functions centered on atom $a$. This is given by the coefficients to the same functions:

$$\vec{\mu}_a^{(\mathrm{cpe})} = \begin{bmatrix} c_{ax} \\ c_{ay} \\ c_{az} \end{bmatrix} \tag{58}$$

where $c_{ax}$ is the coefficient of the dipole function centered on atom $a$ in the $x$-direction, and so on. The total dipole-moment of the molecule in the DFTB2/CPE description is now:

$$\vec{\mu}^{(\mathrm{dftb2/cpe})} = \sum_a \Delta q_a \ \vec{r}_a + \sum_a \vec{\mu}_a^{(\mathrm{cpe})} \tag{59}$$

## DFTB2 and DFTB2/CPE polarizability

The elements of the polarizability tensor is calculated as:

$$\alpha_{ij} = \left(\frac{\partial \mu_i}{\partial F_j}\right)_{\vec{F}=0} = - \left(\frac{\partial^2 E}{\partial F_i \partial F_j}\right)_{\vec{F}=0} \tag{60}$$

where $i$ and $j$ are the $x$, $y$, $z$ Cartesian components and $\mu_i$ is the $i$-component of the DFTB2 or DFTB2/CPE dipole moment. The partial derivatives are calculated numerically by means of the two-point, forward-backtrack finite differences method, e.g.:

$$\alpha_{ij} = \frac{\mu_i\left(F_j + h\right) - \mu_i\left(F_j - h\right)}{2h} \tag{61}$$

The isotropic polarizability used in the fitting routine is calculated as:

$$\alpha_{\text{iso}} = \frac{1}{3}\left(\alpha_{xx} + \alpha_{yy} + \alpha_{zz}\right) \tag{62}$$

The anisotropic polarizability can be calculated using:

$$\alpha_{\text{aniso}}^2 = \frac{1}{2}\left(\left(\alpha_{xx} - \alpha_{yy}\right)^2 + \left(\alpha_{yy} - \alpha_{zz}\right)^2 + \left(\alpha_{zz} - \alpha_{xx}\right)^2 + 6\alpha_{xz}^2 + 6\alpha_{yz}^2 + 6\alpha_{zx}^2\right) \tag{63}$$

## Molecular orientation

Reference dipole moment and polarizability calculations were carried out in Gaussian 09 using the center of nuclear charge as center of the coordinate system (default behavior). Likewise, the CHARMM/DFTB implementation translates the molecule to the center of nuclear charge before the dipole moment is calculated.

# Hamiltonian shifts

## Mulliken population

First we define the density matrix is defined in terms of the coefficient matrix:

$$\rho_{\mu\nu} = \sum_{i}^{\text{occ}} n_i C_{\mu i} C_{\nu i} \tag{64}$$

We can then define the Mulliken population by the density matrix:

$$\begin{aligned} q_a &= \sum_{i}^{\text{occ}} n_i \sum_{\mu \in a} \sum_{b} \sum_{\nu \in b} C_{\mu i} C_{\nu i} S_{\mu\nu} \\ &= \frac{1}{2} \sum_{\mu \in a} \sum_{b} \sum_{\nu \in b} \left( \rho_{\mu\nu} + \rho_{\nu\mu} \right) S_{\mu\nu} \end{aligned} \tag{65}$$

The charge fluctuation is then:

$$\Delta q_a = q_a - q_a^0 \tag{66}$$

Using these definitions we derive the following relation:

$$\frac{\partial q_a}{\partial \rho_{\mu\nu}} = \frac{\partial}{\partial \rho_{\mu\nu}} \left[ \frac{1}{2} \sum_{\mu \in a} \sum_{b} \sum_{\nu \in b} \left( \rho_{\mu\nu} + \rho_{\nu\mu} \right) S_{\mu\nu} \right] = \begin{cases} \frac{1}{2} S_{\mu\nu} & \mu \in a \\ \frac{1}{2} S_{\mu\nu} & \nu \in a \\ S_{\mu\nu} & \mu \in a \quad \text{and} \quad \nu \in a \end{cases} \tag{67}$$

Note that the *Mulliken population* is always a positive number, and the *Mulliken charge* is the negative of the Mulliken population.

## CPE Hamiltonian shift

Using the relation from section in the previous section, the CPE Hamiltonian shift is calculated using the chain rule:

$$\Delta H_{\mu\nu}^{(\mathrm{cpe})} = \frac{\partial E_{\mathrm{cpe}}[\mathbf{q},\mathbf{c}]}{\partial \rho_{\mu\nu}} \tag{68}$$

$$= \sum_i \frac{\partial E_{\mathrm{cpe}}[\mathbf{q},\mathbf{c}]}{\partial c_i}\frac{\partial c_i}{\partial \rho_{\mu\nu}} + \sum_a \frac{\partial E_{\mathrm{cpe}}[\mathbf{q},\mathbf{c}]}{\partial q_a}\frac{\partial q_a}{\partial \rho_{\mu\nu}} \tag{69}$$

$$= \frac{1}{2} S_{\mu\nu}\left(\frac{\partial E_{\mathrm{cpe}}[\mathbf{q},\mathbf{c}]}{\partial q_a} + \frac{\partial E_{\mathrm{cpe}}[\mathbf{q},\mathbf{c}]}{\partial q_b}\right) \qquad \mu \in a, \nu \in b \tag{70}$$

The CPE energy has no parametric dependence on the density, but depends implicitly via the charges and coefficients. The first term is zero due to the CPE energy being variationally minimized with respect to the coefficients. The first part of the second term is in the CPE charge-independent case:

$$\frac{\partial E_{\mathrm{cpe}}[\mathbf{q},\mathbf{c}]}{\partial q_a} = [\mathbf{c}^{\mathbf{T}} \cdot \mathbf{M}]_a \tag{71}$$

In the CPE charge-dependent case:

$$\frac{\partial E_{\mathrm{cpe}}[\mathbf{q},\mathbf{c}]}{\partial q_a} = \mathbf{c}^{\mathbf{T}} \cdot \left(\frac{\partial \mathbf{M}}{\partial q_a}\right) \cdot \mathbf{q} + [\mathbf{c}^{\mathbf{T}} \cdot \mathbf{M}]_a + \frac{1}{2}\mathbf{c}^{T} \cdot \left(\frac{\partial \mathbf{N}}{\partial q_a}\right) \cdot \mathbf{c}. \tag{72}$$

See Giese and York (2012) for further details.[S12]